\documentclass[journal=nalefd,manuscript=letter]{achemso}
\usepackage[version=3]{mhchem} 
\usepackage{epsfig}
\usepackage{color}
\usepackage{amsmath}
\usepackage[colorlinks,linkcolor=blue,citecolor=blue,urlcolor=blue]{hyperref}

\usepackage[english]{babel}
\usepackage{float}
\usepackage{graphicx}

\author{Luka Benić}
\affiliation{Ruđer Bošković Institute, Zagreb, Croatia}
\alsoaffiliation{University of Zagreb, Zagreb, Croatia}

\author{Dino Novko}
\email{dino.novko@gmail.com}
\affiliation{Centre for Advanced Laser Techniques, Institute of Physics, Zagreb, Croatia}

\author{Ivor Lončarić}
\email{ivor.loncaric@gmail.com}
\affiliation{Ruđer Bošković Institute, Zagreb, Croatia}

\title{Fluctuation-driven multi-step charge density wave melting in monolayer TiSe$_2$}

\begin{document}

\begin{abstract}
The exact microscopic origin, symmetry, and thermal melting mechanism of the charge density wave (CDW) phase in TiSe$_2$ remain intensely debated, particularly regarding chiral structural order and multi-step phase transitions. Using large-scale molecular dynamics driven by an accurate machine-learning interatomic potential, we resolve the finite-temperature structural dynamics of the monolayer TiSe$_2$. We demonstrate that CDW melting deviates from conventional mean-field second-order behavior, undergoing a two-step process characterized by an extended fluctuation region ($200-250$\,K) with proliferating domain walls, phase decoherence, component-selective melting, and an overdamped soft optical phonon. Furthermore, anisotropic long-wavelength thermal fluctuations spontaneously stabilize an asymmetric $3Q$ CDW order with $P1$ symmetry. These findings provide a unified microscopic framework for fluctuation-driven transitions in 2D quantum materials, demonstrating that the complex CDW physics of TiSe$_2$ is captured entirely via non-perturbative electron-phonon and phonon-phonon interactions.
\end{abstract}

\maketitle


Despite fifty years of research, the charge density wave (CDW) phase in TiSe$_{2}$ remains heavily debated regarding its microscopic origin, structural symmetry, and melting process\,\cite{disalvo76,rossnagel11}. The CDW is widely attributed to combined exciton condensation\,\cite{disalvo76,cercellier07,kogar2017a} and electron-phonon coupling\,\cite{rossnagel11,hughes1977,yoshida80,calandra11,novko22}, however, the driving electronic mode has not been clearly observed\,\cite{kogar2017a,lin22}, and theoretical studies incorporating electron-hole interactions dispute its existence\,\cite{lian2019,novko2025}. Conversely, the soft CDW phonon mode responsible for periodic lattice distortions (PLDs) was reported in x-ray scattering\,\cite{holt2001,weber2011} and Raman spectroscopy\,\cite{sugai1980} experiments, however, \emph{ab initio} theories with only phonon-driven processes were not able to capture the right experimental transition temperature $T_{\rm CDW}$\,\cite{bianco_tise2,chen2023}.

Furthermore, thermal melting of bulk (monolayer) was traditionally modeled as a conventional second-order transition from the $P\overline{3}c1$ ($P321$) structure to a high-symmetry $P\overline{3}m1$ phase at $T_{\rm CDW}\approx200$\,K\,\cite{disalvo76,rossnagel11}, supported by first-principles calculations\,\cite{bianco_tise2,chen2023,bianco2015}. However, recent experiments have revealed a much more complex melting process. The signatures of CDW fluctuations were shown to persist far beyond $T_{\rm CDW}$\,\cite{cercellier07,woo1976lattice,miyahara1995sts,guo2025,fragkos2026}, while ultrafast electron diffraction (UED) and scanning tunneling microscopy (STM) have identified structural domains, $\pi$ phase slips, and topological defects for the 2D limit of CDW and in monolayers\,\cite{cheng2024,liu2024}, all pointing to a strong-coupling theory of CDW melting\,\cite{mcmillan1977,varma1983}. Additionally, STM, photocurrent, and optical spectroscopy studies of bulk samples have highlighted asymmetric CDW vectors that break threefold rotation and potentially inversion symmetries\,\cite{ishioka2010,xu2020,tyulnev2025}. This suggests emergent chiral, ferroaxial, or nematic orders\,\cite{segovia2025,edwards2026,jiang2026} with separate transition temperatures below $T_{\rm CDW}$\,\cite{xu2020,tyulnev2025,edwards2026,jiang2026}.

Comprehending the microscopic origin of this intricate phase diagram and the multi-step and multi-component melting process of TiSe$_2$ and similar correlated materials containing CDW fluctuations, structural domains, and potentially chiral order is crucial, especially considering the impact that CDW has on unconventional superconductivity and other quantum ordered states\,\cite{chen2019,lv2026}. However, computational demand of large supercells has restricted full melting dynamics simulations to phenomenological and downfolding models\, \cite{mcmillan1976,vodeb2019,cheng2023,berges2024,shen2026}.

Here, we study the structural dynamics of monolayer TiSe$_2$ using large-scale molecular dynamics (MD) driven by an accurate machine-learning interatomic potential. We demonstrate that the melting process of the three-component ($3Q$) CDW order deviates from mean-field conventional second-order transition, and shows various features of the strong-coupling CDW theory\,\cite{mcmillan1977,varma1983}. We show that the CDW PLDs melt in two steps at $T^{\ast}\approx 200$\,K and $T_{\rm CDW}\approx 250$\,K, where the latter matches the monolayer transition temperature\,\cite{chen2015,chen2016,xrd_displ}. 
Our analysis of the CDW order parameter reveals that the region between $T^{\ast}$ and $T_{\rm CDW}$ is dominated by strong CDW fluctuations, translational symmetry-breaking domains, phase decoherence, and component-selective melting of the $3Q$ order. 
Furthermore, we show that the low-temperature $3Q$ order actually consists of three non-equivalent CDW displacement vectors with $P1$ symmetry, that break both twofold and threefold rotational symmetries.
Crucially, we show that long-wavelength thermal fluctuations (acoustic phonons) drive both the anisotropic $3Q$ order and two-step melting, accompanied by an overdamped soft CDW phonon between $T^{\ast}$ and $T_{\rm CDW}$. Because our interatomic potential is entirely based on PBE density-functional theory (DFT), our findings show that electron-phonon and phonon-phonon interactions\,\cite{yoshiyama86,varma1983,mcmillan1977} accurately describe the CDW dynamics in TiSe$_2$ without invoking excitonic correlations.

\begin{figure*}[ht!]
    \centering
    \includegraphics[width=\textwidth]{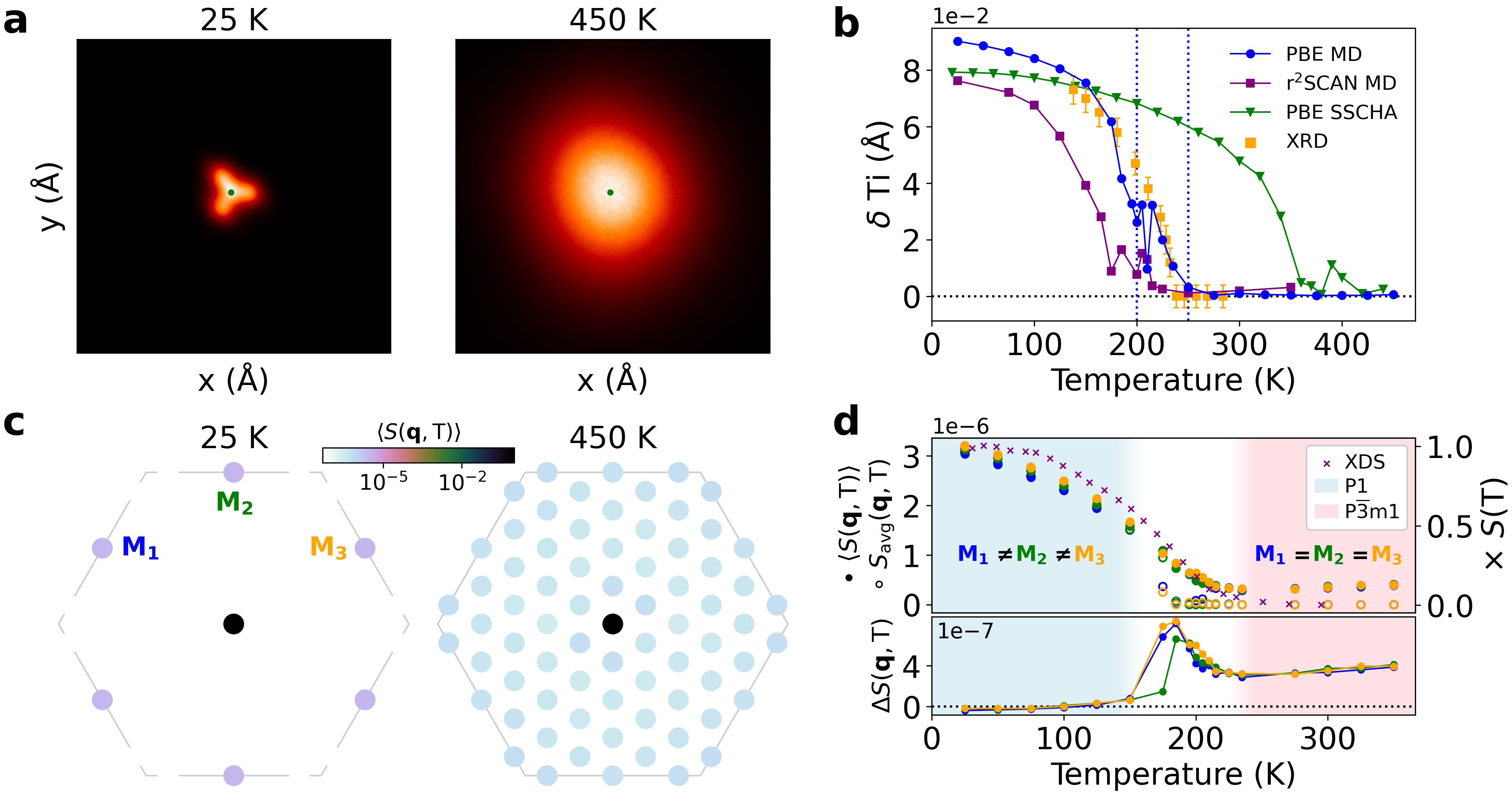}
    \caption{\textbf{Average displacements and structure factors across the CDW transition.} \textbf{a} Ti atomic positions from $56\times56\times1$ MD folded onto normal phase unit cell. Green points indicate the normal phase positions of Ti atoms. \textbf{b} Average displacements of Ti atoms in the $xy$ plane calculated from MD data using the PBE-based MACE potential and the one fine-tuned using r$^{2}$SCAN data, compared to SSCHA and XRD results for the monolayer\,\cite{xrd_displ}. \textbf{c} Time-averaged structure factors calculated from MD data, shown for the irreducible $\mathbf{q}$ points of the $8\times8\times1$ supercell for better visualization. \textbf{d} Top: time-averaged structure factors $\langle S(\mathbf{q},T)\rangle$ Eq.\,(S1) (filled circles) and structure factors of the time-averaged structure $S_{\mathrm{avg}}(\mathbf{q},T)$ Eq.\,(S13) (empty circles) calculated with the $56\times56\times1$ supercell as a function of temperature for $3$ different $\mathbf{q}=\mathrm{M}$ points indicated in \textbf{c} with the same colors. Crosses show x-ray diffuse scattering (XDS) results for bulk~\cite{guo2025} (right axis). Bottom: difference of the two structure factors $\Delta S(\mathbf{q},T)$ given by Eq.\,(S14), indicating CDW fluctuations. The light blue (pink) color indicates the region with the $P1$ ($P\overline{3}m1$) symmetry, while the white represents the CDW fluctuating region.}
    \label{fig:subplots_positions_structure_factors}
\end{figure*}

\textbf{Anisotropic $3Q$ CDW order, and strong-coupling melting to the normal phase.} 
We constructed a MACE interatomic potential for monolayer TiSe$_{2}$ trained on PBE data to perform MD simulations across varying ionic temperatures [see Supporting Information (SI) for computational details]. Figure\,\ref{fig:subplots_positions_structure_factors}a displays the resulting $xy$-plane atomic positions folded onto the normal-phase $1\times 1 \times 1$ unit cell. At low temperatures, the expected $3Q$ CDW phase emerges with seemingly threefold symmetry, featuring three displaced non-equivalent Ti atoms [i.e., $\mathrm{Ti^{(A)}}$, $\mathrm{Ti^{(B)}}$, and $\mathrm{Ti^{(C)}}$] and one fixed at equilibrium [i.e., $\mathrm{Ti^{(0)}}$]\,\cite{disalvo76}. As the temperature increases, strong thermal fluctuations destroy this $3Q$ signature, rendering it barely visible already at $100$\,K (Figs.\,S5, S6, and S7).

Figure\,\ref{fig:subplots_positions_structure_factors}b compares the average $xy$-plane Ti PLDs calculated from our MD and SSCHA\,\cite{Monacelli_2021} simulations (see SI for details) with the experimental x-ray diffraction (XRD) results done on monolayers\,\cite{xrd_displ} (see Fig.\,S8 for Se PLDs). We also show the results obtained with the potential fine-tuned using r$^2$SCAN data. XRD data suggest a conventional second-order melting process, following the mean-field functional form $\delta \mathrm{Ti}\propto\tanh\sqrt{T_{\rm CDW}/T - 1}$. Harmonic DFT calculations replicate this behavior (Fig.\,S9), however, since they completely neglect lattice thermal fluctuations, they severely overestimate the transition with $T_{\mathrm{CDW}}\approx 1150$\,K. By incorporating anharmonicity, both MD and SSCHA results, regardless of functional complexity, capture clear deviations from a mean-field second-order transition. 
Namely, the results for $\delta \mathrm{Ti}$ show that the CDW order is melted for $T_{\mathrm{CDW}}\approx 250$\,K, however, there is a temperature range between $T^{\ast}\approx 200$\,K (or $T^{\ast}\approx185$\,K for r$^2$SCAN) and $T_{\mathrm{CDW}}\approx 250$\,K characterized by strong CDW fluctuations.
This two-step melting profile aligns with experimental observations in few-layer and bulk TiSe$_2$\,\cite{cheng2022,amin2024,chen2016}, and was previously observed in other quantum materials, such as 2H-TaSe$_2$\,\cite{shen2023} and kagome metals\,\cite{park2023}. Notably, the envelope of our PBE MD results matches the XRD data very well. That standard Bragg scattering fails to detect this intermediate step is physically consistent, as the local distortions and defects driving this unconventional behavior manifest primarily in the diffuse scattering regime\,\cite{amin2024}. Note also that the PBE SSCHA results capture quite well the low-temperature region as well as the temperature range of the CDW fluctuating intermediate phase, however, both $T^{\ast}$ and $T_{\rm CDW}$ are overestimated by around $150$\,K with respect to the experimental and MD values (which is consistent with previous SSCHA results\,\cite{bianco_tise2}). This highlights the crucial methodological advantage of large-scale MD atomistic simulations for the strongly-coupled CDW phase.

In Fig.\,\ref{fig:subplots_positions_structure_factors}c we present the time-averaged static structure factors Eq.\,(S1) for three different temperatures and calculated for both Ti and Se atoms (see Fig.\,S10 for the convergence of the averaged results). For better visualization of the CDW signatures, we use the irreducible $\mathbf{q}$ points of the $8\times8\times1$ supercell (see also Fig.\,S11, and Fig.\,S12 for the $56\times56\times1$ supercell). The results clearly show how the three non-equivalent $\mathbf{q}$ signals of CDW ($\mathrm{M_1}$, $\mathrm{M_2}$, and $\mathrm{M_3}$) are melted with thermal fluctuations. 
The corresponding intensities for the three M points are presented in Fig.\,\ref{fig:subplots_positions_structure_factors}d (see also Fig.\,S13 for separate contributions of Ti and Se atoms). The results show that the low-temperature CDW up to around $150$\,K is characterized by three different intensities of the M points, which points to an anisotropic $3Q$ order recently observed in the STM measurements of bulk samples\,\cite{ishioka2010,iavarone2012,hkim2024}. In our study, the disparate intensity signals come from three slightly different displacements (PLDs) of Ti and Se, which then break both the twofold and threefold rotation symmetries of the perfect $3Q$ CDW structure of the single layer ($P321$), with $P1$ as the remaining symmetry. Namely, the three anisotropic Ti-Se bond lengths for the low-temperature CDW phase are 2.49\,\AA, 2.57\,\AA, and 2.65\,\AA ~(see Fig.\,S14 for displacements at $25$\,K), which agrees well with the neutron diffraction estimation on the deviation of the bond lengths in bulk\,\cite{disalvo76,wegner2020}.
Note that in bulk the two layers of the CDW unit cell are related by inversion ($P\overline{3}c1$), so the same monolayer anisotropy alone would preserve inversion and give $P\overline{1}$.
Here we also observe the intermediate fluctuating regime in the structure factor from roughly $150$\,K to $250$\,K, in line with x-ray diffuse scattering results for bulk\,\cite{guo2025}, where the static anisotropic $3Q$ order vanishes, but the normal $P\bar{3}m1$ phase is still not established (see Fig.\,S15 for the precise determination of the deviation from the mean-field result, and Fig.\,S14 for displacements at $195$\,K). 
To quantify these fluctuations, in Fig.\,\ref{fig:subplots_positions_structure_factors}d we also show the structure factor of the time-averaged MD configuration $S_{\mathrm{avg}}$ Eq.\,(S13), which represents the order parameter itself, and the difference $\Delta S=\langle S\rangle-S_{\mathrm{avg}}$ Eq.\,(S14). While $S_{\mathrm{avg}}$ abruptly drops to zero between $175$\,K and $185$\,K, $\langle S\rangle$, which also contains the squared fluctuations, decreases smoothly and reaches its minimum only around $T_{\rm CDW}$, resulting in a pronounced peak in $\Delta S$ that clearly deviates from the background increase due to thermal population of phonons. 

The temperature window for which the anisotropy of the $3Q$ order is quenched is in line with the bulk photocurrent\,\cite{xu2020} and non-linear optical spectroscopy measurements\,\cite{tyulnev2025}, as well as with the recent photoemission study where the full coherent signal of the anisotropic $3Q$ order with $P1$ symmetry in the electronic band structure is established only below 160\,K\,\cite{higashihira2026}. 
Note also the small energy difference between different structures across the temperature phase diagram (see Fig.\,S16), which rationalizes to some extent why the fluctuations are so strong.

\begin{figure*}[ht!]
    \centering
    \includegraphics[width=\textwidth]{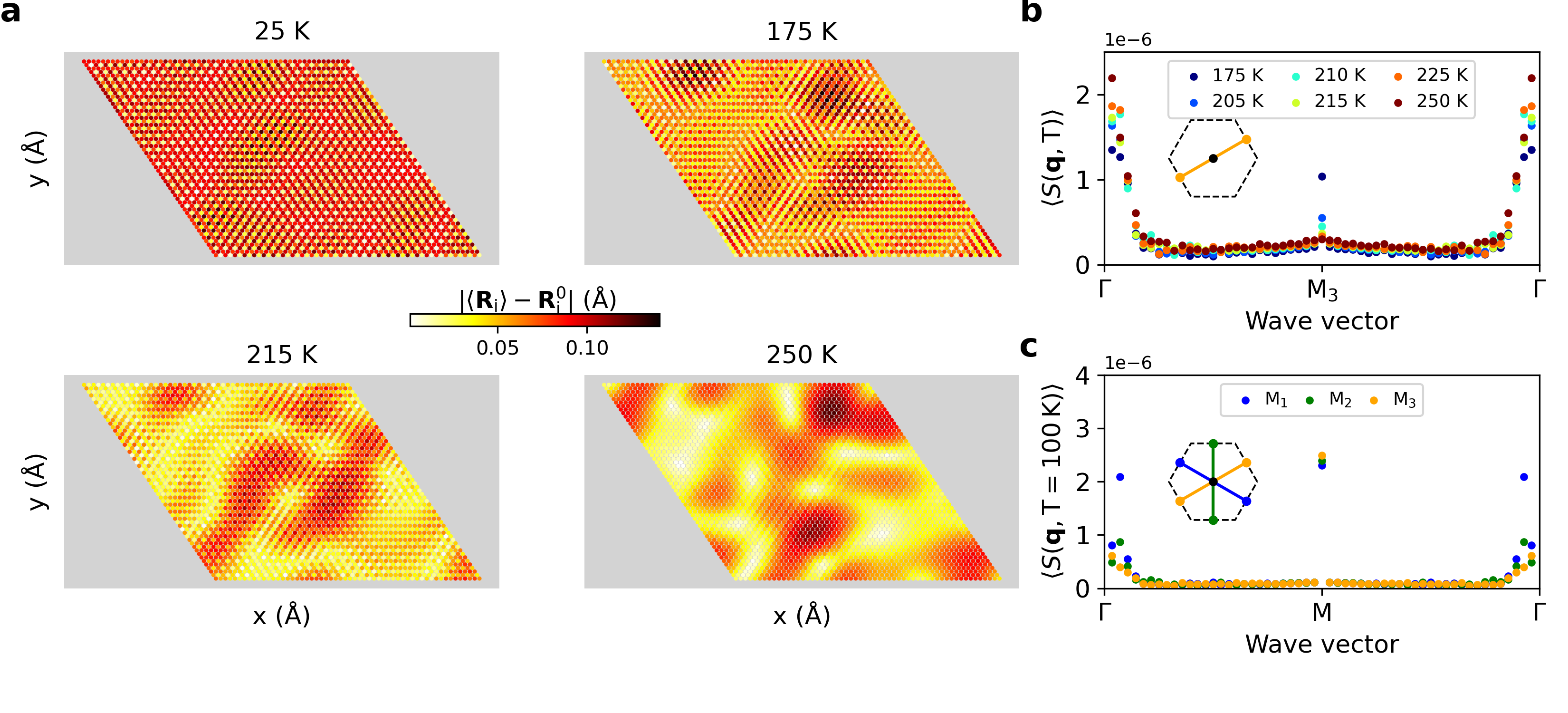}
    \caption{\textbf{Different spatial orderings of CDW phase.} \textbf{a} Distances between time-averaged position of each Ti atom and Ti atoms from normal phase given by Eq.\,(S2) in the $56\times56\times1$ supercell. \textbf{b} Time-averaged structure factor across the $\Gamma-\mathrm{M}-\Gamma$ path for different temperatures at $\mathbf{q}=\mathrm{M_{3}}$. \textbf{c} Time-averaged structure factor at $100$\,K along the $\Gamma-\mathrm{M}-\Gamma$ path for three different M points.}
    \label{fig:subplots_displacements_gmg}
\end{figure*}

\textbf{CDW domains and domain walls induced by thermal fluctuations.} 
To visualize the different CDW spatial orderings with increasing temperature, in Fig.\,\ref{fig:subplots_displacements_gmg}a we plot the displacements between the average MD and normal phase ($P\overline{3}m1$) positions [see Eq.\,(S2)] of Ti atoms in $56\times56\times1$ supercell (see also Fig.\,S17 for more temperatures). For the lowest temperatures considered, we again observe three different displacements of Ti atoms (note the different colors in hexagons consisting of Ti atoms) leading to the anisotropic $3Q$ order with $P1$ symmetry on average (compare with Fig.\,S14). As the temperature increases, separate domains are formed with different displacement patterns. For instance, at $175$\,K we can observe three different local domains: (1) fully anisotropic $3Q$ CDW order with $\delta \mathrm{Ti^{\rm (A)}}>\delta \mathrm{Ti^{\rm (B)}} > \delta \mathrm{Ti^{\rm (C)}}$, (2) nematic order with $\delta \mathrm{Ti^{\rm (A)}}\approx\delta \mathrm{Ti^{\rm (B)}} \gg \delta \mathrm{Ti^{\rm (C)}}$ (note the stripes in Fig.\,\ref{fig:subplots_displacements_gmg}a), and (3) $2\times1$ order with $\delta \mathrm{Ti^{\rm (A)}}\gg\delta \mathrm{Ti^{\rm (B)}} \approx \delta \mathrm{Ti^{\rm (C)}}$. With further increase of temperature (e.g., at $215$\,K) the size of these domains (i.e., correlation length) is reduced, while finally above $250$\,K no CDW order is found (note only the presence of red puddles representing thermal fluctuations). The domain walls that separate these phases are formed at places where thermal fluctuations drop to almost zero value.

Furthermore, in Fig.\,\ref{fig:subplots_displacements_gmg}b we plot the time-averaged structure factors across the $\Gamma-\mathrm{M_3}-\Gamma$ symmetry path for several temperatures below the transition to the normal phase. We observe that the signals at $\mathbf{q}=\mathrm{M}$ are anticorrelated with the thermal fluctuation signal around $\mathbf{q}=\Gamma$ (see also Fig.\,S12). The latter signal comes from the thermal population of long-wavelength acoustic phonons, and it is clearly responsible for the appearance of different CDW domains and finally for the melting of the CDW order. Interestingly, the long-wavelength acoustic phonons were proposed to disrupt the long-range translational order for the pressure-induced melting of CDW in bulk TiSe$_2$\,\cite{snow2003}. Moreover, recent UED experiments on TiSe$_2$ have revealed how the non-thermal excitation of optical phonons can destroy the 2D CDW order by creating 1D domain walls\,\cite{cheng2024}. In fact, it was discussed that any excited collective mode can generate defects and lead to CDW phase transition. Here we show how thermally-excited acoustic phonons play this role under equilibrium conditions.
In addition, in Fig.\,\ref{fig:subplots_displacements_gmg}c we show structure factors across the $\Gamma-\mathrm{M}-\Gamma$ path for $3$ different M points at $100$\,K, where we again observe a strong thermal signal around $\mathbf{q}=\Gamma$ and different intensities at different M points (see also Fig.\,\ref{fig:subplots_positions_structure_factors}d). Interestingly, in the same way as different strengths of $\mathbf{q}\approx \Gamma$ thermal fluctuations at different temperatures impact the signal for particular M point, 
so do the anisotropic long-wavelength thermal fluctuations along $\Gamma-\mathrm{M_1}$, $\Gamma-\mathrm{M_2}$, and $\Gamma-\mathrm{M_3}$ produce a different impact on the structure factors at M$_1$, M$_2$, and M$_3$, as well as on the Ti atom displacements $\delta\mathrm{Ti}^{\rm (A)}$, $\delta\mathrm{Ti}^{\rm (B)}$, and $\delta\mathrm{Ti}^{\rm (C)}$. Note also anisotropy of thermal fluctuations around $\Gamma$ in Fig.\,S12. This means that the anisotropic $3Q$ order with $P1$ symmetry in monolayer TiSe$_2$ comes from anisotropy of long-wavelength thermal fluctuations. We additionally checked this by relaxing the averaged MD structures with anisotropic $3Q$ CDW order without thermal fluctuations, which resulted in the conventional isotropic $3Q$ CDW phase (see Figs.\,S18a and S18b). Our conclusions are in line with Ref.\,\citenum{zenker2013}, where they claim that the anisotropic order in TiSe$_2$ naturally emerges when higher orders of electron-phonon and phonon-phonon interaction are taken into account, and that it is structural in origin. 
Also, the anisotropic spectroscopy signals in bulk revealed in the high harmonic generation\,\cite{tyulnev2025,zhang2025} and recent photoemission study\,\cite{higashihira2026} were explained with a small anisotropy of three components of the CDW order parameter.
\begin{figure*}[ht!]
    \centering
    \includegraphics[width=\textwidth]{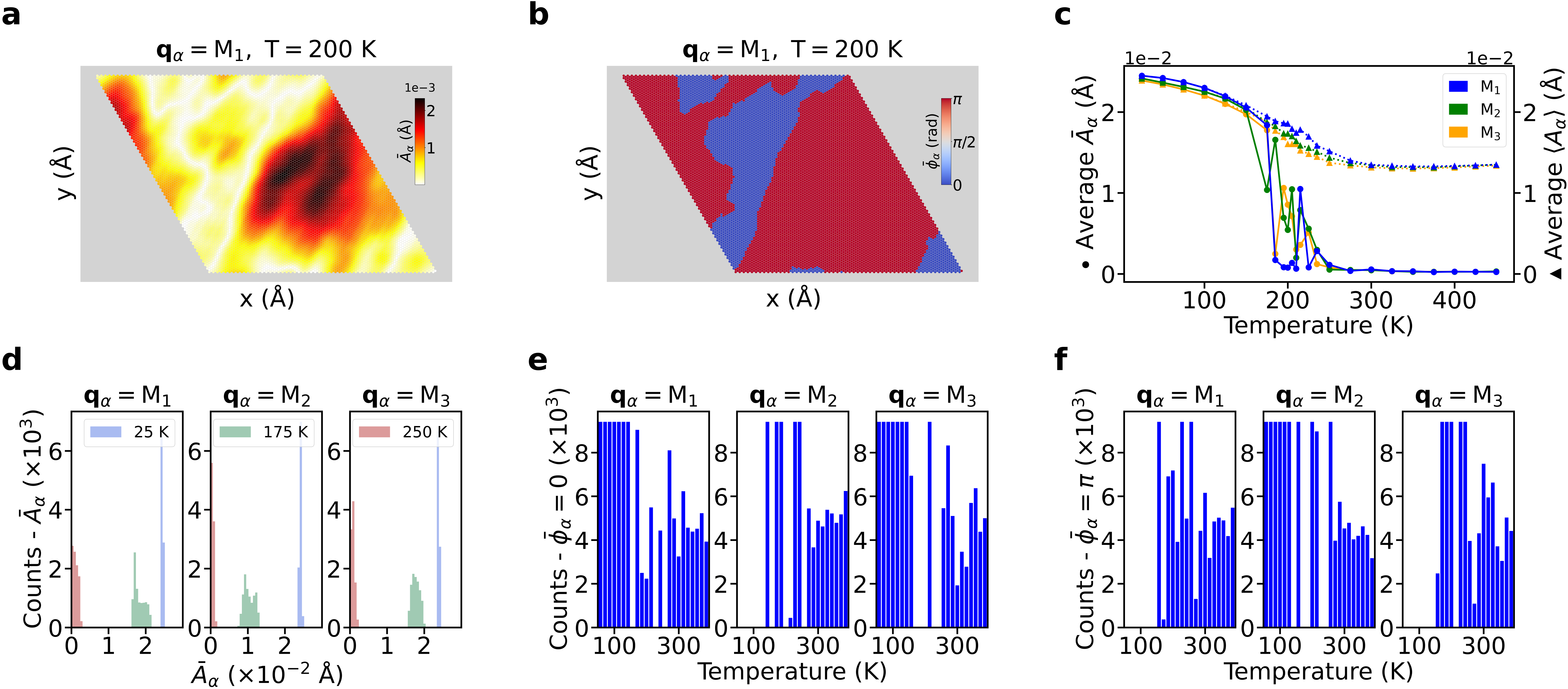}
    \caption{\textbf{Melting process of the CDW order parameter.} \textbf{a} Amplitude $\bar{A}_{\alpha}\left(\mathbf{R}_{i}^{0}\right)$ and \textbf{b} phase $\bar{\phi}_{\alpha}\left(\mathbf{R}_{i}^{0}\right)$ of time-averaged order parameter for $\mathbf{q}_{\alpha}=\mathrm{M}_{1}$ at $200$\,K. \textbf{c} Comparison of $\bar{A}_{\alpha}\left(\mathbf{R}_{i}^{0}\right)$ (dots) and $\langle A_{\alpha}\left(\mathbf{R}_{i}^{0}\right)\rangle$ (triangles) averaged over all atoms as a function of temperature for different $\mathbf{q}=\mathrm{M}$ points. \textbf{d} Distributions of amplitudes $\bar{A}_{\alpha}\left(\mathbf{R}_{i}^{0}\right)$ for three different temperatures and three $\mathbf{q}=\mathrm{M}$ points. Number of atoms with the phase \textbf{e} $\bar{\phi}_{\alpha}\left(\mathbf{R}_{i}^{0}\right)=0$ and \textbf{f} $\bar{\phi}_{\alpha}\left(\mathbf{R}_{i}^{0}\right)=\pi$ as a function of temperature for three $\mathbf{q}=\mathrm{M}$ points.}
    \label{fig:subplots_ampl_phase}
\end{figure*}

\textbf{CDW order parameter and phase decoherence.}
In the weak-coupling theory, the CDW amplitude and phase are simultaneously destroyed at the transition temperature $T_{\rm CDW}$. In the strongly-coupled case, in contrast, the long-range phase coherence is expected to be lost around $T_{\rm CDW}$, while the CDW amplitude can still be finite. Also, different $\mathbf{q}=\mathrm{M_i}$ components could melt independently and phase slips are possible among them. These strong-coupling melting features were discussed for the CDW in NbSe$_2$\,\cite{chatterjee2015,flicker15}, however, for TiSe$_2$ they are rarely studied\,\cite{guo2025,liu2024}, even though the estimations on the ratio between the CDW gap and mean-field transition temperature in bulk and monolayer TiSe$_2$ are way above the weak-coupling value, i.e., $2\Delta/kT=5.5-10.7$\,\cite{rossnagel11,li2007,watson2021}.

To further explore the features of the strong-coupling melting, we have constructed the CDW order parameter $\Psi_{\alpha}$ from the MD results, and analyzed the behavior of CDW amplitude $A_{\alpha}$ and phase $\phi_{\alpha}$ for each of the components $\alpha$ across the CDW transition. The definition of the $\Psi_{\alpha}$ is based on the atomic displacements from the equilibrium $\mathbf{u}_{j}(t)=\mathbf{R}_{j}(t)-\mathbf{R}_{j}^{0}$ projected on the eigenvectors of the three $\mathbf{q}=\mathrm{M}$ soft phonon modes, and is in line with the lock-in analysis used in the STM studies\,\cite{liu2024,xiaolong2021}. Namely, the projected displacements are demodulated with the phase factor of the CDW modulation and averaged with a Gaussian around each atom, giving a local order parameter $\Psi_{\alpha}\left(\mathbf{R}_{i}^{0},t\right)$ (see SI for details). Since the $\mathbf{q}=\mathrm{M}$ point is a half of a reciprocal lattice vector, the order parameter is real and its phase takes only the values $0$ and $\pi$. 

The corresponding results for a particular case of $\mathbf{q}_{\alpha}=\mathrm{M}_1$ and $T=200$\,K are shown in Figs.\,\ref{fig:subplots_ampl_phase}a and \ref{fig:subplots_ampl_phase}b (for other components and temperatures see Figs.\,S19 and S20). This represents a particular directional component of the total domain map (see Fig.\,\ref{fig:subplots_displacements_gmg}), and the phase also shows a domain structure: the supercell is divided into domains with $\bar{\phi}_{\alpha}=0$ and $\bar{\phi}_{\alpha}=\pi$, separated by domain walls where the amplitude $\bar{A}_{\alpha}$ vanishes. Such $\pi$ phase slips are expected for $2\times 2$ commensurate strongly-coupled CDW phase\,\cite{mcmillan1976} and were already reported in the STM measurements of pristine monolayers\,\cite{liu2024} and doped bulk TiSe$_2$\,\cite{hildebrand2016,spera2019,hu2024}.

Figure \ref{fig:subplots_ampl_phase}c shows the amplitudes of the three components as a function of temperature, where dots represent $\bar{A}_{\alpha}$, i.e., the absolute value of the time-averaged order parameter, and triangles $\left\langle A_{\alpha} \right\rangle$, i.e., the time average of its absolute value, both averaged over all atoms. The results show that components of $\bar{A}_{\alpha}$ melt independently between $T^{\ast}$ and $T_{\rm CDW}$, while $\left\langle A_{\alpha} \right\rangle$ decreases only moderately and remains finite above $T_{\rm CDW}$. This behavior is typical for the strong-coupling melting and demonstrates that the CDW is melted with the proliferation of phase fluctuations and not by quenching of the CDW amplitude\,\cite{mcmillan1977}. The same conclusions for $\bar{A}_{\alpha}$ are obtained with the global order parameter, i.e., the projection of the displacements of the whole supercell on the soft phonon mode without the Gaussian (see Fig.\,S21). In panel \ref{fig:subplots_ampl_phase}d we show distributions of $\bar{A}_{\alpha}$ for all three components and for three temperature regimes. At low temperatures the distributions are narrow with a small component anisotropy (see also Fig.\,S22), at $175$\,K they are broader with the M$_2$ amplitude considerably more reduced, and at $T\approx T_{\rm CDW}$ all three amplitudes are almost fully quenched. 

Finally, in Figs.\,\ref{fig:subplots_ampl_phase}e and \ref{fig:subplots_ampl_phase}f we show the number of atoms with $\bar{\phi}_{\alpha}=0$ and $\bar{\phi}_{\alpha}=\pi$ across the temperature diagram for all three components, and resolve three temperature regions: (i) for $T<T^{\ast}$ each component has a single value of the phase for all atoms, i.e., the phase is coherent across the whole supercell (note that $0$ and $\pi$ correspond to equivalent CDW domains, so which of them is realized in a particular MD simulation is arbitrary); (ii) for $T^{\ast}<T<T_{\rm CDW}$ the phase is fluctuating, and domains with both $\bar{\phi}_{\alpha}=0$ and $\bar{\phi}_{\alpha}=\pi$ appear; (iii) for $T>T_{\rm CDW}$ the coherence of the phase is fully lost and both values are present with comparable weights. Intriguingly, different components lose coherence independently: for instance, at $205$\,K only the M$_1$ component is divided into domains, while M$_2$ and M$_3$ still have a single value of the phase (see Fig.\,S20). Therefore, the components behave similarly in the regions (i) and (iii), while in the fluctuating region (ii) they seem to be independent and decoupled.

\begin{figure*}[ht!]
    \centering
    \includegraphics[width=\textwidth]{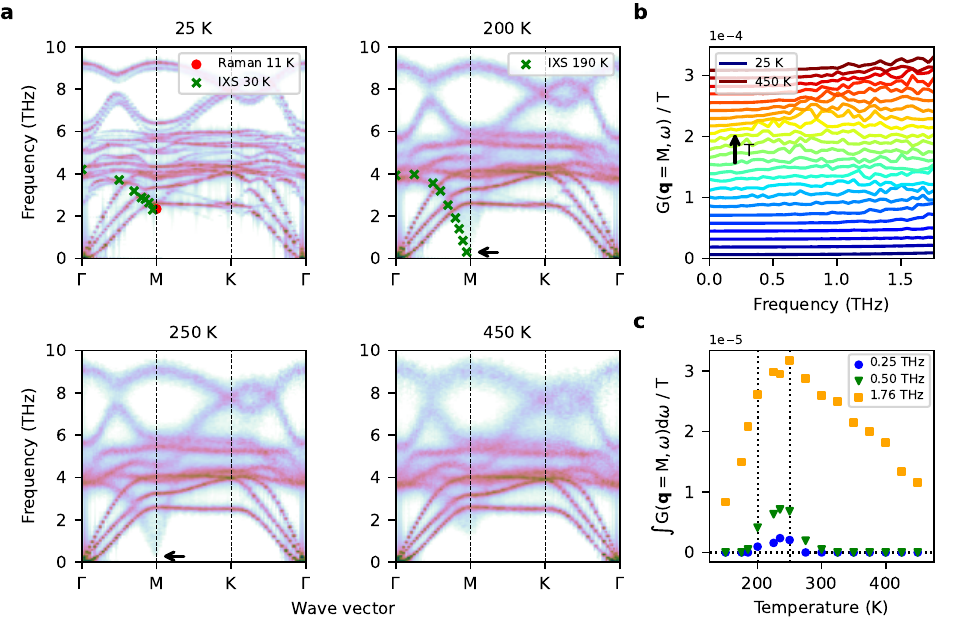}
    \caption{\textbf{Anharmonic phonon dispersions across the CDW transition.} \textbf{a} Phonon dispersions calculated from $56\times56\times1$ MD data along the $\Gamma-\mathrm{M}-\mathrm{K}-\Gamma$ path and compared to Raman \cite{sugai1980} and inelastic x-ray scattering (IXS)\,\cite{weber2011} experiments. Black arrows point to minimum of the Kohn anomaly. \textbf{b} Power spectrum (phonon spectral function) Eq.\,(S3) calculated for the $\mathbf{q}=\mathrm{M}$ point indicated in \textbf{a} and normalized by temperature as a function of frequency for different MD temperatures. Black arrow indicates increase in temperature. \textbf{c} Frequency-integrated power spectrum at $\mathbf{q}=\mathrm{M}$ indicated in \textbf{a} as a function of temperature, calculated with $3$ different upper integration bounds.}
    \label{fig:subplots_dispersions_spectrum}
\end{figure*}

\textbf{Anharmonic phonon dispersion and softening of the optical phonon mode.}
The MD simulations can also provide us with the phonon dynamics by projecting the atomic velocities onto the harmonic phonon modes and calculating their wave-vector projected power spectrum Eq.\,(S3) (i.e., Fourier transform of the velocity autocorrelation function). In Fig.\,\ref{fig:subplots_dispersions_spectrum}a we show the phonon dispersions of monolayer TiSe$_2$ along the $\Gamma-\mathrm{M}-\mathrm{K}-\Gamma$ path for several relevant temperatures and projected on the $1\times 1\times 1$ normal-phase unit cell (see also Fig.\,S23 for more temperatures). As expected, for higher temperatures $T>T_{\rm CDW}$ we observe $9$ phonon bands, $6$ optical and $3$ acoustic ones, while at the lowest temperature the phonon branches are split due to formation of $2\times2\times1$ CDW supercell. As the temperature increases, all phonon dispersions become broader as a consequence of anharmonic phonon-phonon coupling. Furthermore, the Kohn anomaly of the transverse optical (TO) phonon around the $\mathbf{q}=\mathrm{M}$ point\,\cite{weber2011} is already formed at $T=25$\,K. This Kohn anomaly is a consequence of the strong coupling between Se-$p$ and Ti-$d$ states with TO mode and is responsible for the formation of the $2\times2\times 1$ reconstruction of the CDW phase\,\cite{disalvo76,yoshida80,rossnagel11,calandra11,novko22}. 
As it softens, the TO phonon hybridizes with the acoustic phonon branch (note the avoided crossings) forming a hybrid amplitude CDW phonon\,\cite{weber2011}.
Such soft phonon modes are known to be influenced by strong anharmonic coupling\,\cite{mcmillan1977,varma1983,yoshiyama86,bianco_tise2,chen2023}. Our MD results show that this CDW phonon is strongly renormalized and overdamped in the region starting around $200$\,K. In fact, it softens to almost zero value around $250$\,K after which it hardens again, which is expected from the behavior of CDW amplitude phonon mode around the instability point and in accordance with the experiments. Also, softening point agrees well with our result on the transition temperature $T_{\rm CDW}\approx 250$\,K extracted from PLDs, structure factors, and analysis of the order parameter.
We also compare our results with Raman\,\cite{sugai1980} and inelastic x-ray scattering (IXS)\,\cite{weber2011} experiments of bulk TiSe$_2$ and see a very good agreement for $T=25$\,K and $T=200$\,K.
%
In Fig.\,\ref{fig:subplots_dispersions_spectrum}b we show the phonon spectral function normalized by temperature at the $\mathbf{q}=\mathrm{M}$ point around the minimum of the Kohn anomaly. This clearly shows that the CDW phonon mode around the M point is overdamped and almost fully softened, having a finite intensity at zero frequency, for temperatures from $200$\,K to $250$\,K. In IXS measurements of bulk TiSe$_2$, the CDW phonon mode is also fully softened and overdamped in an extended temperature range, i.e., from $192$\,K potentially up to $220$\,K\,\cite{weber2011}. Similarly, infrared optical spectroscopy of bulk TiSe$_2$ revealed a low-frequency CDW-related phonon mode being overdamped from $200$\,K to almost $250$\,K\,\cite{velebit16}.
To take a closer look at this overdamped region, we have integrated the spectral function at the M point using three different upper integration bounds (see Fig.\,\ref{fig:subplots_dispersions_spectrum}c). We see that the strongest signal is again around $250$\,K, while the finite low-frequency intensity is always present between $200$\,K and $250$\,K, which exactly matches the CDW fluctuation region bounded by $T^{\ast}$ and $T_{\rm CDW}$. Such precursor phase with full softening of the CDW-related phonon was also observed in 2H-TaSe$_2$ by means of IXS\,\cite{shen2023}, where the region dominated by the CDW fluctuations extends by $8$\,K from the transition temperature $T_{\rm CDW}$. 

In conclusion, by incorporating full thermal fluctuations, our large-scale MD simulations successfully reproduce the TiSe$_2$ structural phase diagram, revealing previously overlooked strong-coupling features.
Recent experiments proposed domain-wall-mediated CDW disordering and two-step 3D-to-2D melting in TiSe$_2$\,\cite{cheng2024,liu2024,guo2025,cheng2022,amin2024}, alongside low-temperature threefold rotational symmetry breaking\,\cite{tyulnev2025,segovia2025,edwards2026,jiang2026,higashihira2026,vangrondelle2026}. Our simulations establish that two-step melting is intrinsic to the monolayer, with long-wavelength acoustic phonons driving domain formation, phase slips, and overdamping of the soft CDW phonon, while anisotropic thermal fluctuations stabilize fully anisotropic $3Q$ order with $P1$ symmetry. These findings may also inform multi-step CDW melting in 1T-TaS$_2$ and kagome metals\,\cite{dai1991,domrose2023,sung2024,lee2025,park2023,subires2025}.

Further, our success in replicating the structural phase diagram using an interatomic potential trained entirely on semi-local PBE-DFT proves that non-perturbative treatment of lattice fluctuations is sufficient to dictate the complex CDW physics of TiSe$_2$\,\cite{yoshiyama86,varma1983,mcmillan1977} and excitonic correlations are not required\,\cite{pashov2025}. Finally, we believe that these findings are vital for pristine, doped and pressurized samples\,\cite{snow2003,yan2017,lee2021} where enhanced fluctuations are intertwined with emergent superconductivity\,\cite{morosan2006,kusmartseva2009,yan2017,lv2026}. 


\section*{Data availability}
\label{sec:data_availability}
Data is available at \href{https://doi.org/10.5281/zenodo.24517322}{10.5281/zenodo.24517322}. Due to the large size of the files of the MD simulations, we only provide the code for reproducing MD trajectories.

\section*{Code availability}
\label{sec:code_availability}
All relevant code is available at \href{https://doi.org/10.5281/zenodo.24517322}{10.5281/zenodo.24517322}.

\section*{Supporting Information}
\label{sec:supporting_information}
Computational details of the machine-learning interatomic potential (training, fine-tuning, and validation), molecular dynamics simulations, and SSCHA calculations; definitions of the time-averaged structure factors, average atomic displacements, phonon power spectrum, and CDW order parameter; additional results on real-space atomic positions and radial distribution functions, Se and harmonic DFT displacements, convergence and momentum dependence of the structure factors, time-averaged structures with their energies and relaxations, maps and distributions of the local order parameter, global order parameter, phonon dispersions at additional temperatures, harmonic phonon dispersions of the $2\times2\times1$ cell, dependence on the MD time step, SSCHA transition temperature, and comparison with the ARPES band gap (PDF)

\subsection*{Acknowledgments}
L.B. and I.L. acknowledge financial support from the Croatian Science Foundation projects UIP-2020-02-5675 and IP-2026-5909. D.N. acknowledges financial support from the Croatian Science Foundation (Grant no. IP-2025-02-5926 and UIP-2025-02-5952), and from the project ``Podizanje znanstvene izvrsnosti Centra za napredne laserske tehnike (CALTboost)'' financed by the European Union through the National Recovery and Resilience Plan 2021-2026 (NRPP). L.B., D.N., and I.L. acknowledge the European Regional Development Fund for the project ‘Materials for clean energy, advanced sensors and quantum technologies’ (Grants No. PK.1.1.10.0002). This paper was supported by the European Union's NextGenerationEU program.

\subsection*{Author contributions}
I.L. and D.N. conceived and supervised the project. L.B. performed the simulations and created figures with guidance from I.L. and D.N. All authors contributed to the discussion and interpretation of the results and writing of the manuscript.

\bibliography{ref}

@article{yan2017,
  title = {{Influence of Domain Walls in the Incommensurate Charge Density Wave State of Cu Intercalated $1T\text{\ensuremath{-}}{\mathrm{TiSe}}_{2}$}},
  author = {Yan, Shichao and Iaia, Davide and Morosan, Emilia and Fradkin, Eduardo and Abbamonte, Peter and Madhavan, Vidya},
  journal = {Phys. Rev. Lett.},
  volume = {118},
  issue = {10},
  pages = {106405},
  numpages = {5},
  year = {2017},
  month = {Mar},
  publisher = {American Physical Society},
  doi = {10.1103/PhysRevLett.118.106405},
  url = {https://link.aps.org/doi/10.1103/PhysRevLett.118.106405}
}

@article{domrose2023, title={{Light-induced hexatic state in a layered quantum material}}, volume={22}, ISSN={1476-4660}, url={http://dx.doi.org/10.1038/s41563-023-01600-6}, DOI={10.1038/s41563-023-01600-6}, number={11}, journal={Nature Materials}, publisher={Springer Science and Business Media LLC}, author={Domröse, Till and Danz, Thomas and Schaible, Sophie F. and Rossnagel, Kai and Yalunin, Sergey V. and Ropers, Claus}, year={2023}, month=jul, pages={1345–1351} }

@article{dai1991,
  title = {{Weak pinning and hexatic order in a doped two-dimensional charge-density-wave system}},
  author = {Dai, Hongjie and Chen, Huifen and Lieber, Charles M.},
  journal = {Phys. Rev. Lett.},
  volume = {66},
  issue = {24},
  pages = {3183},
  numpages = {0},
  year = {1991},
  month = {Jun},
  publisher = {American Physical Society},
  doi = {10.1103/PhysRevLett.66.3183},
  url = {https://link.aps.org/doi/10.1103/PhysRevLett.66.3183}
}

@article{lee2025, title={{Observation of a hidden charge density wave liquid}}, volume={22}, ISSN={1745-2481}, url={http://dx.doi.org/10.1038/s41567-025-03108-z}, DOI={10.1038/s41567-025-03108-z}, number={1}, journal={Nature Physics}, publisher={Springer Science and Business Media LLC}, author={Lee, Joshua S. H. and Sutter, Thomas M. and Karapetrov, Goran and Musumeci, Pietro and Kogar, Anshul}, year={2025}, month=dec, pages={68} }

@article{sung2024, title={{Endotaxial stabilization of 2D charge density waves with long-range order}}, volume={15}, ISSN={2041-1723}, url={http://dx.doi.org/10.1038/s41467-024-45711-3}, DOI={10.1038/s41467-024-45711-3}, number={1}, journal={Nature Communications}, publisher={Springer Science and Business Media LLC}, author={Sung, Suk Hyun and Agarwal, Nishkarsh and El Baggari, Ismail and Kezer, Patrick and Goh, Yin Min and Schnitzer, Noah and Shen, Jeremy M. and Chiang, Tony and Liu, Yu and Lu, Wenjian and Sun, Yuping and Kourkoutis, Lena F. and Heron, John T. and Sun, Kai and Hovden, Robert}, year={2024}, pages={1403}}

@article{subires2025, title={{Frustrated charge density wave and quasi-long-range bond-orientational order in the magnetic kagome FeGe}}, volume={16}, ISSN={2041-1723}, url={http://dx.doi.org/10.1038/s41467-025-58725-2}, DOI={10.1038/s41467-025-58725-2}, number={1}, journal={Nature Communications}, publisher={Springer Science and Business Media LLC}, author={Subires, D. and Kar, A. and Korshunov, A. and Fuller, C. A. and Jiang, Yi and Hu, H. and Călugăru, D. and McMonagle, C. and Yi, C. and Roychowdhury, S. and Schnelle, W. and Shekhar, C. and Strempfer, J. and Jana, A. and Vobornik, I. and Dai, J. and Tallarida, M. and Chernyshov, D. and Bosak, A. and Felser, C. and Bernevig, B. Andrei and Blanco-Canosa, S.}, year={2025}, pages={4091} }

@article{cheng2024, title={{Ultrafast formation of topological defects in a two-dimensional charge density wave}}, volume={20}, ISSN={1745-2481}, url={http://dx.doi.org/10.1038/s41567-023-02279-x}, DOI={10.1038/s41567-023-02279-x}, number={1}, journal={Nature Physics}, publisher={Springer Science and Business Media LLC}, author={Cheng, Yun and Zong, Alfred and Wu, Lijun and Meng, Qingping and Xia, Wei and Qi, Fengfeng and Zhu, Pengfei and Zou, Xiao and Jiang, Tao and Guo, Yanfeng and van Wezel, Jasper and Kogar, Anshul and Zuerch, Michael W. and Zhang, Jie and Zhu, Yimei and Xiang, Dao}, year={2024}, month=jan, pages={54} }

@article{segovia2025,
  title = {{Doping-induced nematic and stripe orders within the charge density wave state of ${\mathrm{TiSe}}_{2}$}},
  author = {Mu\~noz-Segovia, Daniel and Venderbos, J\"orn W. F. and Grushin, Adolfo G. and de Juan, Fernando},
  journal = {Phys. Rev. B},
  volume = {112},
  issue = {16},
  pages = {165119},
  numpages = {22},
  year = {2025},
  month = {Oct},
  publisher = {American Physical Society},
  doi = {10.1103/1nxh-3v88},
  url = {https://link.aps.org/doi/10.1103/1nxh-3v88}
}

@article{chen2019,
  title = {{Discommensuration-driven superconductivity in the charge density wave phases of transition-metal dichalcogenides}},
  author = {Chen, Chuan and Su, Lei and Castro Neto, A. H. and Pereira, Vitor M.},
  journal = {Phys. Rev. B},
  volume = {99},
  issue = {12},
  pages = {121108},
  numpages = {5},
  year = {2019},
  month = {Mar},
  publisher = {American Physical Society},
  doi = {10.1103/PhysRevB.99.121108},
  url = {https://link.aps.org/doi/10.1103/PhysRevB.99.121108}
}

@article{hu2024,
  title = {{From domain walls and the stripe phase to full suppression of charge density waves in superconducting $1T\text{\ensuremath{-}}\mathrm{T}{\mathrm{i}}_{1\ensuremath{-}x}\mathrm{T}{\mathrm{a}}_{x}\mathrm{S}{\mathrm{e}}_{2}$}},
  author = {Hu, Q. and Venturini, R. and Vaskivskyi, Y. and Lipi\ifmmode \check{c}\else \v{c}\fi{}, J. and Jagli\ifmmode \check{c}\else \v{c}\fi{}i\ifmmode \acute{c}\else \'{c}\fi{}, Z. and Mihailovic, D.},
  journal = {Phys. Rev. B},
  volume = {110},
  issue = {16},
  pages = {165156},
  numpages = {7},
  year = {2024},
  month = {Oct},
  publisher = {American Physical Society},
  doi = {10.1103/PhysRevB.110.165156},
  url = {https://link.aps.org/doi/10.1103/PhysRevB.110.165156}
}

@article{chatterjee2015, title={{Emergence of coherence in the charge-density wave state of 2H-NbSe$_{2}$}}, volume={6}, ISSN={2041-1723}, url={http://dx.doi.org/10.1038/ncomms7313}, DOI={10.1038/ncomms7313}, number={1}, journal={Nature Communications}, publisher={Springer Science and Business Media LLC}, author={Chatterjee, U. and Zhao, J. and Iavarone, M. and Di Capua, R. and Castellan, J. P. and Karapetrov, G. and Malliakas, C. D. and Kanatzidis, M. G. and Claus, H. and Ruff, J. P. C. and Weber, F. and van Wezel, J. and Campuzano, J. C. and Osborn, R. and Randeria, M. and Trivedi, N. and Norman, M. R. and Rosenkranz, S.}, year={2015}, pages={6313} }

@article{snow2003,
  title = {{Quantum Melting of the Charge-Density-Wave State in $1T\mathrm{\text{\ensuremath{-}}}{\mathrm{T}\mathrm{i}\mathrm{S}\mathrm{e}}_{2}$}},
  author = {Snow, C. S. and Karpus, J. F. and Cooper, S. L. and Kidd, T. E. and Chiang, T.-C.},
  journal = {Phys. Rev. Lett.},
  volume = {91},
  issue = {13},
  pages = {136402},
  numpages = {4},
  year = {2003},
  month = {Sep},
  publisher = {American Physical Society},
  doi = {10.1103/PhysRevLett.91.136402},
  url = {https://link.aps.org/doi/10.1103/PhysRevLett.91.136402}
}

@article{guo2025,
  title = {{In-Plane Anisotropy of Charge Density Wave Fluctuations in $1T\text{\ensuremath{-}}{\mathrm{TiSe}}_{2}$}},
  author = {Guo, Xuefei and Kogar, Anshul and Henke, Jans and Flicker, Felix and de Juan, Fernando and Sun, Stella X.-L. and Khayr, Issam and Peng, Yingying and Lee, Sangjun and Krogstad, Matthew J. and Rosenkranz, Stephan and Osborn, Raymond and Ruff, Jacob P. C. and Lioi, David B. and Karapetrov, Goran and Campbell, Daniel J. and Paglione, Johnpierre and van Wezel, Jasper and Chiang, Tai C. and Abbamonte, Peter},
  journal = {Phys. Rev. Lett.},
  volume = {135},
  issue = {13},
  pages = {136102},
  numpages = {6},
  year = {2025},
  month = {Sep},
  publisher = {American Physical Society},
  doi = {10.1103/j8vm-wb65},
  url = {https://link.aps.org/doi/10.1103/j8vm-wb65}
}

@article{ishioka2010,
  title = {{Chiral Charge-Density Waves}},
  author = {Ishioka, J. and Liu, Y. H. and Shimatake, K. and Kurosawa, T. and Ichimura, K. and Toda, Y. and Oda, M. and Tanda, S.},
  journal = {Phys. Rev. Lett.},
  volume = {105},
  issue = {17},
  pages = {176401},
  numpages = {4},
  year = {2010},
  month = {Oct},
  publisher = {American Physical Society},
  doi = {10.1103/PhysRevLett.105.176401},
  url = {https://link.aps.org/doi/10.1103/PhysRevLett.105.176401}
}

@article{tyulnev2025, title={{High harmonic spectroscopy reveals anisotropy of the charge-density-wave phase transition in TiSe$_{2}$}}, volume={6}, ISSN={2662-4443}, url={http://dx.doi.org/10.1038/s43246-025-00873-5}, DOI={10.1038/s43246-025-00873-5}, number={1}, journal={Communications Materials}, publisher={Springer Science and Business Media LLC}, author={Tyulnev, Igor and Zhang, Lin and Vamos, Lenard and Poborska, Julita and Bhattacharya, Utso and Chhajlany, Ravindra W. and Grass, Tobias and Mañas-Valero, Samuel and Coronado, Eugenio and Lewenstein, Maciej and Biegert, Jens}, year={2025}, pages={152} }

@article{iavarone2012,
  title = {{Evolution of the charge density wave state in Cu${}_{x}$TiSe${}_{2}$}},
  author = {Iavarone, M. and Di Capua, R. and Zhang, X. and Golalikhani, M. and Moore, S. A. and Karapetrov, G.},
  journal = {Phys. Rev. B},
  volume = {85},
  issue = {15},
  pages = {155103},
  numpages = {6},
  year = {2012},
  month = {Apr},
  publisher = {American Physical Society},
  doi = {10.1103/PhysRevB.85.155103},
  url = {https://link.aps.org/doi/10.1103/PhysRevB.85.155103}
}

@article{zenker2013,
  title = {{Chiral charge order in 1$T$-TiSe$_{2}$: Importance of lattice degrees of freedom}},
  author = {Zenker, B. and Fehske, H. and Beck, H. and Monney, C. and Bishop, A. R.},
  journal = {Phys. Rev. B},
  volume = {88},
  issue = {7},
  pages = {075138},
  numpages = {12},
  year = {2013},
  month = {Aug},
  publisher = {American Physical Society},
  doi = {10.1103/PhysRevB.88.075138},
  url = {https://link.aps.org/doi/10.1103/PhysRevB.88.075138}
}

@article{shen2026, title={{Melting of charge density waves in low dimensions}}, volume={9}, ISSN={2590-2385}, url={http://dx.doi.org/10.1016/j.matt.2026.102665}, DOI={10.1016/j.matt.2026.102665}, number={4}, journal={Matter}, publisher={Elsevier BV}, author={Shen, Jeremy M. and Stangel, Alex and Sung, Suk Hyun and Agarwal, Nishkarsh and Ye, Gaihua and Nnokwe, Cynthia and Zhao, Liuyan and Zhang, Yang and He, Rui and El Baggari, Ismail and Sun, Kai and Hovden, Robert}, year={2026}, month=apr, pages={102665} }

@article{fragkos2026, title={{Electron-phonon-dominated charge-density-wave fluctuations in TiSe2 accessed by ultrafast nonequilibrium dynamics}}, ISSN={2399-3650}, url={http://dx.doi.org/10.1038/s42005-026-02521-x}, DOI={10.1038/s42005-026-02521-x}, journal={Communications Physics}, publisher={Springer Science and Business Media LLC}, author={Fragkos, Sotirios and Orio, Hibiki and Girotto Erhardt, Nina and Jabed, Akib and Sasi, Sarath and Courtade, Quentin and Masilamani, Muthu P. T. and Ünzelmann, Maximilian and Diekmann, Florian and Hildebrand, Baptiste and Descamps, Dominique and Petit, Stéphane and Boschini, Fabio and Minár, Ján and Mairesse, Yann and Reinert, Friedrich and Rossnagel, Kai and Novko, Dino and Beaulieu, Samuel and Schusser, Jakub}, volume={9}, number={1}, pages={86}, year={2026}, month=feb }

@article{holt2001,
  title = {{X-Ray Studies of Phonon Softening in ${\mathrm{TiSe}}_{2}$}},
  author = {Holt, M. and Zschack, P. and Hong, Hawoong and Chou, M. Y. and Chiang, T.-C.},
  journal = {Phys. Rev. Lett.},
  volume = {86},
  issue = {17},
  pages = {3799--3802},
  numpages = {0},
  year = {2001},
  month = {Apr},
  publisher = {American Physical Society},
  doi = {10.1103/PhysRevLett.86.3799},
  url = {https://link.aps.org/doi/10.1103/PhysRevLett.86.3799}
}

@article{weber2011,
  title = {{Electron-Phonon Coupling and the Soft Phonon Mode in ${\mathrm{TiSe}}_{2}$}},
  author = {Weber, F. and Rosenkranz, S. and Castellan, J.-P. and Osborn, R. and Karapetrov, G. and Hott, R. and Heid, R. and Bohnen, K.-P. and Alatas, A.},
  journal = {Phys. Rev. Lett.},
  volume = {107},
  issue = {26},
  pages = {266401},
  numpages = {5},
  year = {2011},
  month = {Dec},
  publisher = {American Physical Society},
  doi = {10.1103/PhysRevLett.107.266401},
  url = {https://link.aps.org/doi/10.1103/PhysRevLett.107.266401}
}

@article{disalvo76,
  title = {{Electronic properties and superlattice formation in the semimetal ${\mathrm{TiSe}}_{2}$}},
  author = {Di Salvo, F. J. and Moncton, D. E. and Waszczak, J. V.},
  journal = {Phys. Rev. B},
  volume = {14},
  issue = {10},
  pages = {4321},
  numpages = {0},
  year = {1976},
  month = {Nov},
  publisher = {American Physical Society},
  doi = {10.1103/PhysRevB.14.4321},
  url = {https://link.aps.org/doi/10.1103/PhysRevB.14.4321}
}

@article{cercellier07,
  title = {{Evidence for an Excitonic Insulator Phase in $1T\mathrm{\text{\ensuremath{-}}}{\mathrm{TiSe}}_{2}$}},
  author = {Cercellier, H. and Monney, C. and Clerc, F. and Battaglia, C. and Despont, L. and Garnier, M. G. and Beck, H. and Aebi, P. and Patthey, L. and Berger, H. and Forr\'o, L.},
  journal = {Phys. Rev. Lett.},
  volume = {99},
  issue = {14},
  pages = {146403},
  numpages = {4},
  year = {2007},
  month = {Oct},
  publisher = {American Physical Society},
  doi = {10.1103/PhysRevLett.99.146403},
  url = {https://link.aps.org/doi/10.1103/PhysRevLett.99.146403}
}

@article{rossnagel11,
	doi = {10.1088/0953-8984/23/21/213001},
	url = {https://doi.org/10.1088/0953-8984/23/21/213001},
	year = 2011,
	month = {may},
	publisher = {{IOP} Publishing},
	volume = {23},
	number = {21},
	pages = {213001},
	author = {K Rossnagel},
	title = {{On the origin of charge-density waves in select layered transition-metal dichalcogenides}},
	journal = {Journal of Physics: Condensed Matter}
}

@article{hughes1977,
doi = {10.1088/0022-3719/10/11/009},
url = {https://dx.doi.org/10.1088/0022-3719/10/11/009},
year = {1977},
month = {jun},
publisher = {},
volume = {10},
number = {11},
pages = {L319},
author = {H P Hughes},
title = {{Structural distortion in TiSe2 and related materials-a possible Jahn-Teller effect?}},
journal = {Journal of Physics C: Solid State Physics}
}

@article{yoshida80,
author = {Yoshida ,Yukimasa and Motizuki ,Kazuko},
title = {{Electron Lattice Interaction and Lattice Instability of 1T-TiSe$_{2}$}},
journal = {Journal of the Physical Society of Japan},
volume = {49},
number = {3},
pages = {898},
year = {1980},
doi = {10.1143/JPSJ.49.898},
URL = {https://doi.org/10.1143/JPSJ.49.898}
}

@article{calandra11,
  title = {{Charge-Density Wave and Superconducting Dome in ${\mathrm{TiSe}}_{2}$ from Electron-Phonon Interaction}},
  author = {Calandra, Matteo and Mauri, Francesco},
  journal = {Phys. Rev. Lett.},
  volume = {106},
  issue = {19},
  pages = {196406},
  numpages = {4},
  year = {2011},
  month = {May},
  publisher = {American Physical Society},
  doi = {10.1103/PhysRevLett.106.196406},
  url = {https://link.aps.org/doi/10.1103/PhysRevLett.106.196406}
}

@article{yoshiyama86,
	doi = {10.1088/0022-3719/19/28/011},
	url = {https://doi.org/10.1088/0022-3719/19/28/011},
	year = 1986,
	month = {oct},
	publisher = {{IOP} Publishing},
	volume = {19},
	number = {28},
	pages = {5591},
	author = {H Yoshiyama and Y Takaoka and N Suzuki and K Motizuki},
	title = {{Effects on lattice fluctuations on the charge-density-wave transition in transition-metal dichalcogenides}},
	journal = {Journal of Physics C: Solid State Physics}
}

@article{varma1983,
  title = {{Strong-Coupling Theory of Charge-Density-Wave Transitions}},
  author = {Varma, C. M. and Simons, A. L.},
  journal = {Phys. Rev. Lett.},
  volume = {51},
  issue = {2},
  pages = {138--141},
  numpages = {0},
  year = {1983},
  month = {Jul},
  publisher = {American Physical Society},
  doi = {10.1103/PhysRevLett.51.138},
  url = {https://link.aps.org/doi/10.1103/PhysRevLett.51.138}
}

@article{mcmillan1977,
  title = {{Microscopic model of charge-density waves in $2H\ensuremath{-}\mathrm{Ta}{\mathrm{Se}}_{2}$}},
  author = {McMillan, W. L.},
  journal = {Phys. Rev. B},
  volume = {16},
  issue = {2},
  pages = {643--650},
  numpages = {0},
  year = {1977},
  month = {Jul},
  publisher = {American Physical Society},
  doi = {10.1103/PhysRevB.16.643},
  url = {https://link.aps.org/doi/10.1103/PhysRevB.16.643}
}

@article{kogar2017a,
author = {Anshul Kogar  and Melinda S. Rak  and Sean Vig  and Ali A. Husain  and Felix Flicker  and Young Il Joe  and Luc Venema  and Greg J. MacDougall  and Tai C. Chiang  and Eduardo Fradkin  and Jasper van Wezel  and Peter Abbamonte },
title = {{Signatures of exciton condensation in a transition metal dichalcogenide}},
journal = {Science},
volume = {358},
number = {6368},
pages = {1314},
year = {2017},
doi = {10.1126/science.aam6432},
URL = {https://www.science.org/doi/abs/10.1126/science.aam6432}
}

@article{lin22,
  title = {{Dramatic Plasmon Response to the Charge-Density-Wave Gap Development in $1T\text{\ensuremath{-}}{\mathrm{TiSe}}_{2}$}},
  author = {Lin, Zijian and Wang, Cuixiang and Balassis, A. and Echeverry, J. P. and Vasenko, A. S. and Silkin, V. M. and Chulkov, E. V. and Shi, Youguo and Zhang, Jiandi and Guo, Jiandong and Zhu, Xuetao},
  journal = {Phys. Rev. Lett.},
  volume = {129},
  issue = {18},
  pages = {187601},
  numpages = {7},
  year = {2022},
  month = {Oct},
  publisher = {American Physical Society},
  doi = {10.1103/PhysRevLett.129.187601},
  url = {https://link.aps.org/doi/10.1103/PhysRevLett.129.187601}
}

@article{novko2025,
  title = {{Excitons and optical response in the excitonic insulator candidate ${\mathrm{TiSe}}_{2}$}},
  author = {Novko, Dino},
  journal = {Phys. Rev. B},
  volume = {112},
  issue = {24},
  pages = {L241114},
  numpages = {7},
  year = {2025},
  month = {Dec},
  publisher = {American Physical Society},
  doi = {10.1103/bs8n-wt4h},
  url = {https://link.aps.org/doi/10.1103/bs8n-wt4h}
}

@article{novko22,
  title = {{Electron correlations rule the phonon-driven instability in single-layer ${\mathrm{TiSe}}_{2}$}},
  author = {Novko, Dino and Torbatian, Zahra and Lon\ifmmode \check{c}\else \v{c}\fi{}ari\ifmmode \acute{c}\else \'{c}\fi{}, Ivor},
  journal = {Phys. Rev. B},
  volume = {106},
  issue = {24},
  pages = {245108},
  numpages = {7},
  year = {2022},
  month = {Dec},
  publisher = {American Physical Society},
  doi = {10.1103/PhysRevB.106.245108},
  url = {https://link.aps.org/doi/10.1103/PhysRevB.106.245108}
}

@article{sugai1980,
title = {{Raman studies of lattice dynamics in 1T-TiSe$_{2}$}},
journal = {Solid State Communications},
volume = {35},
number = {5},
pages = {433-436},
year = {1980},
issn = {0038-1098},
doi = {https://doi.org/10.1016/0038-1098(80)90175-1},
url = {https://www.sciencedirect.com/science/article/pii/0038109880901751},
author = {S. Sugai and K. Murase and S. Uchida and S. Tanaka}
}

@article{velebit16,
  title = {{Scattering-dominated high-temperature phase of $1T\text{\ensuremath{-}}\mathrm{TiS}{\mathrm{e}}_{2}$: An optical conductivity study}},
  author = {Velebit, K. and Pop\ifmmode \check{c}\else \v{c}\fi{}evi\ifmmode \acute{c}\else \'{c}\fi{}, P. and Batisti\ifmmode \acute{c}\else \'{c}\fi{}, I. and Eichler, M. and Berger, H. and Forr\'o, L. and Dressel, M. and Bari\ifmmode \check{s}\else \v{s}\fi{}i\ifmmode \acute{c}\else \'{c}\fi{}, N. and Tuti\ifmmode \check{s}\else \v{s}\fi{}, E.},
  journal = {Phys. Rev. B},
  volume = {94},
  issue = {7},
  pages = {075105},
  numpages = {9},
  year = {2016},
  month = {Aug},
  publisher = {American Physical Society},
  doi = {10.1103/PhysRevB.94.075105},
  url = {https://link.aps.org/doi/10.1103/PhysRevB.94.075105}
}

@article{shen2023, title={{Precursor region with full phonon softening above the charge-density-wave phase transition in 2H-TaSe$_{2}$}}, volume={14}, ISSN={2041-1723}, url={http://dx.doi.org/10.1038/s41467-023-43094-5}, DOI={10.1038/s41467-023-43094-5}, number={1}, journal={Nature Communications}, publisher={Springer Science and Business Media LLC}, author={Shen, Xingchen and Heid, Rolf and Hott, Roland and Haghighirad, Amir-Abbas and Salzmann, Björn and dos Reis Cantarino, Marli and Monney, Claude and Said, Ayman H. and Frachet, Mehdi and Murphy, Bridget and Rossnagel, Kai and Rosenkranz, Stephan and Weber, Frank}, year={2023}, pages={7282} }

@article{hkim2024,
author = {Kim, Hyeonjung and Jin, Kyung-Hwan and Yeom, Han Woong},
title = {{Electronically Seamless Domain Wall of Chiral Charge Density Wave in 1T-TiSe2}},
journal = {Nano Letters},
volume = {24},
number = {45},
pages = {14323-14328},
year = {2024},
doi = {10.1021/acs.nanolett.4c03970},
URL = { https://doi.org/10.1021/acs.nanolett.4c03970}
}

@article{pashov2025, title={{TiSe$_{2}$ is a band insulator created by lattice fluctuations, not an excitonic insulator}}, volume={11}, ISSN={2057-3960}, url={http://dx.doi.org/10.1038/s41524-025-01631-4}, DOI={10.1038/s41524-025-01631-4}, number={1}, journal={npj Computational Materials}, publisher={Springer Science and Business Media LLC}, author={Pashov, Dimitar and Larsen, Ross E. and Watson, Matthew D. and Acharya, Swagata and van Schilfgaarde, Mark}, year={2025}, pages={152} }

@misc{Batatia2022Design,
  title = {{The Design Space of E(3)-Equivariant Atom-Centered Interatomic Potentials}},
  author = {Batatia, Ilyes and Batzner, Simon and Kov{\'a}cs, D{\'a}vid P{\'e}ter and Musaelian, Albert and Simm, Gregor N. C. and Drautz, Ralf and Ortner, Christoph and Kozinsky, Boris and Cs{\'a}nyi, G{\'a}bor},
  year = {2022},
  number = {arXiv:2205.06643},
  eprint = {2205.06643},
  eprinttype = {arxiv},
  doi = {10.48550/arXiv.2205.06643},
  archiveprefix = {arXiv}
}

@inproceedings{NEURIPS2022_4a36c3c5,
 author = {Batatia, Ilyes and Kovacs, David P and Simm, Gregor and Ortner, Christoph and Csanyi, Gabor},
 booktitle = {{Advances in Neural Information Processing Systems}},
 editor = {S. Koyejo and S. Mohamed and A. Agarwal and D. Belgrave and K. Cho and A. Oh},
 pages = {11423--11436},
 publisher = {Curran Associates, Inc.},
 title = {{MACE: Higher Order Equivariant Message Passing Neural Networks for Fast and Accurate Force Fields}},
 url = {https://proceedings.neurips.cc/paper_files/paper/2022/file/4a36c3c51af11ed9f34615b81edb5bbc-Paper-Conference.pdf},
 volume = {35},
 year = {2022}
}

@misc{izmailov2019averagingweightsleadswider,
      title={{Averaging Weights Leads to Wider Optima and Better Generalization}}, 
      author={Pavel Izmailov and Dmitrii Podoprikhin and Timur Garipov and Dmitry Vetrov and Andrew Gordon Wilson},
      year={2019},
      eprint={1803.05407},
      archivePrefix={arXiv},
      primaryClass={cs.LG},
      url={https://arxiv.org/abs/1803.05407}, 
}

@misc{athiwaratkun2019consistentexplanationsunlabeleddata,
      title={{There Are Many Consistent Explanations of Unlabeled Data: Why You Should Average}}, 
      author={Ben Athiwaratkun and Marc Finzi and Pavel Izmailov and Andrew Gordon Wilson},
      year={2019},
      eprint={1806.05594},
      archivePrefix={arXiv},
      primaryClass={cs.LG},
      url={https://arxiv.org/abs/1806.05594}, 
}

@Article{ISI:000175131400009,
Author = {S. R. Bahn and K. W. Jacobsen},
Title = {{An object-oriented scripting interface to a legacy electronic structure code}},
JournalFull = {COMPUTING IN SCIENCE \& ENGINEERING},
Year = {2002},
Volume = {4},
Number = {3},
Pages = {56-66},
Month = {MAY-JUN},
Publisher = {IEEE COMPUTER SOC},
Address = {10662 LOS VAQUEROS CIRCLE, PO BOX 3014, LOS ALAMITOS, CA 90720-1314 USA},
Type = {Article},
Language = {English},
Affiliation = {Bahn, SR (Reprint Author), Tech Univ Denmark, Dept Phys, CAMP, Bldg 307, DK-2800 Lyngby, Denmark.
Tech Univ Denmark, Dept Phys, CAMP, DK-2800 Lyngby, Denmark.},
ISSN = {1521-9615},
Keywords-Plus = {MULTISCALE SIMULATION; GOLD ATOMS},
Subject-Category = {Computer Science, Interdisciplinary Applications},
Author-Email = {bahn@fysik.dtu.dk kwj@fysik.dtu.dk},
Number-of-Cited-References = {19},
Journal-ISO = {Comput. Sci. Eng.},
Journal = {Comput. Sci. Eng.},
Doc-Delivery-Number = {543YL},
Unique-ID = {ISI:000175131400009},
DOI = {10.1109/5992.998641},
}

@article{ase-paper,
  author={Ask Hjorth Larsen and Jens Jørgen Mortensen and Jakob Blomqvist and Ivano E Castelli and Rune Christensen and Marcin
Dułak and Jesper Friis and Michael N Groves and Bjørk Hammer and Cory Hargus and Eric D Hermes and Paul C Jennings and Peter
Bjerre Jensen and James Kermode and John R Kitchin and Esben Leonhard Kolsbjerg and Joseph Kubal and Kristen
Kaasbjerg and Steen Lysgaard and Jón Bergmann Maronsson and Tristan Maxson and Thomas Olsen and Lars Pastewka and Andrew
Peterson and Carsten Rostgaard and Jakob Schiøtz and Ole Schütt and Mikkel Strange and Kristian S Thygesen and Tejs
Vegge and Lasse Vilhelmsen and Michael Walter and Zhenhua Zeng and Karsten W Jacobsen},
  title={{The atomic simulation environment—a Python library for working with atoms}},
  journal={Journal of Physics: Condensed Matter},
  volume={29},
  number={27},
  pages={273002},
  doi={10.1088/1361-648X/aa680e},
  url={http://stacks.iop.org/0953-8984/29/i=27/a=273002},
  year={2017}
}

@article{bussi,
    author = {Bussi, Giovanni and Donadio, Davide and Parrinello, Michele},
    title = {{Canonical sampling through velocity rescaling}},
    journal = {The Journal of Chemical Physics},
    volume = {126},
    number = {1},
    pages = {014101},
    year = {2007},
    month = {01},
    issn = {0021-9606},
    doi = {10.1063/1.2408420},
    url = {https://doi.org/10.1063/1.2408420},
}

@article{andersen,
    author = {Andersen, Hans C.},
    title = {{Molecular dynamics simulations at constant pressure and/or temperature}},
    journal = {The Journal of Chemical Physics},
    volume = {72},
    number = {4},
    pages = {2384-2393},
    year = {1980},
    month = {02},
    issn = {0021-9606},
    doi = {10.1063/1.439486},
    url = {https://doi.org/10.1063/1.439486},
    eprint = {https://pubs.aip.org/aip/jcp/article-pdf/72/4/2384/18920903/2384\_1\_online.pdf},
}

@article{PhysRevB.47.558,
  title = {{Ab initio molecular dynamics for liquid metals}},
  author = {Kresse, G. and Hafner, J.},
  journal = {Phys. Rev. B},
  volume = {47},
  issue = {1},
  pages = {558--561},
  numpages = {0},
  year = {1993},
  month = {Jan},
  publisher = {American Physical Society},
  doi = {10.1103/PhysRevB.47.558},
  url = {https://link.aps.org/doi/10.1103/PhysRevB.47.558}
}

@article{KRESSE199615,
title = {{Efficiency of ab-initio total energy calculations for metals and semiconductors using a plane-wave basis set}},
journal = {Computational Materials Science},
volume = {6},
number = {1},
pages = {15-50},
year = {1996},
issn = {0927-0256},
doi = {https://doi.org/10.1016/0927-0256(96)00008-0},
url = {https://www.sciencedirect.com/science/article/pii/0927025696000080},
author = {G. Kresse and J. Furthmüller}
}

@article{PhysRevB.54.11169,
  title = {{Efficient iterative schemes for ab initio total-energy calculations using a plane-wave basis set}},
  author = {Kresse, G. and Furthm\"uller, J.},
  journal = {Phys. Rev. B},
  volume = {54},
  issue = {16},
  pages = {11169--11186},
  numpages = {0},
  year = {1996},
  month = {Oct},
  publisher = {American Physical Society},
  doi = {10.1103/PhysRevB.54.11169},
  url = {https://link.aps.org/doi/10.1103/PhysRevB.54.11169}
}

@article{G_Kresse_1994,
doi = {10.1088/0953-8984/6/40/015},
url = {https://dx.doi.org/10.1088/0953-8984/6/40/015},
year = {1994},
month = {oct},
publisher = {},
volume = {6},
number = {40},
pages = {8245},
author = {G Kresse and J Hafner},
title = {{Norm-conserving and ultrasoft pseudopotentials for first-row and transition elements}},
journal = {Journal of Physics: Condensed Matter}
}

@article{PhysRevB.59.1758,
  title = {{From ultrasoft pseudopotentials to the projector augmented-wave method}},
  author = {Kresse, G. and Joubert, D.},
  journal = {Phys. Rev. B},
  volume = {59},
  issue = {3},
  pages = {1758--1775},
  numpages = {0},
  year = {1999},
  month = {Jan},
  publisher = {American Physical Society},
  doi = {10.1103/PhysRevB.59.1758},
  url = {https://link.aps.org/doi/10.1103/PhysRevB.59.1758}
}

@article{PhysRevLett.77.3865,
  title = {{Generalized Gradient Approximation Made Simple}},
  author = {Perdew, John P. and Burke, Kieron and Ernzerhof, Matthias},
  journal = {Phys. Rev. Lett.},
  volume = {77},
  issue = {18},
  pages = {3865--3868},
  numpages = {0},
  year = {1996},
  month = {Oct},
  publisher = {American Physical Society},
  doi = {10.1103/PhysRevLett.77.3865},
  url = {https://link.aps.org/doi/10.1103/PhysRevLett.77.3865}
}

@misc{cueq,
  title        = {{NVIDIA/cuEquivariance}},
  howpublished = {\url{https://github.com/NVIDIA/cuEquivariance}},
}

@article{Monacelli_2021,
doi = {10.1088/1361-648X/ac066b},
url = {https://dx.doi.org/10.1088/1361-648X/ac066b},
year = {2021},
month = {jul},
publisher = {IOP Publishing},
volume = {33},
number = {36},
pages = {363001},
author = {Monacelli, Lorenzo and Bianco, Raffaello and Cherubini, Marco and Calandra, Matteo and Errea, Ion and Mauri, Francesco},
title = {{The stochastic self-consistent harmonic approximation: calculating vibrational properties of materials with full quantum and anharmonic effects}},
journal = {Journal of Physics: Condensed Matter}
}

@article{bianco_tise2,
author = {Zhou, Jianqiang Sky and Monacelli, Lorenzo and Bianco, Raffaello and Errea, Ion and Mauri, Francesco and Calandra, Matteo},
title = {{Anharmonicity and Doping Melt the Charge Density Wave in Single-Layer TiSe$_{2}$}},
journal = {Nano Letters},
volume = {20},
number = {7},
pages = {4809-4815},
year = {2020},
doi = {10.1021/acs.nanolett.0c00597},
URL = {https://doi.org/10.1021/acs.nanolett.0c00597},
eprint = {https://doi.org/10.1021/acs.nanolett.0c00597}
}

@article{xrd_displ,
  title = {{X-ray study of the charge-density-wave transition in single-layer $\mathrm{TiS}{\mathrm{e}}_{2}$}},
  author = {Fang, Xin-Yue and Hong, Hawoong and Chen, Peng and Chiang, T.-C.},
  journal = {Phys. Rev. B},
  volume = {95},
  issue = {20},
  pages = {201409},
  numpages = {4},
  year = {2017},
  month = {May},
  publisher = {American Physical Society},
  doi = {10.1103/PhysRevB.95.201409},
  url = {https://link.aps.org/doi/10.1103/PhysRevB.95.201409}
}

@article{Giannozzi2017,
       author = {Giannozzi, P and Andreussi, O and Brumme, T and Bunau, O and Buongiorno Nardelli, M and Calandra, M and Car, R and Cavazzoni, C and Ceresoli, D and Cococcioni, M and Colonna, N and Carnimeo, I and Dal Corso, A and de Gironcoli, S and Delugas, P and DiStasio, R A and Ferretti, A and Floris, A and Fratesi, G and Fugallo, G and Gebauer, R and Gerstmann, U and Giustino, F and Gorni, T and Jia, J and Kawamura, M and Ko, H-Y and Kokalj, A and Küçükbenli, E and Lazzeri, M and Marsili, M and Marzari, N and Mauri, F and Nguyen, N L and Nguyen, H-V and Otero-de-la-Roza, A and Paulatto, L and Poncé, S and Rocca, D and Sabatini, R and Santra, B and Schlipf, M and Seitsonen, A P and Smogunov, A and Timrov, I and Thonhauser, T and Umari, P and Vast, N and Wu, X and Baroni, S},
        title = {{Advanced capabilities for materials modelling with \textsc{Quantum ESPRESSO}}},
      journal = {Journal of Physics: Condensed Matter},
       volume = {29},
       number = {46},
        pages = {465901},
         year = {2017},
          doi = {10.1088/1361-648X/aa8f79},
         url = {https://iopscience.iop.org/article/10.1088/1361-648X/aa8f79}
}

@article{freud2020,
    title = {{freud: A Software Suite for High Throughput
             Analysis of Particle Simulation Data}},
    author = {Vyas Ramasubramani and
              Bradley D. Dice and
              Eric S. Harper and
              Matthew P. Spellings and
              Joshua A. Anderson and
              Sharon C. Glotzer},
    journal = {Computer Physics Communications},
    volume = {254},
    pages = {107275},
    year = {2020},
    issn = {0010-4655},
    doi = {https://doi.org/10.1016/j.cpc.2020.107275},
    url = {http://www.sciencedirect.com/science/article/pii/S0010465520300916},
}

@article{amin2024,
  title = {{Two-step charge density wave transition and hidden transient phase in $1T\text{\ensuremath{-}}{\mathrm{TiSe}}_{2}$}},
  author = {Amin, R. and Pandey, J. and Beyer, K. and Petkov, V.},
  journal = {Phys. Rev. B},
  volume = {109},
  issue = {14},
  pages = {144106},
  numpages = {8},
  year = {2024},
  month = {Apr},
  publisher = {American Physical Society},
  doi = {10.1103/PhysRevB.109.144106},
  url = {https://link.aps.org/doi/10.1103/PhysRevB.109.144106}
}

@article{chen2016,
author = {Chen, P. and Chan, Y.-H. and Wong, M.-H. and Fang, X.-Y. and Chou, M. Y. and Mo, S.-K. and Hussain, Z. and Fedorov, A.-V. and Chiang, T.-C.},
title = {{Dimensional Effects on the Charge Density Waves in Ultrathin Films of TiSe$_{2}$}},
journal = {Nano Letters},
volume = {16},
number = {10},
pages = {6331},
year = {2016},
doi = {10.1021/acs.nanolett.6b02710},
URL = {https://doi.org/10.1021/acs.nanolett.6b02710}
}

@article{dynaphopy,
title = {{DynaPhoPy: A code for extracting phonon quasiparticles from molecular dynamics simulations}},
journal = {Computer Physics Communications},
volume = {221},
pages = {221-234},
year = {2017},
issn = {0010-4655},
doi = {https://doi.org/10.1016/j.cpc.2017.08.017},
url = {https://www.sciencedirect.com/science/article/pii/S0010465517302631},
author = {Abel Carreras and Atsushi Togo and Isao Tanaka},
}

@article{berges2024,
	title={{Ab initio electron-lattice downfolding: Potential energy landscapes, anharmonicity, and molecular dynamics in charge density wave materials}},
	author={Arne Schobert and Jan Berges and Erik G. C. P. van Loon and Michael A. Sentef and Sergey Brener and Mariana Rossi and Tim O. Wehling},
	journal={SciPost Phys.},
	volume={16},
	pages={046},
	year={2024},
	publisher={SciPost},
	doi={10.21468/SciPostPhys.16.2.046},
	url={https://scipost.org/10.21468/SciPostPhys.16.2.046},
}

@misc{elphmod_code,
  author       = {Berges, J. and Schobert, A. and van Loon, E. G. C. P. and R{\"o}sner, M. and Wehling, T. O.},
  title        = {{elphmod: Python modules for electron-phonon models (v0.32)}},
  year         = {2025},
  publisher    = {Zenodo},
  doi          = {10.5281/zenodo.17702034},
  url          = {https://doi.org/10.5281/zenodo.17702034}
}

@article{xu2020, title={{Spontaneous gyrotropic electronic order in a transition-metal dichalcogenide}}, volume={578}, ISSN={1476-4687}, url={http://dx.doi.org/10.1038/s41586-020-2011-8}, DOI={10.1038/s41586-020-2011-8}, number={7796}, journal={Nature}, publisher={Springer Science and Business Media LLC}, author={Xu, Su-Yang and Ma, Qiong and Gao, Yang and Kogar, Anshul and Zong, Alfred and Mier Valdivia, Andrés M. and Dinh, Thao H. and Huang, Shin-Ming and Singh, Bahadur and Hsu, Chuang-Han and Chang, Tay-Rong and Ruff, Jacob P. C. and Watanabe, Kenji and Taniguchi, Takashi and Lin, Hsin and Karapetrov, Goran and Xiao, Di and Jarillo-Herrero, Pablo and Gedik, Nuh}, year={2020}, month=feb, pages={545} }

@article{chen2015, title={{Charge density wave transition in single-layer titanium diselenide}}, volume={6}, ISSN={2041-1723}, url={http://dx.doi.org/10.1038/ncomms9943}, DOI={10.1038/ncomms9943}, number={1}, journal={Nature Communications}, publisher={Springer Science and Business Media LLC}, author={Chen, P and Chan, Y. -H. and Fang, X. -Y. and Zhang, Y and Chou, M Y and Mo, S. -K. and Hussain, Z and Fedorov, A. -V. and Chiang, T. -C.}, year={2015}, pages={8943} }

@article{togo2023,
doi = {10.1088/1361-648X/acd831},
url = {https://dx.doi.org/10.1088/1361-648X/acd831},
year = {2023},
month = {jun},
publisher = {IOP Publishing},
volume = {35},
number = {35},
pages = {353001},
author = {Togo, Atsushi and Chaput, Laurent and Tadano, Terumasa and Tanaka, Isao},
title = {{Implementation strategies in phonopy and phono3py}},
journal = {Journal of Physics: Condensed Matter}
}

@misc{edwards2026,
      title={{Ferroaxial and nematic transitions in the charge density wave phase of 1T-TiSe$_2$}}, 
      author={Sarah Edwards and Elliott Rosenberg and Ilaria Maccari and Jiaqin Wen and Chaowei Hu and Xiaodong Xu and Jong-Woo Kim and Philip J. Ryan and Rafael M. Fernandes and Fernando de Juan and Maria N. Gastiasoro and Jiun-Haw Chu},
      year={2026},
      eprint={2603.14614},
      archivePrefix={arXiv},
      primaryClass={cond-mat.str-el},
      url={https://arxiv.org/abs/2603.14614}, 
}

@misc{jiang2026,
      title={{Evidence for ferroaxial order in 1T-TiSe$_2$ via elastoresistivity measurements}}, 
      author={Qianni Jiang and Ezra Day-Roberts and Benito Gonzalez and Awadhesh Das and Darius H. Torchinsky and Turan Birol and Rafael M. Fernandes and Ian R. Fisher},
      year={2026},
      eprint={2603.14613},
      archivePrefix={arXiv},
      primaryClass={cond-mat.str-el},
      url={https://arxiv.org/abs/2603.14613}, 
}

@article{zhang2025,
  title = {{Revealing the anisotropic charge-density-wave order of ${\mathrm{TiSe}}_{2}$ through high harmonic generation}},
  author = {Zhang, Lin and Tyulnev, Igor and Vamos, Lenard and Poborska, Julita and Bhattacharya, Utso and Chhajlany, Ravindra W. and Grass, Tobias and Biegert, Jens and Lewenstein, Maciej},
  journal = {Phys. Rev. B},
  volume = {112},
  issue = {8},
  pages = {085147},
  numpages = {9},
  year = {2025},
  month = {Aug},
  publisher = {American Physical Society},
  doi = {10.1103/812z-nl4m},
  url = {https://link.aps.org/doi/10.1103/812z-nl4m}
}

@article{lee2021, title={{Metal-to-insulator transition in Pt-doped TiSe2 driven by emergent network of narrow transport channels}}, volume={6}, ISSN={2397-4648}, url={http://dx.doi.org/10.1038/s41535-020-00305-2}, DOI={10.1038/s41535-020-00305-2}, number={1}, journal={npj Quantum Materials}, publisher={Springer Science and Business Media LLC}, author={Lee, Kyungmin and Choe, Jesse and Iaia, Davide and Li, Juqiang and Zhao, Junjing and Shi, Ming and Ma, Junzhang and Yao, Mengyu and Wang, Zhenyu and Huang, Chien-Lung and Ochi, Masayuki and Arita, Ryotaro and Chatterjee, Utpal and Morosan, Emilia and Madhavan, Vidya and Trivedi, Nandini}, year={2021}, pages={8} }

@article{park2023, title={{Condensation of preformed charge density waves in kagome metals}}, volume={14}, ISSN={2041-1723}, url={http://dx.doi.org/10.1038/s41467-023-43170-w}, DOI={10.1038/s41467-023-43170-w}, number={1}, journal={Nature Communications}, publisher={Springer Science and Business Media LLC}, author={Park, Changwon and Son, Young-Woo}, year={2023}, pages={7309} }

@article{cheng2022, title={{Light-induced dimension crossover dictated by excitonic correlations}}, volume={13}, ISSN={2041-1723}, url={http://dx.doi.org/10.1038/s41467-022-28309-5}, DOI={10.1038/s41467-022-28309-5}, number={1}, journal={Nature Communications}, publisher={Springer Science and Business Media LLC}, author={Cheng, Yun and Zong, Alfred and Li, Jun and Xia, Wei and Duan, Shaofeng and Zhao, Wenxuan and Li, Yidian and Qi, Fengfeng and Wu, Jun and Zhao, Lingrong and Zhu, Pengfei and Zou, Xiao and Jiang, Tao and Guo, Yanfeng and Yang, Lexian and Qian, Dong and Zhang, Wentao and Kogar, Anshul and Zuerch, Michael W. and Xiang, Dao and Zhang, Jie}, year={2022}, pages={963} }

@article{cheng2023,
  title = {{Machine learning for phase ordering dynamics of charge density waves}},
  author = {Cheng, Chen and Zhang, Sheng and Chern, Gia-Wei},
  journal = {Phys. Rev. B},
  volume = {108},
  issue = {1},
  pages = {014301},
  numpages = {15},
  year = {2023},
  month = {Jul},
  publisher = {American Physical Society},
  doi = {10.1103/PhysRevB.108.014301},
  url = {https://link.aps.org/doi/10.1103/PhysRevB.108.014301}
}

@article{mcmillan1976,
  title = {{Theory of discommensurations and the commensurate-incommensurate charge-density-wave phase transition}},
  author = {McMillan, W. L.},
  journal = {Phys. Rev. B},
  volume = {14},
  issue = {4},
  pages = {1496--1502},
  numpages = {0},
  year = {1976},
  month = {Aug},
  publisher = {American Physical Society},
  doi = {10.1103/PhysRevB.14.1496},
  url = {https://link.aps.org/doi/10.1103/PhysRevB.14.1496}
}

@article{lian2019,
  title = {{Charge density wave hampers exciton condensation in $1T\text{\ensuremath{-}}{\mathrm{TiSe}}_{2}$}},
  author = {Lian, Chao and Ali, Zulfikhar A. and Wong, Bryan M.},
  journal = {Phys. Rev. B},
  volume = {100},
  issue = {20},
  pages = {205423},
  numpages = {10},
  year = {2019},
  month = {Nov},
  publisher = {American Physical Society},
  doi = {10.1103/PhysRevB.100.205423},
  url = {https://link.aps.org/doi/10.1103/PhysRevB.100.205423}
}

@article{chen2023,
  title = {{Development of an $ab initio$ method for exciton condensation and its application to ${\mathrm{TiSe}}_{2}$}},
  author = {Chen, Hsiao-Yi and Nomoto, Takuya and Arita, Ryotaro},
  journal = {Phys. Rev. Res.},
  volume = {5},
  issue = {4},
  pages = {043183},
  numpages = {20},
  year = {2023},
  month = {Nov},
  publisher = {American Physical Society},
  doi = {10.1103/PhysRevResearch.5.043183},
  url = {https://link.aps.org/doi/10.1103/PhysRevResearch.5.043183}
}

@article{bianco2015,
  title = {{Electronic and vibrational properties of ${\mathrm{TiSe}}_{2}$ in the charge-density-wave phase from first principles}},
  author = {Bianco, Raffaello and Calandra, Matteo and Mauri, Francesco},
  journal = {Phys. Rev. B},
  volume = {92},
  issue = {9},
  pages = {094107},
  numpages = {19},
  year = {2015},
  month = {Sep},
  publisher = {American Physical Society},
  doi = {10.1103/PhysRevB.92.094107},
  url = {https://link.aps.org/doi/10.1103/PhysRevB.92.094107}
}

@article{woo1976lattice,
  title = {{Superlattice formation in titanium diselenide}},
  author = {Woo, K. C. and Brown, F. C. and McMillan, W. L. and Miller, R. J. and Schaffman, M. J. and Sears, M. P.},
  journal = {Phys. Rev. B},
  volume = {14},
  issue = {8},
  pages = {3242--3247},
  numpages = {0},
  year = {1976},
  month = {Oct},
  publisher = {American Physical Society},
  doi = {10.1103/PhysRevB.14.3242},
  url = {https://link.aps.org/doi/10.1103/PhysRevB.14.3242}
}

@article{miyahara1995sts,
doi = {10.1088/0953-8984/7/13/006},
url = {https://dx.doi.org/10.1088/0953-8984/7/13/006},
year = {1995},
volume = {7},
number = {13},
pages = {2553},
author = {Y Miyahara and H Bando and H Ozaki},
title = {{Tunnelling spectroscopic study of the CDW energy gap in TiSe2}},
journal = {Journal of Physics: Condensed Matter}
}

@misc{lv2026,
      title={{Nematic-fluctuation-mediated superconductivity in CuxTiSe2}}, 
      author={Xingyu Lv and Yang Fu and Shangjie Tian and Ying Ma and Shouguo Wang and Cedomir Petrovic and Xiao Zhang and Hechang Lei},
      year={2026},
      eprint={2601.00723},
      archivePrefix={arXiv},
      primaryClass={cond-mat.supr-con},
      url={https://arxiv.org/abs/2601.00723}, 
}

@article{liu2024,
author = {Liu, Cheng-Yen and Zhao, Meng and Wang, Zhongjie and Gao, Chun-Lei},
title = {{P-d Correlation-Determined Charge Order Stiffness and Corresponding Quantum Melting in Monolayer 1T-TiSe2}},
journal = {ACS Nano},
volume = {18},
number = {47},
pages = {32858},
year = {2024},
doi = {10.1021/acsnano.4c11704},
URL = { https://doi.org/10.1021/acsnano.4c11704}
}

@article{vodeb2019,
doi = {10.1088/1367-2630/ab3057},
url = {https://doi.org/10.1088/1367-2630/ab3057},
year = {2019},
month = {aug},
publisher = {IOP Publishing},
volume = {21},
number = {8},
pages = {083001},
author = {Vodeb, Jaka and Kabanov, Viktor V and Gerasimenko, Yaroslav A and Venturini, Rok and Jan Ravnik and van Midden, Marion A and Zupanic, Erik and Sutar, Petra and Mihailovic, Dragan},
title = {{Configurational electronic states in layered transition metal dichalcogenides}},
journal = {New Journal of Physics}
}

@article{kusmartseva2009,
  title = {{Pressure Induced Superconductivity in Pristine $1T\mathrm{\text{\ensuremath{-}}}{\mathrm{TiSe}}_{2}$}},
  author = {Kusmartseva, A. F. and Sipos, B. and Berger, H. and Forr\'o, L. and Tuti\ifmmode \check{s}\else \v{s}\fi{}, E.},
  journal = {Phys. Rev. Lett.},
  volume = {103},
  issue = {23},
  pages = {236401},
  numpages = {4},
  year = {2009},
  month = {Nov},
  publisher = {American Physical Society},
  doi = {10.1103/PhysRevLett.103.236401},
  url = {https://link.aps.org/doi/10.1103/PhysRevLett.103.236401}
}

@article{morosan2006,
	doi = {10.1038/nphys360},
	url = {https://doi.org/10.1038/nphys360},
	year = 2006,
	month = {jul},
	publisher = {Springer Science and Business Media {LLC}},
	volume = {2},
	number = {8},
	pages = {544},
	author = {E. Morosan and H. W. Zandbergen and B. S. Dennis and J. W. G. Bos and Y. Onose and T. Klimczuk and A. P. Ramirez and N. P. Ong and R. J. Cava},
	title = {{Superconductivity in {CuxTiSe}2}},
	journal = {Nature Physics}
}

@article{flicker15, title={{Charge order from orbital-dependent coupling evidenced by NbSe2}}, volume={6}, ISSN={2041-1723}, url={http://dx.doi.org/10.1038/ncomms8034}, DOI={10.1038/ncomms8034}, number={1}, journal={Nature Communications}, publisher={Springer Science and Business Media LLC}, author={Flicker, Felix and van Wezel, Jasper}, year={2015}, pages={7034} }

@article{spera2019,
  title = {{Energy-dependent spatial texturing of charge order in $1T\ensuremath{-}{\mathrm{Cu}}_{x}{\mathrm{TiSe}}_{2}$}},
  author = {Spera, M. and Scarfato, A. and Giannini, E. and Renner, Ch.},
  journal = {Phys. Rev. B},
  volume = {99},
  issue = {15},
  pages = {155133},
  numpages = {7},
  year = {2019},
  month = {Apr},
  publisher = {American Physical Society},
  doi = {10.1103/PhysRevB.99.155133},
  url = {https://link.aps.org/doi/10.1103/PhysRevB.99.155133}
}

@article{hildebrand2016,
  title = {{Short-range phase coherence and origin of the $1T\ensuremath{-}{\mathrm{TiSe}}_{2}$ charge density wave}},
  author = {Hildebrand, B. and Jaouen, T. and Didiot, C. and Razzoli, E. and Monney, G. and Mottas, M.-L. and Ubaldini, A. and Berger, H. and Barreteau, C. and Beck, H. and Bowler, D. R. and Aebi, P.},
  journal = {Phys. Rev. B},
  volume = {93},
  issue = {12},
  pages = {125140},
  numpages = {5},
  year = {2016},
  month = {Mar},
  publisher = {American Physical Society},
  doi = {10.1103/PhysRevB.93.125140},
  url = {https://link.aps.org/doi/10.1103/PhysRevB.93.125140}
}

@article{xiaolong2021,
author = {Xiaolong Liu  and Yi Xue Chong  and Rahul Sharma  and J. C. Séamus Davis },
title = {{Discovery of a Cooper-pair density wave state in a transition-metal dichalcogenide}},
journal = {Science},
volume = {372},
number = {6549},
pages = {1447-1452},
year = {2021},
doi = {10.1126/science.abd4607},
URL = {https://www.science.org/doi/abs/10.1126/science.abd4607}
}

@article{watson2021,
doi = {10.1088/2053-1583/abafec},
url = {https://doi.org/10.1088/2053-1583/abafec},
year = {2020},
month = {oct},
publisher = {IOP Publishing},
volume = {8},
number = {1},
pages = {015004},
author = {Watson, Matthew D and Rajan, Akhil and Antonelli, Tommaso and Underwood, Kaycee and Marković, Igor and Mazzola, Federico and Clark, Oliver J and Siemann, Gesa-Roxanne and Biswas, Deepnarayan and Hunter, Andrew and Jandura, Sven and Reichstetter, Janika and McLaren, Martin and Le Fèvre, Patrick and Vinai, Giovanni and King, Philip D C},
title = {{Strong-coupling charge density wave in monolayer TiSe2}},
journal = {2D Materials}
}

@article{li2007,
  title = {{Semimetal-to-Semimetal Charge Density Wave Transition in $1T\mathrm{\text{\ensuremath{-}}}{\mathrm{TiSe}}_{2}$}},
  author = {Li, G. and Hu, W. Z. and Qian, D. and Hsieh, D. and Hasan, M. Z. and Morosan, E. and Cava, R. J. and Wang, N. L.},
  journal = {Phys. Rev. Lett.},
  volume = {99},
  issue = {2},
  pages = {027404},
  numpages = {4},
  year = {2007},
  month = {Jul},
  publisher = {American Physical Society},
  doi = {10.1103/PhysRevLett.99.027404},
  url = {https://link.aps.org/doi/10.1103/PhysRevLett.99.027404}
}

@article{wegner2020,
  title = {{Evidence for pseudo--Jahn-Teller distortions in the charge density wave phase of $1T\text{\ensuremath{-}}{\mathrm{TiSe}}_{2}$}},
  author = {Wegner, A. and Zhao, J. and Li, J. and Yang, J. and Anikin, A. A. and Karapetrov, G. and Esfarjani, K. and Louca, D. and Chatterjee, U.},
  journal = {Phys. Rev. B},
  volume = {101},
  issue = {19},
  pages = {195145},
  numpages = {6},
  year = {2020},
  month = {May},
  publisher = {American Physical Society},
  doi = {10.1103/PhysRevB.101.195145},
  url = {https://link.aps.org/doi/10.1103/PhysRevB.101.195145}
}

@article{higashihira2026, title={{Unexpected stabilization of a single-q charge density wave in pristine 1T-TiSe2}}, ISSN={2041-1723}, url={http://dx.doi.org/10.1038/s41467-026-78072-0}, DOI={10.1038/s41467-026-78072-0}, journal={Nature Communications}, publisher={Springer Science and Business Media LLC}, author={Higashihira Han, Kimoon and Hwang, Dongju and Kim, Dongsin and Ahn, Yeojin and Cha, Jaehun and Gim, Seonggeon and Lee, Gyubin and Jho, Mingi and Lim, Chan-young and Ryu, Sae Hee and Jozwiak, Chris and Bostwick, Aaron and Rotenberg, Eli and Denlinger, Jonathan D. and Hashimoto, Makoto and Lu, Donghui and Choi, Young Woo and Cho, Gil Young and Kim, Yeongkwan}, volume={17}, pages={11220}, year={2026}}

@Article{harris2020array,
 title         = {{Array programming with NumPy}},
 author        = {Charles R. Harris and K. Jarrod Millman and St{\'{e}}fan J.
                 van der Walt and Ralf Gommers and Pauli Virtanen and David
                 Cournapeau and Eric Wieser and Julian Taylor and Sebastian
                 Berg and Nathaniel J. Smith and Robert Kern and Matti Picus
                 and Stephan Hoyer and Marten H. van Kerkwijk and Matthew
                 Brett and Allan Haldane and Jaime Fern{\'{a}}ndez del
                 R{\'{i}}o and Mark Wiebe and Pearu Peterson and Pierre
                 G{\'{e}}rard-Marchant and Kevin Sheppard and Tyler Reddy and
                 Warren Weckesser and Hameer Abbasi and Christoph Gohlke and
                 Travis E. Oliphant},
 year          = {2020},
 month         = sep,
 journal       = {Nature},
 volume        = {585},
 number        = {7825},
 pages         = {357--362},
 doi           = {10.1038/s41586-020-2649-2},
 publisher     = {Springer Science and Business Media {LLC}},
 url           = {https://doi.org/10.1038/s41586-020-2649-2}
}

@misc{vangrondelle2026,
      title={{Infrared evidence for strong $C_{3}$ symmetry breaking in 1$T$-TiSe$_{2}$}}, 
      author={Esther van Grondelle and Kai Rossnagel and Jasper van Wezel and Erik van Heumen},
      year={2026},
      eprint={2609.24705},
      archivePrefix={arXiv},
      primaryClass={cond-mat.str-el},
      url={https://arxiv.org/abs/2609.24705}, 
      doi={10.48550/arXiv.2609.24705}
}

@article{lawler2010,
  author    = {Lawler, M. J. and Fujita, K. and Lee, Jhinhwan and Schmidt, A. R. and Kohsaka, Y. and Kim, Chung Koo and Eisaki, H. and Uchida, S. and Davis, J. C. and Sethna, J. P. and Kim, Eun-Ah},
  title     = {{Intra-unit-cell electronic nematicity of the high-{$T_c$} copper-oxide pseudogap states}},
  journal   = {Nature},
  volume    = {466},
  number    = {7304},
  pages     = {347},
  year      = {2010},
  month     = jul,
  doi       = {10.1038/nature09169},
  url       = {https://www.nature.com/articles/nature09169}
}

@article{mesaros2011,
author = {A. Mesaros  and K. Fujita  and H. Eisaki  and S. Uchida  and J. C. Davis  and S. Sachdev  and J. Zaanen  and M. J. Lawler  and Eun-Ah Kim },
title = {{Topological Defects Coupling Smectic Modulations to Intra–Unit-Cell Nematicity in Cuprates}},
journal = {Science},
volume = {333},
number = {6041},
pages = {426},
year = {2011},
doi = {10.1126/science.1201082},
URL = {https://www.science.org/doi/abs/10.1126/science.1201082},
eprint = {https://www.science.org/doi/pdf/10.1126/science.1201082}
}

@article{gu2023,
  author    = {Gu, Qiangqiang and Carroll, Joseph P. and Wang, Shuqiu and Ran, Sheng and Broyles, Christopher and Siddiquee, Hasan and Butch, Nicholas P. and Saha, Shanta R. and Paglione, Johnpierre and Davis, J. C. S{\'e}amus and Liu, Xiaolong},
  title     = {{Detection of a pair density wave state in {UTe$_2$}}},
  journal   = {Nature},
  volume    = {618},
  number    = {7967},
  pages     = {921},
  year      = {2023},
  month     = jun,
  doi       = {10.1038/s41586-023-05919-7},
  url       = {https://www.nature.com/articles/s41586-023-05919-7}
}

@article{fujita2014,
  author    = {Fujita, Kazuhiro and Hamidian, Mohammad H. and Edkins, Stephen D. and Kim, Chung Koo and Kohsaka, Yuhki and Azuma, Masaki and Takano, Mikio and Takagi, Hidenori and Eisaki, Hiroshi and Uchida, Shin-ichi and Allais, Andrea and Lawler, Michael J. and Kim, Eun-Ah and Sachdev, Subir and Davis, J. C. S{\'e}amus},
  title     = {{Direct phase-sensitive identification of a d-form factor density wave in underdoped cuprates}},
  journal   = {Proceedings of the National Academy of Sciences},
  volume    = {111},
  number    = {30},
  pages     = {E3026--E3032},
  year      = {2014},
  month     = jul,
  doi       = {10.1073/pnas.1406297111},
  url       = {https://www.pnas.org/doi/10.1073/pnas.1406297111}
}

\end{document}


\maketitle
\newpage


\section{Methods}
\subsection{Machine learning interatomic potential}

To construct a dataset for monolayer TiSe$_{2}$, we performed DFT calculations using the Vienna Ab initio Simulation Package (VASP) software version 6.4.2\,\cite{PhysRevB.47.558, KRESSE199615, PhysRevB.54.11169, G_Kresse_1994, PhysRevB.59.1758} with the Perdew-Burke-Ernzerhof (PBE) functional\,\cite{PhysRevLett.77.3865}. For the plane-wave basis set energy cutoff we used $270$ eV. To obtain accurate results, we used fine sampling of the Brillouin zone of $\mathrm{k}=93\times93\times1$ for the unit cell, and for the larger supercells it was scaled accordingly. Also, we have used the Methfessel-Paxton scheme of order $1$ with a smearing value of $0.0043$ eV. To collect data, we performed molecular dynamics with the Andersen thermostat\,\cite{andersen} spanning temperatures from $10$ to $500$\,K and the Bayesian-learning algorithm for on-the-fly machine learning as implemented in VASP. Our dataset consists of $2180$ different structures in total, with supercell sizes of $2\times2\times1$, $3\times3\times1$, $4\times4\times1$ and $8\times8\times1$. Lattice constant used for calculations was $3.54$~\AA. The dataset was randomly split into $80\%$ training and $20\%$ test sets. To train Machine Learning (ML) interatomic potential for monolayer TiSe$_{2}$ we used MACE framework\,\cite{Batatia2022Design, NEURIPS2022_4a36c3c5}, where most of the training hyperparameters were kept to the default values (see Appendix A of Ref.\,\citenum{NEURIPS2022_4a36c3c5}).
The cutoff radius was set to $8$~\AA~and we used $8$ Bessel basis functions, $128$ embedding channels and $l_{\rm max}=2$, namely, we included equivariant messages up to the second order. In the weighted loss function, energy ($\lambda_{\rm E}$) and forces ($\lambda_{\rm F}$) weights were set to $1$ and $100$, respectively. The model was trained for $2000$ epochs, with batch size of $5$. $80$\% of the original structures were taken as a training set, where $5$\% of them were taken for the validation set, and the other $20$\% of the original structures comprised a test set. Stochastic weight averaging\,\cite{izmailov2019averagingweightsleadswider, athiwaratkun2019consistentexplanationsunlabeleddata} was performed for the last 20\% of epochs, and in this part of training $\lambda_{E}$ weight was set to $1000$ and $\lambda_{F}$ weight was kept at $100$.

We also fine-tuned MACE interatomic potential trained on PBE data using r$^{2}$SCAN data. We built up a dataset of r$^{2}$SCAN data by sampling $171$ representative points from the PBE dataset using the Farthest Point Sampling Algorithm.

\begin{figure}[H]
    \centering
    \includegraphics[scale=0.9]{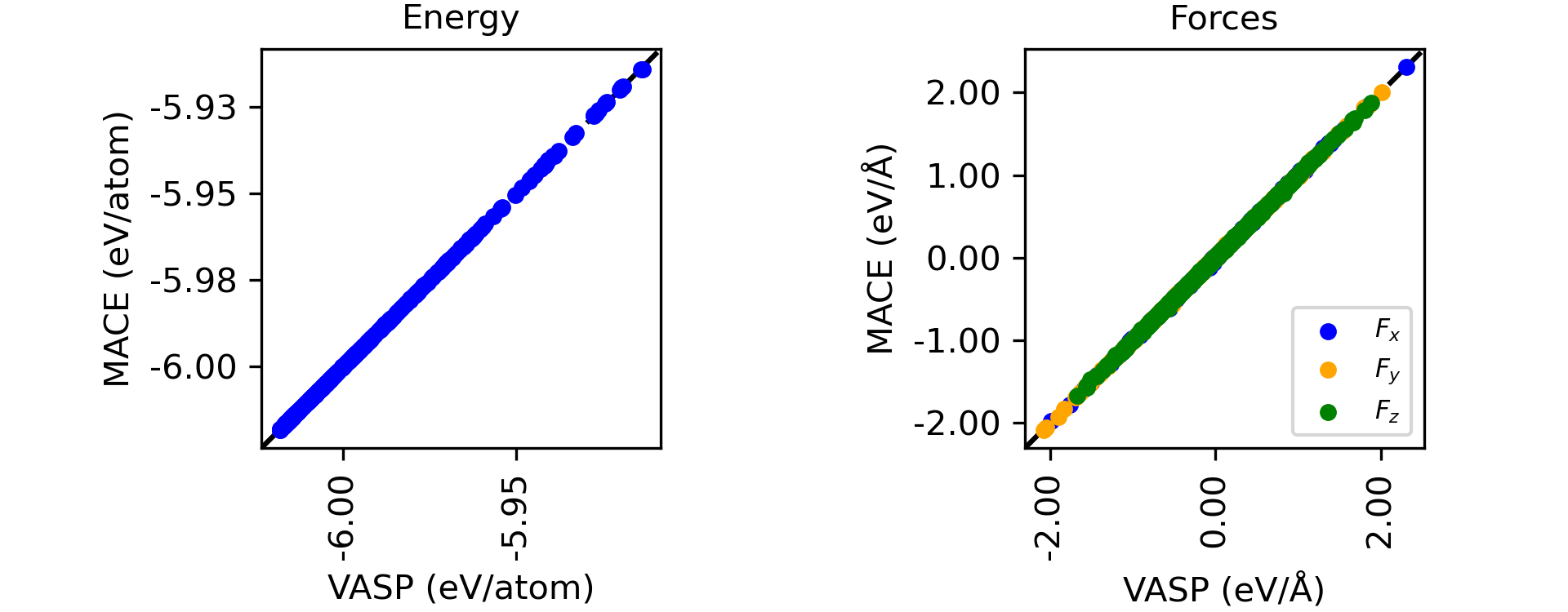}
    \caption{Comparison of MACE ML potential trained using PBE data and DFT results on the test set.}
    \label{fig:MACE_vs_DFT}
\end{figure}

\begin{figure}[H]
    \centering
    \includegraphics[scale=0.9]{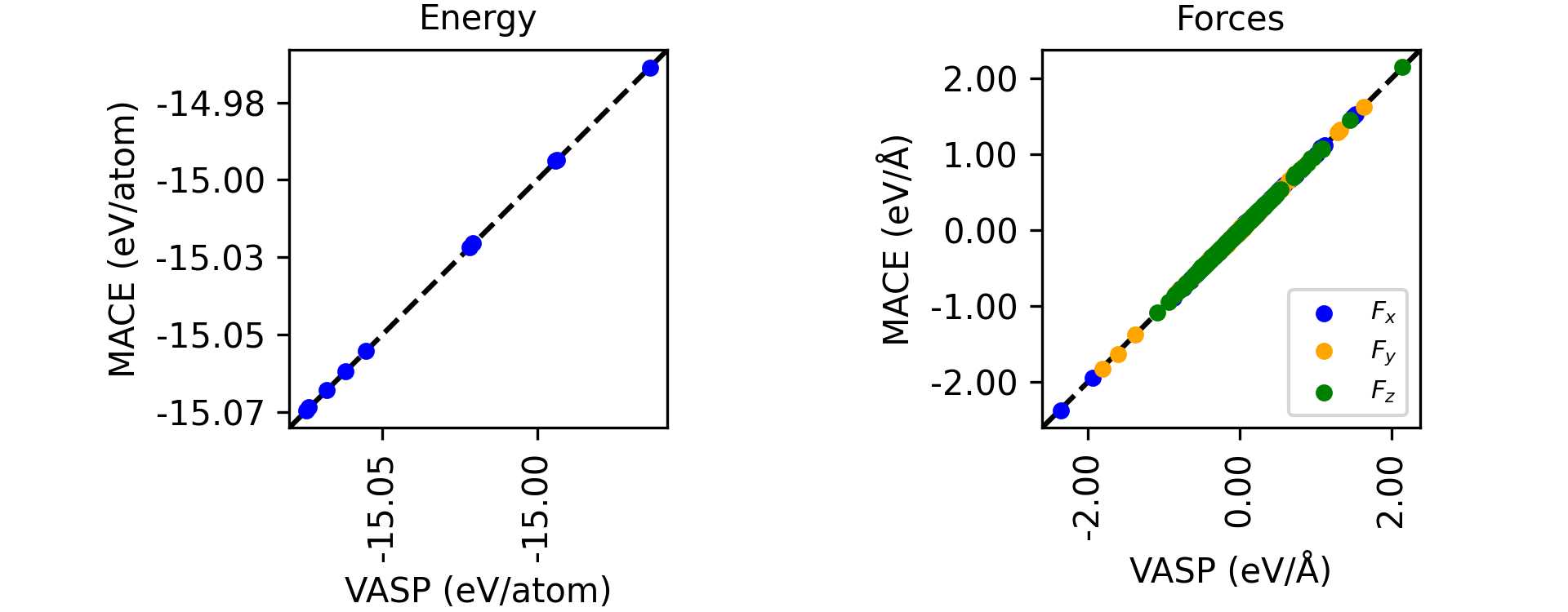}
    \caption{Comparison of MACE ML potential fine-tuned using r$^{2}$SCAN data and DFT results on the test set.}
    \label{fig:MACE_vs_DFT_r2scan}
\end{figure}

In Fig.\,\ref{fig:MACE_vs_DFT} we report the comparison of the MACE ML potential trained using PBE data and the DFT results on the test set and in Table \ref{table:mace_vs_dft} we report metrics evaluated on the test set.
\begin{table}[H]
\caption{Performance of the MACE model trained using PBE data evaluated on the test set.}
\label{table:mace_vs_dft}
\centering
\begin{tabular}{|c|c|c|c|}
\hline
\text{quantity} & \text{MAE} & \text{RMSE} & \text{R}$^{2}$ \\
\hline\hline
\text{energy}&$0.05$ \text{meV/atom}&$0.07$ \text{meV/atom}&$1.00$\\
\hline
\text{F$_{x}$}&$1.89$ \text{meV/Å}&$3.05$ \text{meV/Å}&$1.00$\\
\hline
\text{F$_{y}$}&$1.90$ \text{meV/Å}&$3.01$ \text{meV/Å}&$1.00$\\
\hline
\text{F$_{z}$}&$1.59$ \text{meV/Å}&$2.65$ \text{meV/Å}&$1.00$\\
\hline
\end{tabular}
\end{table}
On the other hand, in Fig.\,\ref{fig:MACE_vs_DFT_r2scan} we report the comparison of the MACE ML potential trained using PBE data and fine-tuned using r$^{2}$SCAN data and the DFT results on the test set and in Table \ref{table:mace_vs_dft_r2scan} we report metrics evaluated on the test set.
\begin{table}[H]
\caption{Performance of the MACE model fine-tuned using r$^{2}$SCAN data evaluated on the test set.}
\label{table:mace_vs_dft_r2scan}
\centering
\begin{tabular}{|c|c|c|c|}
\hline
\text{quantity} & \text{MAE} & \text{RMSE} & \text{R}$^{2}$ \\
\hline\hline
\text{energy}&$0.18$ \text{meV/atom}&$0.19$ \text{meV/atom}&$1.00$\\
\hline
\text{F$_{x}$}&$5.20$ \text{meV/Å}&$6.80$ \text{meV/Å}&$1.00$\\
\hline
\text{F$_{y}$}&$5.59$ \text{meV/Å}&$7.69$ \text{meV/Å}&$1.00$\\
\hline
\text{F$_{z}$}&$3.86$ \text{meV/Å}&$5.13$ \text{meV/Å}&$1.00$\\
\hline
\end{tabular}
\end{table}

\begin{figure}[H]
    \centering
    \includegraphics[scale=0.525]{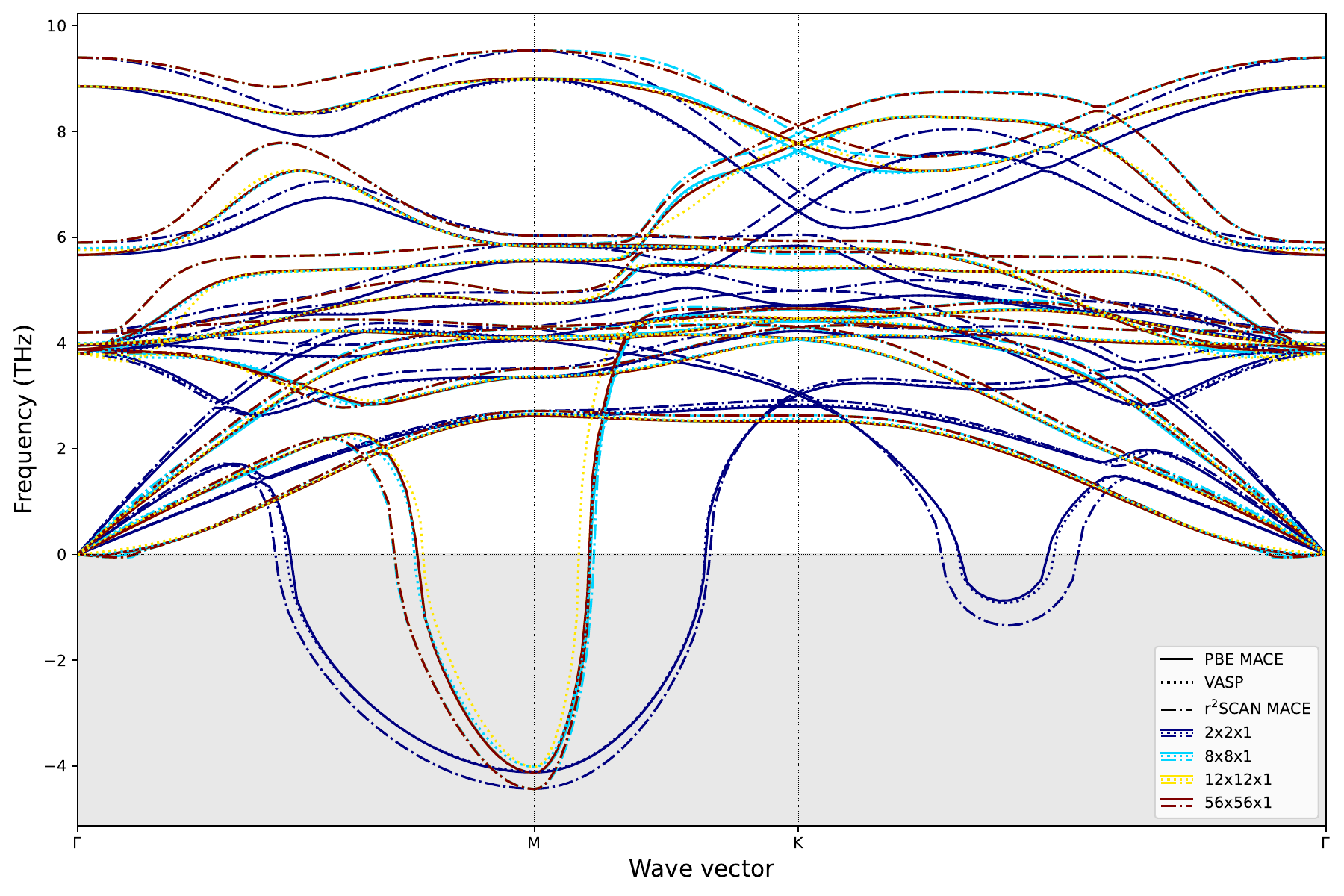}
    \caption{Comparison of harmonic phonon dispersions (at $T\approx 0$\,K) calculated using DFT and MACE interatomic potentials, one trained using only PBE data (solid lines) and the PBE based one fine-tuned using r$^{2}$SCAN data (dash-dotted lines).}
    \label{fig:MACE_vs_DFT_diff_cells}
\end{figure}

To further estimate the precision of our ML interatomic potential, we calculated phonon dispersions in the harmonic limit, as shown in Fig.\,\ref{fig:MACE_vs_DFT_diff_cells}. We see that for supercells larger than $8\times8\times1$ ML potential converges, as expected from the construction of the dataset.
%
\subsection{Molecular dynamics}
\begin{figure}[H]
    \centering
    \includegraphics[width=\textwidth]{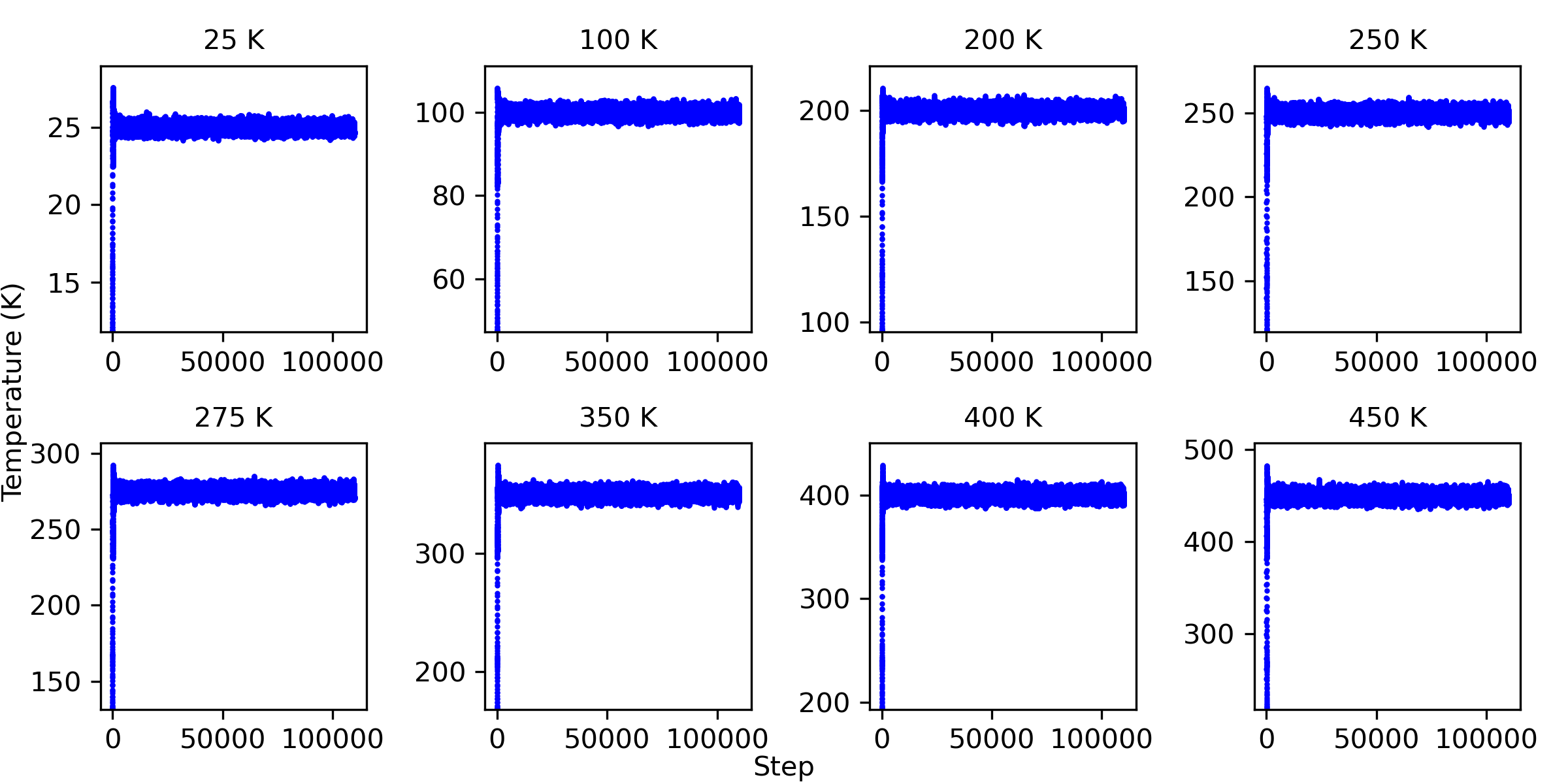}
    \caption{MD thermalization for multiple different temperatures.}
    \label{fig:thermalisation}
\end{figure}

Molecular Dynamics (MD) simulations were performed using Bussi NVT\,\cite{bussi} MD implemented in Atomic Simulation Environment (ASE) \cite{ISI:000175131400009, ase-paper} combined with our MACE interatomic potentials. To make MD simulations more efficient, we used NVIDIA's cuEquivariance package \cite{cueq}. We ran multiple MD simulations covering temperatures from $25$\,K to $450$\,K. All MDs ran with a time step of $1.5$ fs and thermostat time constant of $50$ fs. We set center-of-mass momentum to $0$ every $10$ steps. We ran MDs in a $56\times56\times1$ supercell, which we construct from normal phase unit cell. Using a $56\times56\times1$ supercell gives a total of $9408$ atoms in our MD simulations, since there are $3$ atoms in the unit cell of the monolayer TiSe$_{2}$. We initialize velocity distributions with the Maxwell-Boltzmann distribution. Each MD ran for $110000$ steps and we take last $100000$ steps for postprocessing, because in that time window all MDs are thermalized as can be seen in Fig.\,\ref{fig:thermalisation}.

\subsection{Time-averaged structure factors}
The structure factor for a given $\mathbf{q}$ point, temperature $T$ and time step $t$ is defined as\,\cite{berges2024}
\begin{align*}
    S(\mathbf{q}, T, t) = \frac{1}{N^{2}_{\mathrm{atoms}}}\left|\sum_{j}e^{-i\mathbf{q}\cdot\mathbf{R}_{j}(t)}\right|^{2}~,
\end{align*}
where $N_{\mathrm{atoms}}$ is the total number of atoms in the MD simulation cell and the sum runs over all atoms. Using the equation from above, we define the time-averaged structure factor as
\begin{align}
    \langle S(\mathbf{q}, T)\rangle = \frac{1}{N_{\mathrm{steps}}}\sum_{t}S(\mathbf{q}, T, t) = \frac{1}{N_{\mathrm{steps}}N^{2}_{\mathrm{atoms}}}\sum_{t}\left|\sum_{j}e^{-i\mathbf{q}\cdot\mathbf{R}_{j}(t)}\right|^{2}~.\label{eq:time_averaged_structure_factor}
\end{align}

\subsection{Average atomic displacements}
\label{subsec:average_atomic_displ}
To calculate the average atomic displacements in the $xy$ plane in Fig.\,1b, we folded the atomic positions from the original $56\times56\times1$ supercell onto a normal phase $2\times2\times1$ supercell, which is the smallest supercell compatible with the CDW phase at the $\mathbf{q}=\mathrm{M}$ point. This serves as a method of averaging, the main reason for this can be seen in Fig.\,2a. Namely, we see that for all temperatures there are regions where atoms are highly displaced even though globally they follow the expected behavior where the system transitions from CDW to normal phase as the temperature increases. The distance between the time-averaged position of the $i$-th atom from the MD simulation cell $\langle\mathbf{R}_{i}\rangle$ and the $i$-th atom from the normal phase reference structure $\mathbf{R}^{0}_{i}$ in Fig.\,2a is calculated as
\begin{align}
    \left|\langle\mathbf{R}_{i}\rangle-\mathbf{R}^{0}_{i}\right| = \left|\frac{1}{N_{\mathrm{steps}}}\sum_{t}\mathbf{R}_{i}(t) - \mathbf{R}^{0}_{i}\right|~,\label{eq:average_distance}
\end{align}
where $N_{\mathrm{steps}}$ is the total number of MD time steps, $\mathbf{R}_{i}(t)$ is the position of the $i$-th atom from the MD simulation cell at time step $t$ and the sum runs over all time steps. Furthermore, when we perform the folding onto a $2\times2\times1$ supercell, we average the atomic positions over time. After averaging, we find which average of atomic positions is closest to its reference atomic position from the $2\times2\times1$ normal phase supercell. Then we shift all of the average positions by a vector that sets the smallest average exactly at the positions of the reference atom. Afterward, we take all averages of the atomic positions except the one that is set at the position of its reference atom from the normal phase and calculate their distances from the respective positions from the reference $2\times2\times1$ normal phase supercell. This is done because we know that at low temperatures exactly one atom should remain still in a $2\times2\times1$ supercell when the system is in the CDW phase. We average these distances to obtain the average atomic displacements in Fig.\,1b, and perform this procedure for each temperature.

\subsection{Phonon dispersions}
To calculate phonon dispersions from the MD data we used the DynaPhoPy framework\,\cite{dynaphopy}, which projects the atomic velocities from MD to harmonic phonon modes. To calculate the harmonic force constants needed for these calculations, we used Phonopy\,\cite{togo2023}. Furthermore, we generate a set of $\mathbf{q}$ points along the $\Gamma-\mathrm{M}-\mathrm{K}-\Gamma$ path also using Phonopy. For each of these $\mathbf{q}$ points we calculate the wave-vector projected power spectrum function of the atomic velocities, which is defined as\,\cite{dynaphopy}
\begin{align}
    G(\mathbf{q}, \omega) = 2\sum_{j \alpha}\int^{+\infty}_{-\infty}\langle\langle v^{\mathbf{q}*}_{j \alpha}(0)v^{\mathbf{q}}_{j \alpha}(t) \rangle\rangle e^{i\omega t}d t~,\label{eq:power_spectrum}
\end{align}
where $\langle\langle v^{\mathbf{q}*}_{j \alpha}(0)v^{\mathbf{q}}_{j \alpha}(t) \rangle\rangle$ is the velocity autocorrelation function, with $v^{\mathbf{q}}_{j \alpha}(t)$ being the $\alpha$-th component of the $j$-th atomic velocity projected onto a wave vector $\mathbf{q}$ at time $t$. We see that Eq.\,\eqref{eq:power_spectrum} is the Fourier transform of the velocity autocorrelation function. Furthermore, it can be shown\,\cite{dynaphopy}
\begin{align}
    G(\omega) = \sum_{\mathbf{q}} G(\mathbf{q}, \omega) \propto g(\omega)~,\label{eq:pdos}
\end{align}
where $G(\omega)$ is the full power spectrum function and $g(\omega)$ is the phonon density of states.

\subsection{CDW order parameter}
For each of three $\mathbf{q}=\mathrm{M}$ points, we construct an order parameter:
    \begin{equation}
        \Psi_{\alpha}\left(\mathbf{R}_{i}^{0},t\right)=\frac{1}{\mathcal{N}_{i}}\sum_{j}G_{ij}\mathbf{e}^{*}_{\alpha j}\cdot\mathbf{u}_{j}(t)e^{-i\mathbf{q}_{\alpha}\cdot\mathbf{R}_{j}^{0}},
        \label{eq:complex_order_par}
    \end{equation}
where $\mathbf{R}_{i}^{0}$ is the equilibrium position of $i$-th atom in $xy$ plane and the sum runs over all atoms, $\mathbf{q}_{\alpha}$ corresponds to different $\mathbf{q}=\mathrm{M}$ points with $\mathbf{e}^{*}_{\alpha j}$ being a complex conjugate of an eigenvector of the soft phonon mode corresponding to $\mathbf{q}_{\alpha}$ and $j$-th atom, and $\mathbf{u}_{j}(t)$ is the displacement of $j$-th atom at time $t$ given by $\mathbf{u}_{j}(t)=\mathbf{R}_{j}(t)-\mathbf{R}_{j}^{0}$, with $\mathbf{R}_{j}(t)$ being the position of $j$-th atom at time $t$. The eigenvectors are obtained from the unstable (soft) phonon mode of the normal phase at $\mathbf{q}_{\alpha}$, calculated in the harmonic approximation with Phonopy\,\cite{togo2023} from the DFT force constants, which we obtained with VASP using the finite displacement method in an $8\times8\times1$ supercell, with which the $\mathbf{q}=\mathrm{M}$ points are commensurate. Phonopy gives the eigenvector $\boldsymbol{\epsilon}_{\alpha j}$ of the mass-weighted dynamical matrix, which is normalized over the unit cell and defined up to one overall phase factor. We first remove the mass weighting, $\mathbf{d}_{\alpha j}=\boldsymbol{\epsilon}_{\alpha j}/\sqrt{m_{j}}$, where $m_{j}$ is the mass of the $j$-th atom, which gives the displacement pattern of the mode. Then we fix the overall phase by multiplying the whole vector with a single phase factor, chosen so that the largest component of the Ti atom is real and positive. Finally, we normalize the vector of each of the three atoms in the unit cell separately, $\mathbf{e}_{\alpha j}=\mathbf{d}_{\alpha j}/\left|\mathbf{d}_{\alpha j}\right|$, so that $\left|\mathbf{e}_{\alpha j}\right|=1$ for every atom. The relative phases between the atoms are kept. In the convention of Phonopy the dynamical matrix is built with the phase factors that contain the full positions of the atoms, so that the eigenvector component of the atom at the position $\boldsymbol{\tau}_{j}$ inside the unit cell carries the phase factor $e^{-i\mathbf{q}_{\alpha}\cdot\boldsymbol{\tau}_{j}}$, i.e., $\mathbf{e}_{\alpha j}\,e^{i\mathbf{q}_{\alpha}\cdot\boldsymbol{\tau}_{j}}$ is real. Therefore, the components of the Ti atom, which is at the origin of the unit cell, are real, while the components of the two Se atoms, which are at the in-plane positions $\pm\boldsymbol{\tau}_{\mathrm{Se}}$, are complex and are complex conjugates of each other. With this normalization $\mathbf{e}^{*}_{\alpha j}\cdot\mathbf{u}_{j}(t)$ is the displacement of the $j$-th atom along its direction in the CDW mode. The projection on the eigenvector selects the displacement pattern of the CDW mode, while the multiplication with the phase factor $e^{-i\mathbf{q}_{\alpha}\cdot\mathbf{R}_{j}^{0}}$ shifts every Fourier component of the displacement field by $\mathbf{q}_{\alpha}$ (demodulation). In this way the CDW modulation at $\mathbf{q}_{\alpha}$ is moved to zero wave vector, where it becomes a slowly varying envelope, while the long-wavelength (acoustic) displacements are moved to $\mathbf{q}_{\alpha}$, where their sign alternates from one row of unit cells to the next, so that they are removed by a spatial average. This is the principle of the lock-in analysis used in the STM studies\,\cite{liu2024,xiaolong2021,lawler2010,mesaros2011,fujita2014,gu2023}. To include influence of neighboring atoms on each individual atom, we include a Gaussian $G_{ij}$ into the sum, given by
\begin{equation}
    G_{ij}=G\left(\left|\mathbf{R}_{i}^{0}-\mathbf{R}_{j}^{0}\right|\right)=\frac{1}{2\pi\sigma^{2}}e^{-\frac{\left|\mathbf{R}_{i}^{0}-\mathbf{R}_{j}^{0}\right|^{2}}{2\sigma^{2}}},\qquad \mathcal{N}_{i}=\sum_{j}G_{ij},
    \label{eq:gaussian}
\end{equation}
where $\sigma$ is the Gaussian width and we use $\sigma=4$ \AA, and $\mathcal{N}_{i}$ is the normalization, i.e., the sum of the Gaussian weights of all atoms around the $i$-th atom. The Gaussian is normalized to unit area in the $xy$ plane with the prefactor $1/(2\pi\sigma^{2})$, so both $G_{ij}$ and $\mathcal{N}_{i}$ have the units of inverse area. Only their ratio enters Eq.\,\eqref{eq:complex_order_par}: the weights $G_{ij}/\mathcal{N}_{i}$ are dimensionless and their sum over $j$ is equal to one, so that the order parameter is the Gaussian-weighted average of the demodulated displacements and has the units of length.
We calculate Eq.\,\eqref{eq:complex_order_par} at the positions $\mathbf{R}_{i}^{0}$ of all atoms (both Ti and Se).

Since $\mathbf{q}_{\alpha}=\mathrm{M}$ is a half of a reciprocal lattice vector, $\mathbf{q}_{\alpha}$ and $-\mathbf{q}_{\alpha}$ represent the same modulation and the phase factor $e^{-i\mathbf{q}_{\alpha}\cdot\mathbf{R}}$ equals $\pm1$ for every lattice vector $\mathbf{R}$. The phases of the eigenvector components compensate the positions of the atoms inside the unit cell, so that Eq.\,\eqref{eq:complex_order_par} is real. Therefore, although we write it in the general form with the amplitude and the phase, the order parameter is real and its phase can only take the values $0$ and $\pi$, i.e., it is the sign of the order parameter. These two values correspond to the same distortion shifted by one lattice constant.

Once we calculate $\Psi_{\alpha}\left(\mathbf{R}_{i}^{0},t\right)$ for each $\mathbf{R}_{i}^{0}$ and $t$ we average it over time to get
\begin{equation}
    \overline{\Psi}_{\alpha}\left(\mathbf{R}_{i}^{0}\right)=\langle\Psi_{\alpha}\left(\mathbf{R}_{i}^{0},t\right)\rangle=\frac{1}{N_{\mathrm{steps}}}\sum_{t}\Psi_{\alpha}\left(\mathbf{R}_{i}^{0},t\right).
    \label{eq:avg_order_param}
\end{equation} 
Furthermore, we can write 
\begin{equation}
    \overline{\Psi}_{\alpha}\left(\mathbf{R}_{i}^{0}\right)=\bar{A}_{\alpha}\left(\mathbf{R}_{i}^{0}\right)e^{i\bar{\phi}_{\alpha}\left(\mathbf{R}_{i}^{0}\right)},
    \label{eq:avg_order_param_polar}
\end{equation}
from this we calculate the average amplitude $\bar{A}_{\alpha}\left(\mathbf{R}_{i}^{0}\right)$ by taking the absolute value of Eq.~\eqref{eq:avg_order_param_polar} and the average phase $\bar{\phi}_{\alpha}\left(\mathbf{R}_{i}^{0}\right)\in\{0,\pi\}$ by using the angle function implemented in \textit{numpy} \cite{harris2020array} package, for each $\mathbf{R}_{i}^{0}$ and $\mathbf{q}_{\alpha}$. When plotting the phases, we add a constant $0$, $\pi$ and $\pi$ to $\bar{\phi}_{\alpha}$ for $\mathrm{M}_{1}$, $\mathrm{M}_{2}$ and $\mathrm{M}_{3}$, respectively, so that the threefold rotation maps the three components onto each other without a change of sign. Note that the overall sign of each component is arbitrary, since different combinations of the signs of the three components are equivalent CDW domains, and MD simulations at different temperatures can end up in different ones.

Another way of calculating time average of the amplitude is to write Eq.\,\eqref{eq:complex_order_par} as
\begin{equation}
    \Psi_{\alpha}\left(\mathbf{R}_{i}^{0},t\right) = A_{\alpha}\left(\mathbf{R}_{i}^{0},t\right)e^{i\phi_{\alpha}\left(\mathbf{R}_{i}^{0},t\right)},
    \label{eq:order_param_polar}
\end{equation}
from which we can extract the amplitude for each $\mathbf{R}_{i}^{0}$ and $t$. Then we calculate time average as
\begin{equation}
    \langle A_{\alpha}\left(\mathbf{R}_{i}^{0}\right)\rangle = \frac{1}{N_{\mathrm{steps}}}\sum_{t}A_{\alpha}\left(\mathbf{R}_{i}^{0},t\right).\label{eq:avg_ampl_2}
\end{equation}
In $\bar{A}_{\alpha}\left(\mathbf{R}_{i}^{0}\right)$ the changes of the sign of the order parameter in time cancel, while $\langle A_{\alpha}\left(\mathbf{R}_{i}^{0}\right)\rangle$ also contains the fluctuations of the order parameter. As functions of temperature (Fig.\,3c in the main text) we show $\bar{A}_{\alpha}\left(\mathbf{R}_{i}^{0}\right)$ and $\langle A_{\alpha}\left(\mathbf{R}_{i}^{0}\right)\rangle$ averaged over all atoms.

In addition, we calculate the global order parameter without the Gaussian, which does not depend on the position $\mathbf{R}_{i}^{0}$,
\begin{equation}
    \Psi^{\mathrm{glob}}_{\alpha}(t)=\frac{1}{N_{\mathrm{atoms}}}\sum_{j}\mathbf{e}^{*}_{\alpha j}\cdot\mathbf{u}_{j}(t)e^{-i\mathbf{q}_{\alpha}\cdot\mathbf{R}_{j}^{0}}.
    \label{eq:global_order_par}
\end{equation}
This is the Fourier component of the displacements at $\mathbf{q}_{\alpha}$ projected on the eigenvector of the soft phonon mode. The local and the global order parameters are directly connected. Both are averages of the same quantity, the demodulated displacement $\zeta_{\alpha j}(t)=\mathbf{e}^{*}_{\alpha j}\cdot\mathbf{u}_{j}(t)e^{-i\mathbf{q}_{\alpha}\cdot\mathbf{R}_{j}^{0}}$: in Eq.\,\eqref{eq:complex_order_par} it is averaged with the Gaussian weights around $\mathbf{R}_{i}^{0}$, while in Eq.\,\eqref{eq:global_order_par} it is averaged with equal weights over the whole supercell. Therefore, Eq.\,\eqref{eq:global_order_par} is the $\sigma\rightarrow\infty$ limit of Eq.\,\eqref{eq:complex_order_par}, in which all the weights become equal, $G_{ij}/\mathcal{N}_{i}\rightarrow1/N_{\mathrm{atoms}}$. In both cases the sum is divided by the sum of the weights, so that both order parameters are averages of $\zeta_{\alpha j}(t)$ and have the same units. The corresponding amplitudes of the global order parameter are
\begin{equation}
    \bar{A}^{\mathrm{glob}}_{\alpha}=\left|\frac{1}{N_{\mathrm{steps}}}\sum_{t}\Psi^{\mathrm{glob}}_{\alpha}(t)\right|,\qquad \langle A^{\mathrm{glob}}_{\alpha}\rangle=\frac{1}{N_{\mathrm{steps}}}\sum_{t}\left|\Psi^{\mathrm{glob}}_{\alpha}(t)\right|.
    \label{eq:global_avg_ampl}
\end{equation}
The difference with respect to the local order parameter is that here the absolute value is taken after the sum over the whole supercell, so the domains of opposite sign cancel, while in the local case the absolute value is taken at each atom, so all domains contribute.

\section{Additional results}
\begin{figure}[H]
    \centering
    \includegraphics[scale=0.625]{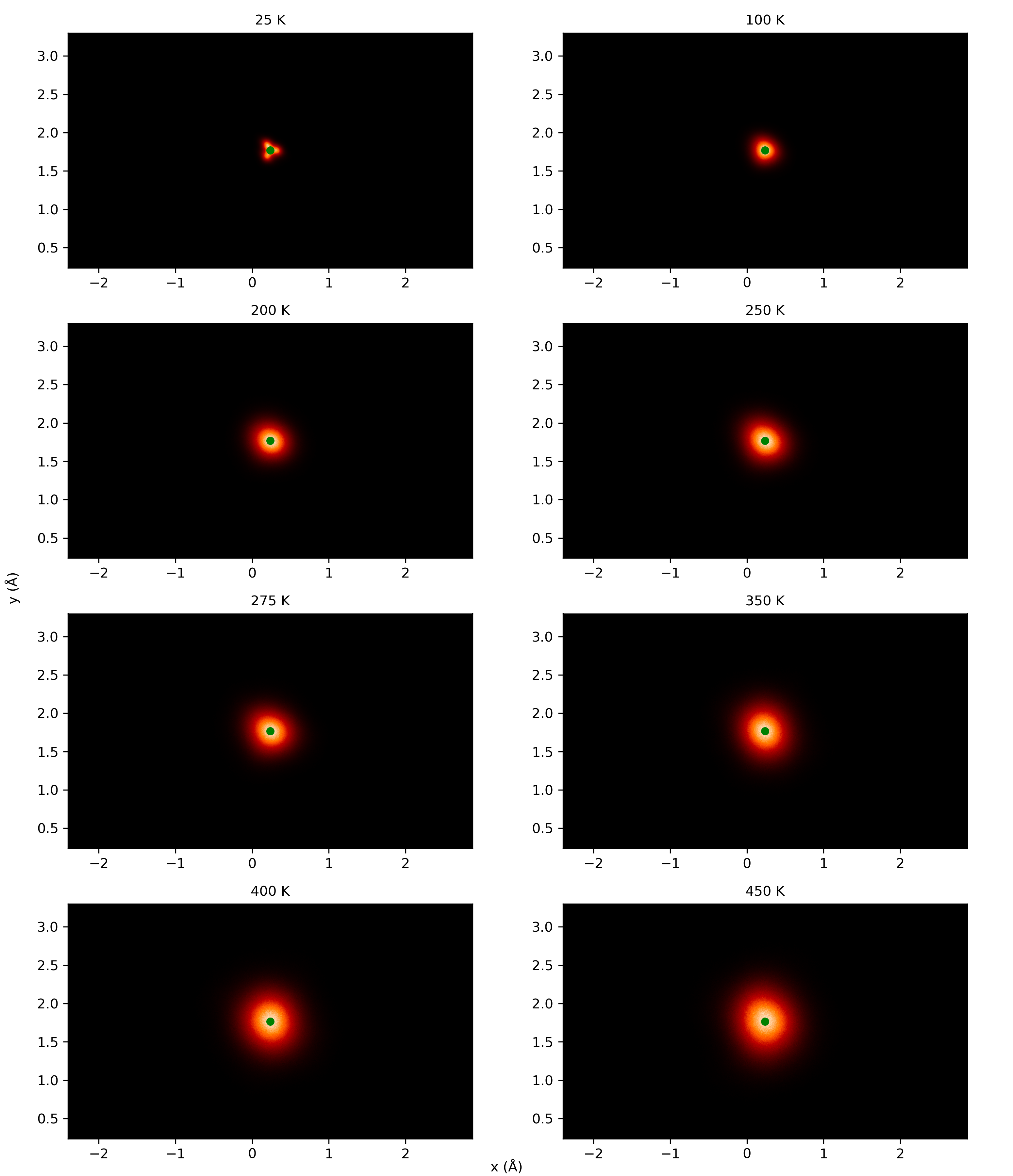}
    \caption{Average positions of Ti atoms from $56\times56\times1$ MDs folded onto normal phase ($1\times 1\times 1$) unit cell. Green points indicate reference normal phase unit cell positions of Ti atoms.}
    \label{fig:positions_unit_cell}
\end{figure}

\begin{figure}[H]
    \centering
    \includegraphics[scale=0.475]{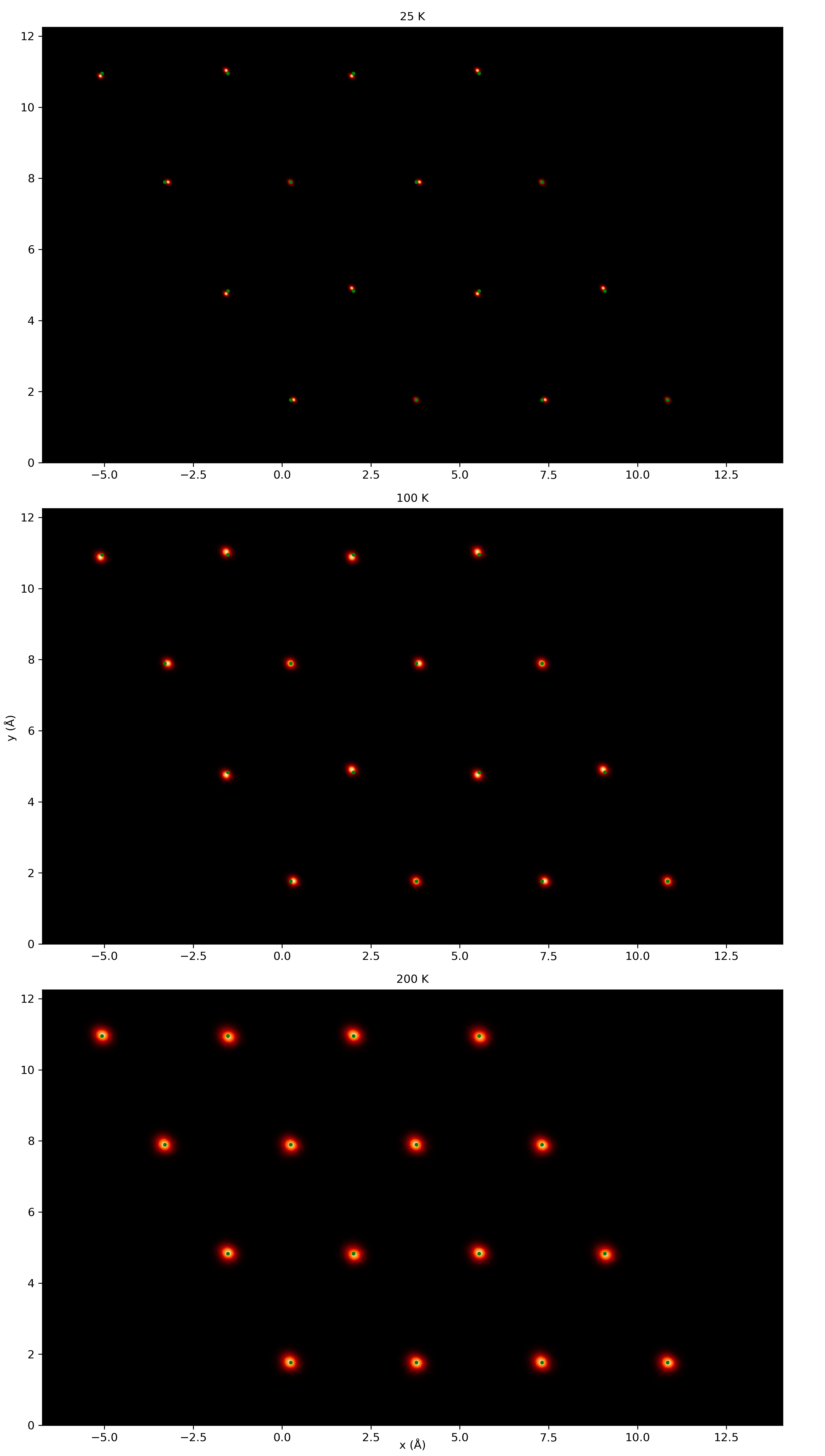}
    \caption{Average positions of Ti atoms from $56\times56\times1$ MDs folded onto normal phase $4\times4\times1$ supercell. Green points indicate reference normal phase $4\times4\times1$ supercell positions of Ti atoms.}
    \label{fig:positions_4x4x1_cell}
\end{figure}

\begin{figure}[H]
    \centering
    \includegraphics[width=\textwidth]{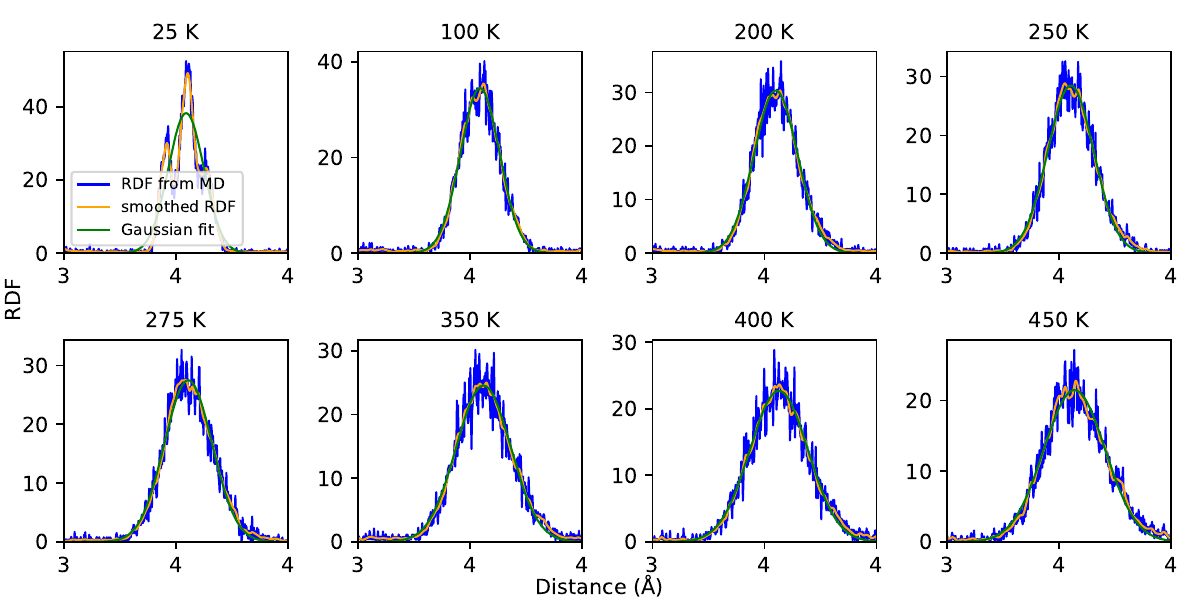}
    \caption{RDF results for Ti atoms from MD simulations at several different temperatures.}
    \label{fig:rdf}
\end{figure}

\begin{figure}[H]
    \centering
    \includegraphics[width=0.75\columnwidth]{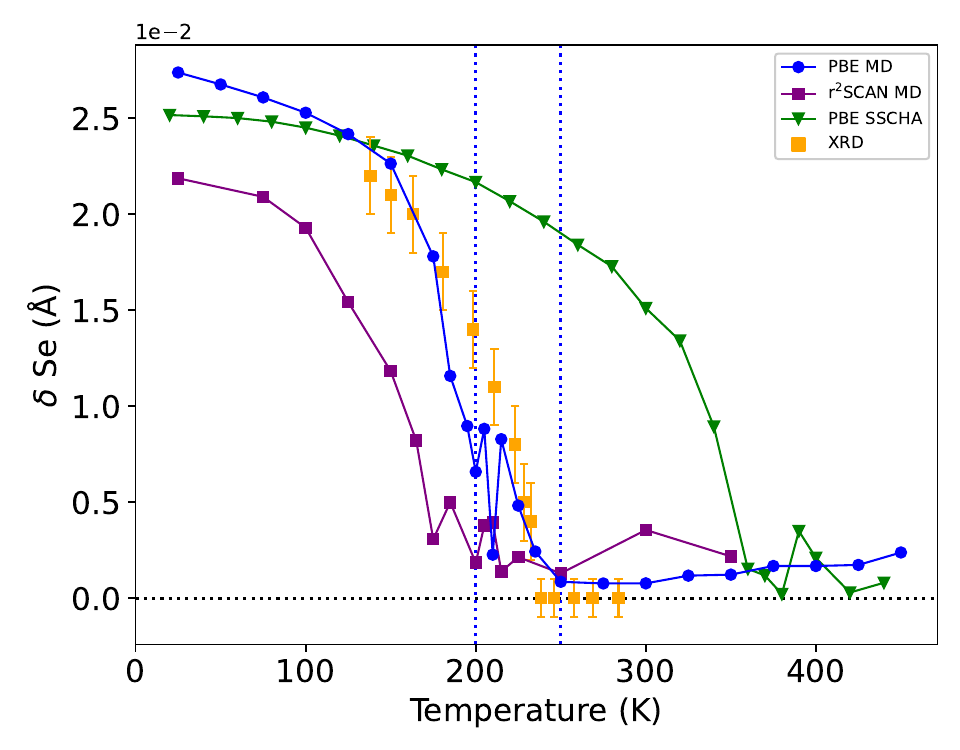}
    \caption{Average displacements of Se atoms in the $xy$ plane calculated using SSCHA and from MD data compared to XRD experiments\,\cite{xrd_displ}.}
    \label{fig:se_displacements}
\end{figure}

\begin{figure}[H]
    \centering
    \includegraphics[scale=0.5]{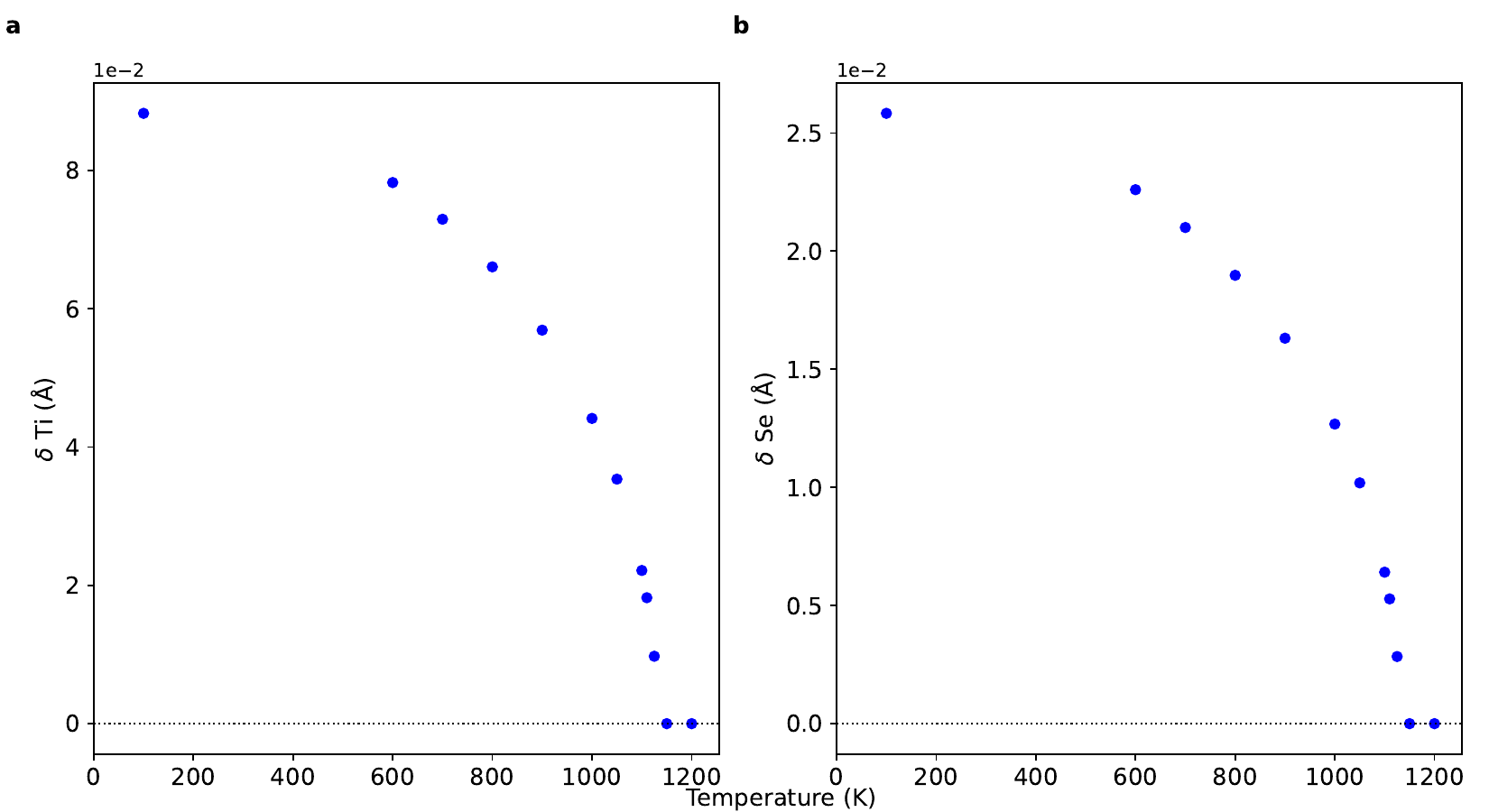}
    \caption{Average displacements in the $xy$ plane calculated with harmonic DFT in $2\times2\times1$ supercell for both Ti atoms in \textbf{a} and Se atoms in \textbf{b}.}
    \label{fig:qe_displacements}
\end{figure}

\begin{figure}[H]
    \centering
    \includegraphics[width=\textwidth]{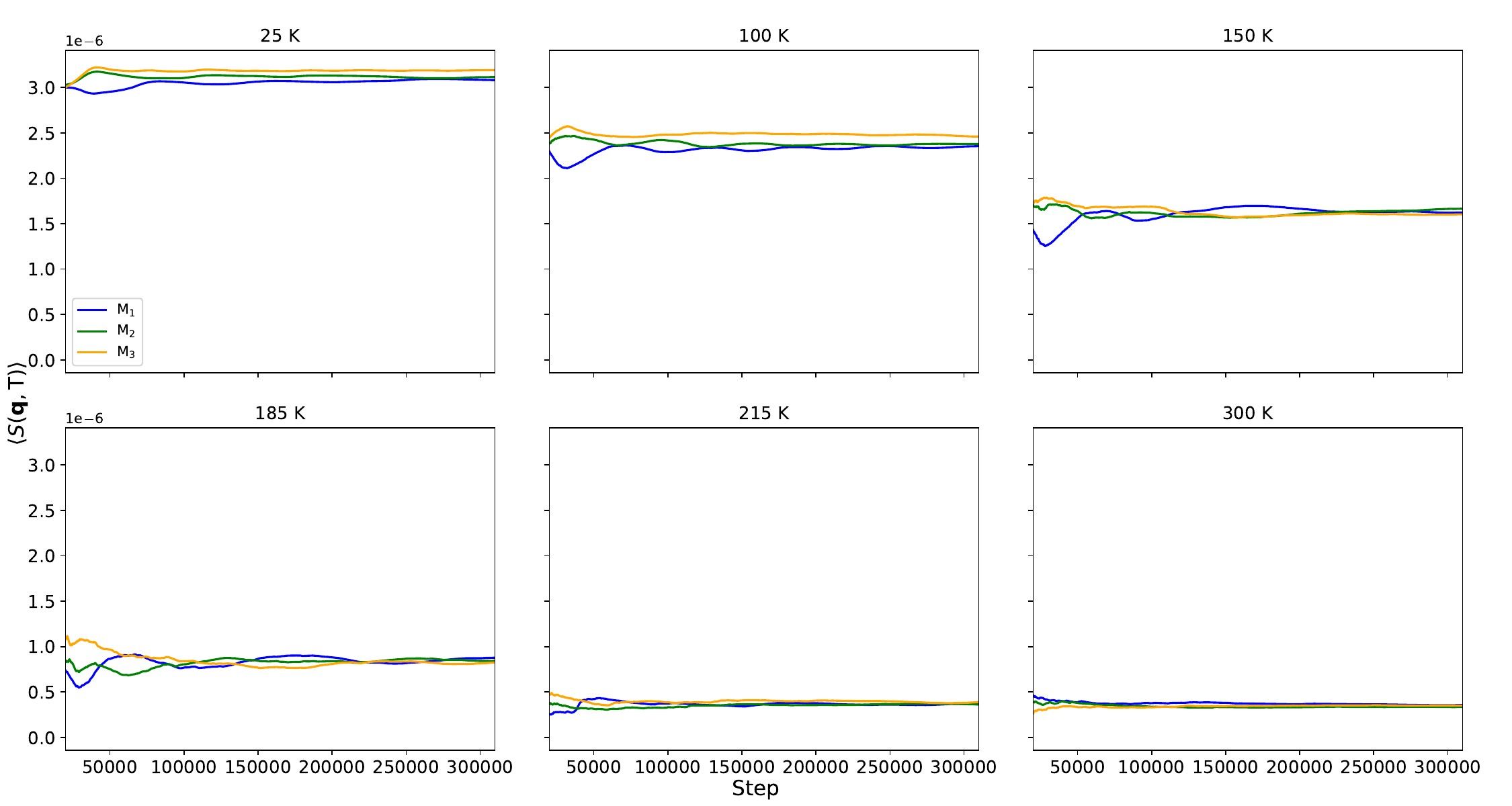}
    \caption{Running average of structure factors for three different $\mathbf{q}=\mathrm{M}$ points calculated for six different temperatures.}
    \label{fig:running_avg}
\end{figure}

\begin{figure}[H]
    \centering
    \includegraphics[scale=0.65]{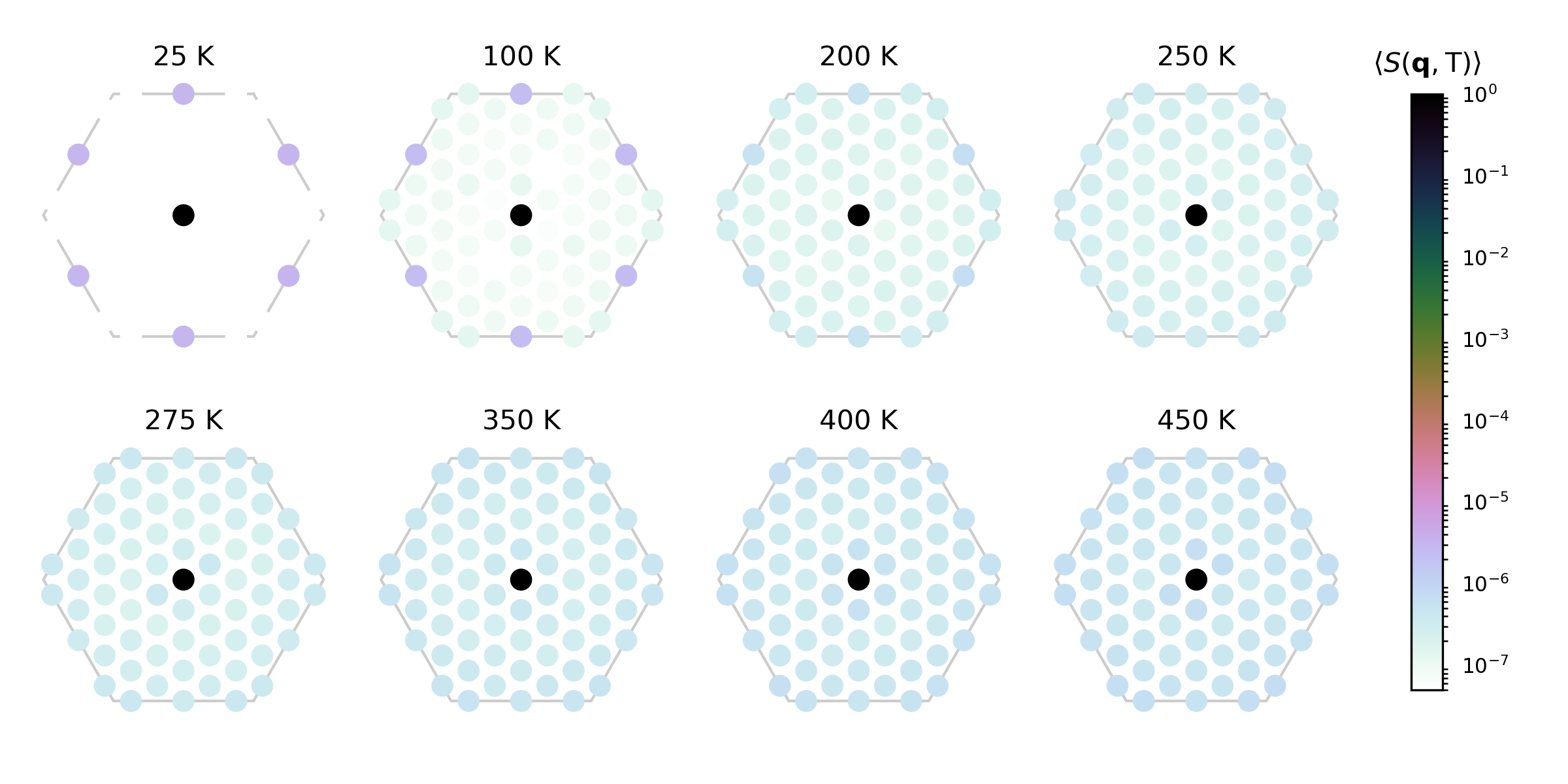}
    \caption{Time-averaged structure factors from $56\times56\times1$ MD data calculated using irreducible $\mathbf{q}$ points of $8\times8\times1$ supercell. Time-averaged structure factors were calculated using Eq.\,\eqref{eq:time_averaged_structure_factor}.}
    \label{fig:structure_factors_8x8x1}
\end{figure}

In Fig.\,\ref{fig:positions_unit_cell} we show average positions of Ti atoms from $56\times56\times1$ MDs folded onto the normal phase unit cell. We see that a distinct CDW shape of the distribution of average positions of Ti atoms at $25$\,K transitions to a spherically symmetric shape as temperature increases and the system transitions to the normal phase. 
\newpage
In Fig.\,\ref{fig:positions_4x4x1_cell} we show the average positions of Ti atoms from $56\times56\times1$ folded onto the normal phase $4\times4\times1$ supercell. We can see that at $25$\,K we have a distinct CDW pattern which disappears as the system transitions to the normal phase with an increase in temperature. Also, one can observe both in Fig.\,\ref{fig:positions_unit_cell} and Fig.\,\ref{fig:positions_4x4x1_cell} that increase in temperature causes spreading of the average positions of Ti atoms, due to the stronger thermal effects.
In Fig.\,\ref{fig:rdf} we report results for Radial Distribution Functions (RDFs) for Ti atoms from MD simulations at several different temperatures. The results show only the first and largest peak in the RDF for each temperature. One can see that at low temperatures there are $3$ distinct peaks, which transition to a Gaussian shape as the temperature increases and the system transitions to the normal phase. The RDFs were calculated from the MD data using the freud software\,\cite{freud2020}.
In Fig.\,\ref{fig:se_displacements} we show the results for the average displacements of Se atoms in the $xy$ plane calculated with SSCHA and from the MD data (from MD simulations run both with PBE based MACE interatomic potential and the one fine-tuned with r$^{2}$SCAN data) compared to the XRD experiments\,\cite{xrd_displ}. We used our MACE interatomic potential based on PBE data to run SSCHA calculations. We performed SSCHA calculations in $4\times4\times1$ supercell made from a $2\times2\times1$ CDW structure, which we discuss later in the text. Most of the SSCHA parameters were kept at their default values. The initial harmonic dynamical matrices for the SSCHA calculations were obtained with the ASE, with a finite difference step of $0.01$ \AA, where we have relaxed the initial normal phase unit structure with the convergence criterion of the forces on atoms during relaxation set to $10^{-8}$ eV/\AA. Relaxation in SSCHA was performed with $10000$ configurations and, to generate ensembles, we used $20000$ configurations. The meaningful factor was set to $10^{-5}$. Both the dynamical matrix and the structure minimization steps were set at a value of $0.05$. The average displacements were calculated from the MD data in the same way as explained in the text above. We can see that both SSCHA and MDs indicate that the phase transition between CDW and normal phases is not purely of the second order. The reason is that both SSCHA and MD calculations show distortions before $T_{\mathrm{CDW}}$ after which the average displacements go to their minimal values. These results align with some recent studies\,\cite{amin2024}, which claim that XRD experiments are agnostic to local distortions that occur in the system below $T_{\mathrm{CDW}}$. In addition, SSCHA overestimates $T_{\mathrm{CDW}}$ compared to experiments and calculations using MD data.
\newpage
In Fig.\,\ref{fig:qe_displacements} we show results of DFT calculations with \textsc{Quantum ESPRESSO} (QE) package\,\cite{Giannozzi2017} of average displacements in $2\times2\times1$ supercell of both Ti and Se atoms in the $xy$ plane. As we can see, these results highly overestimate $T_{\mathrm{CDW}}$ and indicate that at that level of theory, phase transition in monolayer TiSe$_{2}$ between CDW and normal phases is of the conventional second order.
In Fig.\,\ref{fig:running_avg} we report running averages for MD simulations using six different temperatures which we ran for $310000$ steps using our PBE data based MACE interatomic potential, we removed the first $10000$ steps which we treat as a thermalization period. The results indicate that for temperatures well below $T_{\mathrm{CDW}}$ structure factors of three different $\mathbf{q}=\mathrm{M}$ points are distinct, for temperatures well above $T_{\mathrm{CDW}}$ they converge to the same value and in the temperature range around $T_{\mathrm{CDW}}$ there is no clear ordering of these structure factors.
A similar analysis as in Fig.\,\ref{fig:positions_unit_cell} and Fig.\,\ref{fig:positions_4x4x1_cell} was performed in the reciprocal space. In Fig.\,\ref{fig:structure_factors_8x8x1} we show time-averaged structure factors from $56\times56\times1$ MDs calculated using irreducible $\mathbf{q}$ points of $8\times8\times1$ supercell (as also done in Fig.\,1c in the main text), which shows that at low temperatures there are distinct signals at $\mathbf{q}=\mathrm{M}$ points and as temperature increases this signal disappears. Also, around $\mathbf{q}=\mathrm{\Gamma}$ there are enhanced signals for all temperatures, due to thermal fluctuations of the atoms in the system. For efficient calculations of time-averaged structure factors, we modified the code from Schobert et al. work\,\cite{berges2024}, which relies on the \textit{elphmod} package\,\cite{elphmod_code}.

\begin{figure}[H]
    \centering
    \includegraphics[scale=0.65]{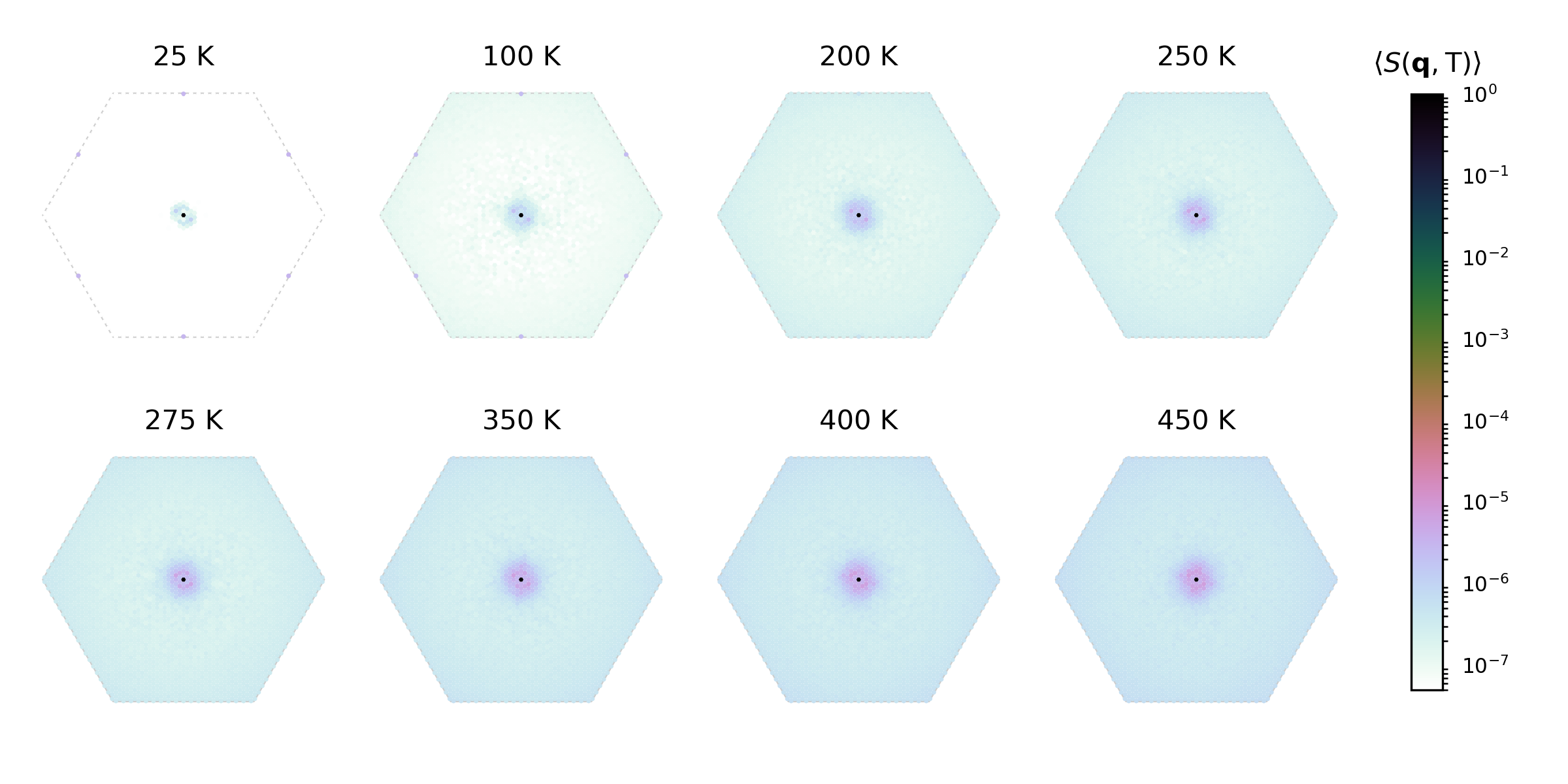}
    \caption{Time-averaged structure factors from $56\times56\times1$ MDs calculated using irreducible $\mathbf{q}$ points of $56\times56\times1$ supercell. Time-averaged structure factors were calculated using Eq.\,\eqref{eq:time_averaged_structure_factor}.}
    \label{fig:structure_factors_56x56x1}
\end{figure}

\begin{figure}[H]
    \centering
    \includegraphics[width=\textwidth]{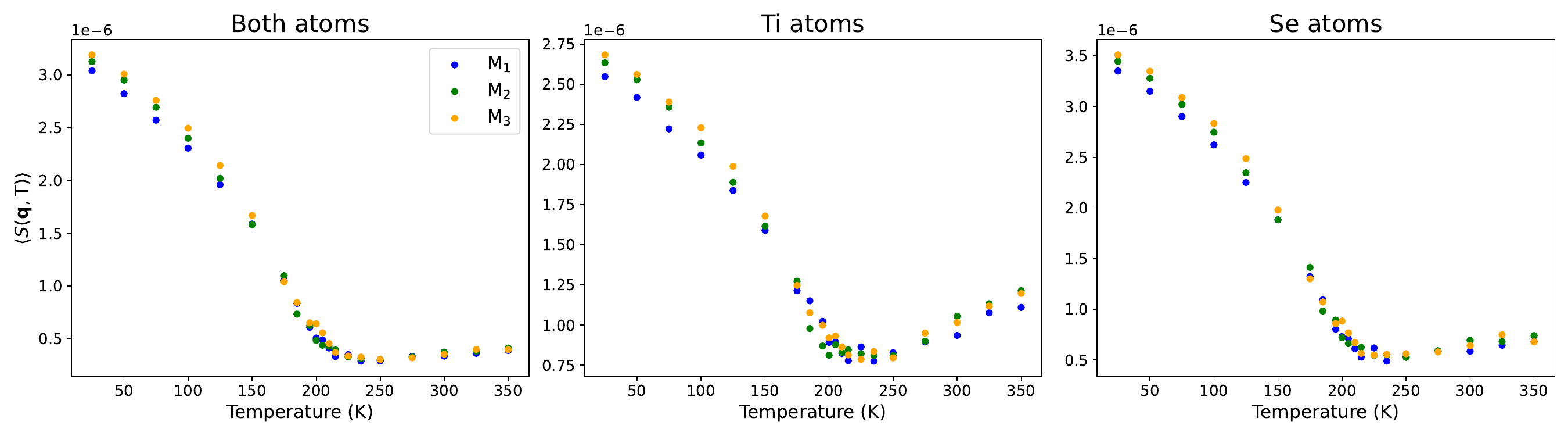}
    \caption{Time-averaged structure factors Eq.\,\eqref{eq:time_averaged_structure_factor} as a function of temperature calculated for $3$ different $\mathbf{q}=\mathrm{M}$ points.}
    \label{fig:subplots_3M}
\end{figure}

Furthermore, in Fig.\,\ref{fig:structure_factors_56x56x1} we show the same results as in the previous image, but we used the irreducible $\mathbf{q}$ points of the $56\times56\times1$ supercell which enhances the effects discussed above.
\newpage
In Fig.\,\ref{fig:subplots_3M} we show the results for the time-averaged structure factors Eq.\,\eqref{eq:time_averaged_structure_factor} firstly for both atoms used in the calculations and then only for the Ti and Se atoms at three different $\mathbf{q}=\mathrm{M}$ points. We see that in all three cases we see a behavior similar to that of Fig.\,1 from the main text.

\begin{figure}[H]
    \centering
    \includegraphics[width=\columnwidth]{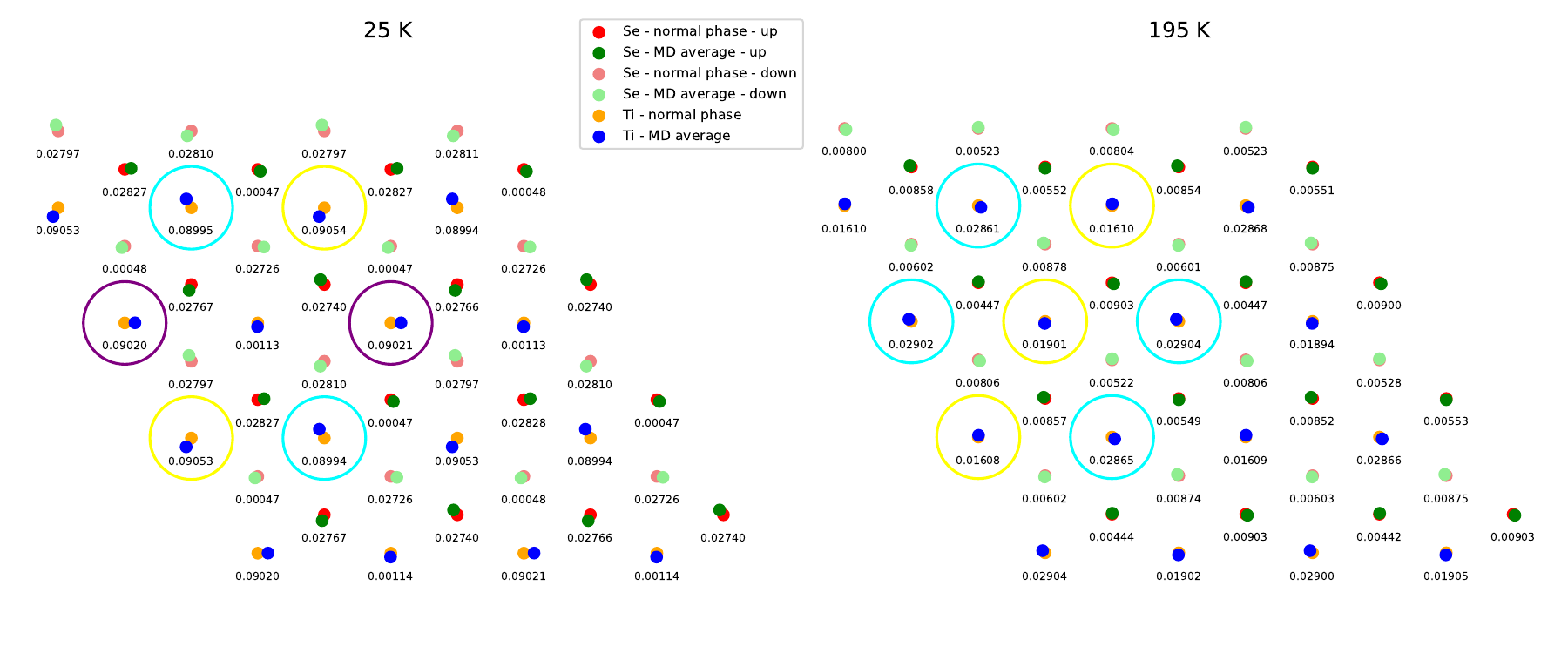}
    \caption{Periodic lattice distortions (Ti and Se atom displacements) of the MD time-averaged structures at $25$\,K and $195$\,K in the $xy$ plane with respect to the normal phase positions.}
    \label{fig:positions}
\end{figure}

\begin{figure}[H]
    \centering
    \includegraphics[scale=0.65]{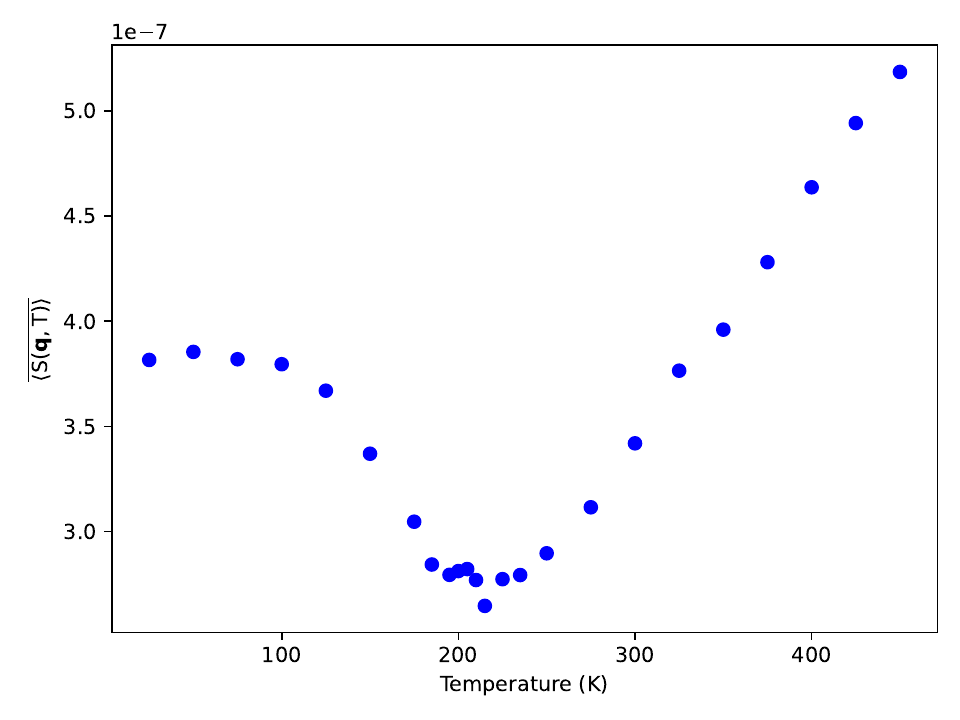}
    \caption{Average of time-averaged structure factors Eq.\,\eqref{eq:time_averaged_structure_factor} of $9$ different $\mathbf{q}$ points around $\mathbf{q}=\mathrm{M_{3}}$ from Fig.\,1c in the main text and Fig.\,\ref{fig:struct_factors}a, including $\mathbf{q}=\mathrm{M_{3}}$ point.}
    \label{fig:average_around_M}
\end{figure}

\begin{figure}[H]
    \centering
    \includegraphics[scale=0.5]{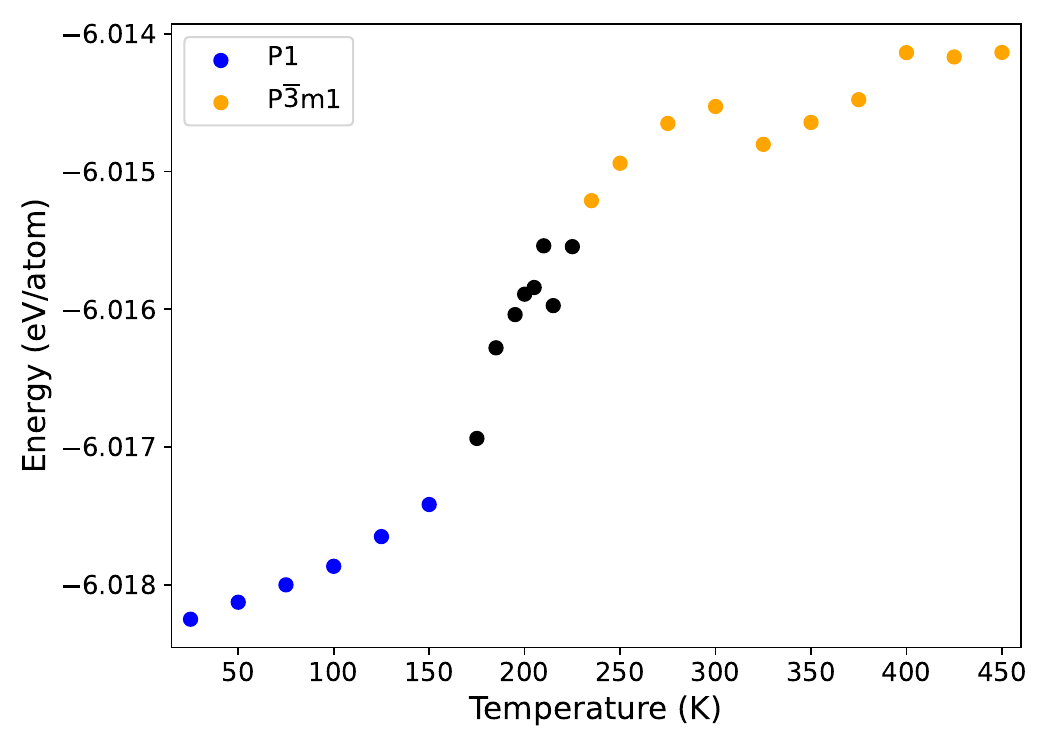}
    \caption{Energies calculated for averaged $4\times4\times1$ structures obtained from MD simulations at different temperatures (indicated on the $x$ axis). Colors indicate symmetries of these structures as in Fig.\,1d.}
    \label{fig:energy_diff}
\end{figure}

\begin{figure}[H]
    \centering
    \includegraphics[scale=0.65]{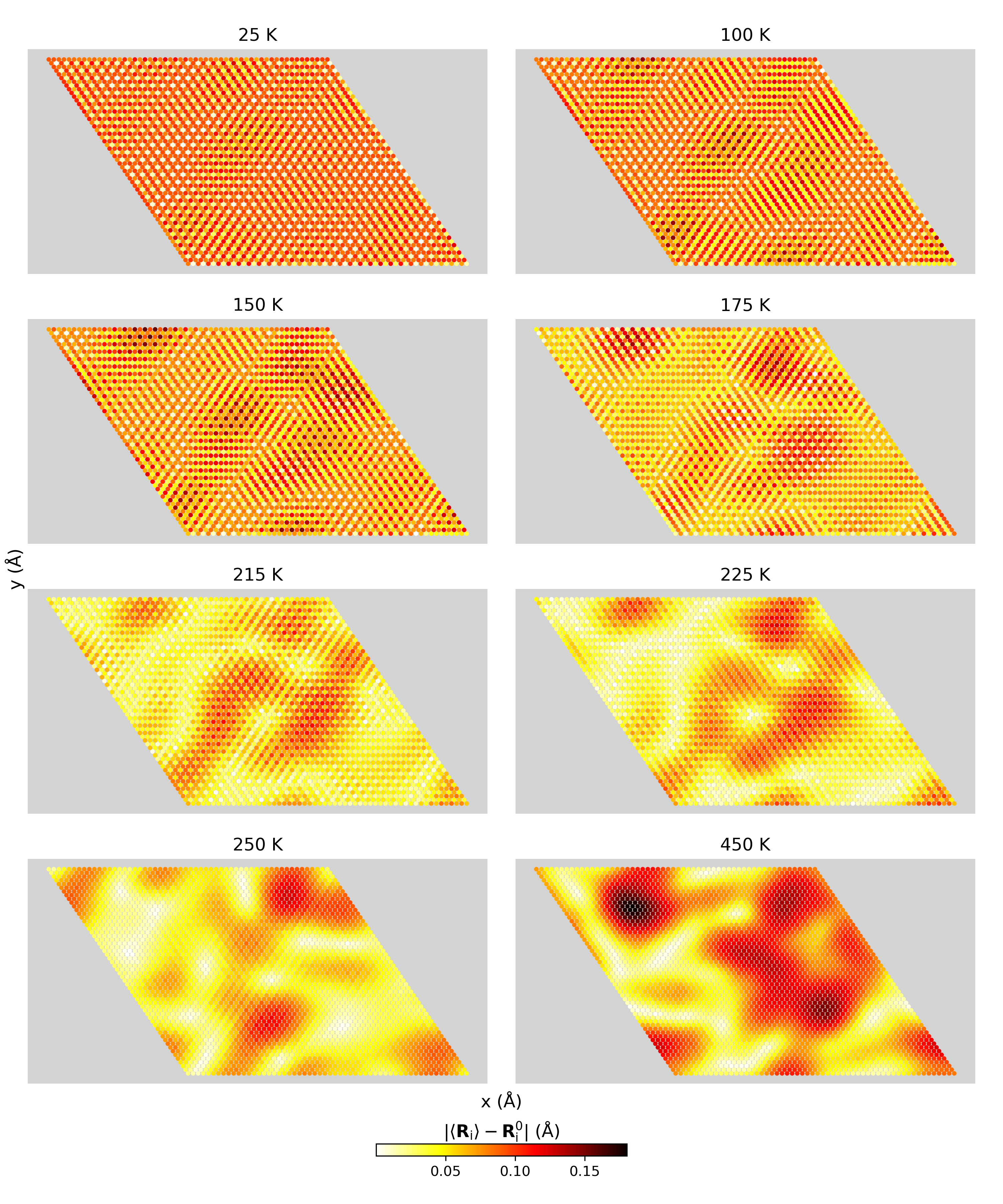}
    \caption{Distances between the time-averaged positions of each Ti atom and Ti atoms from normal phase for several different temperatures. Notice that a different color scale was used than in Fig.\,2 in the main text.}
    \label{fig:displ}
\end{figure}

\begin{figure}[H]
    \centering
    \includegraphics[scale=0.5]{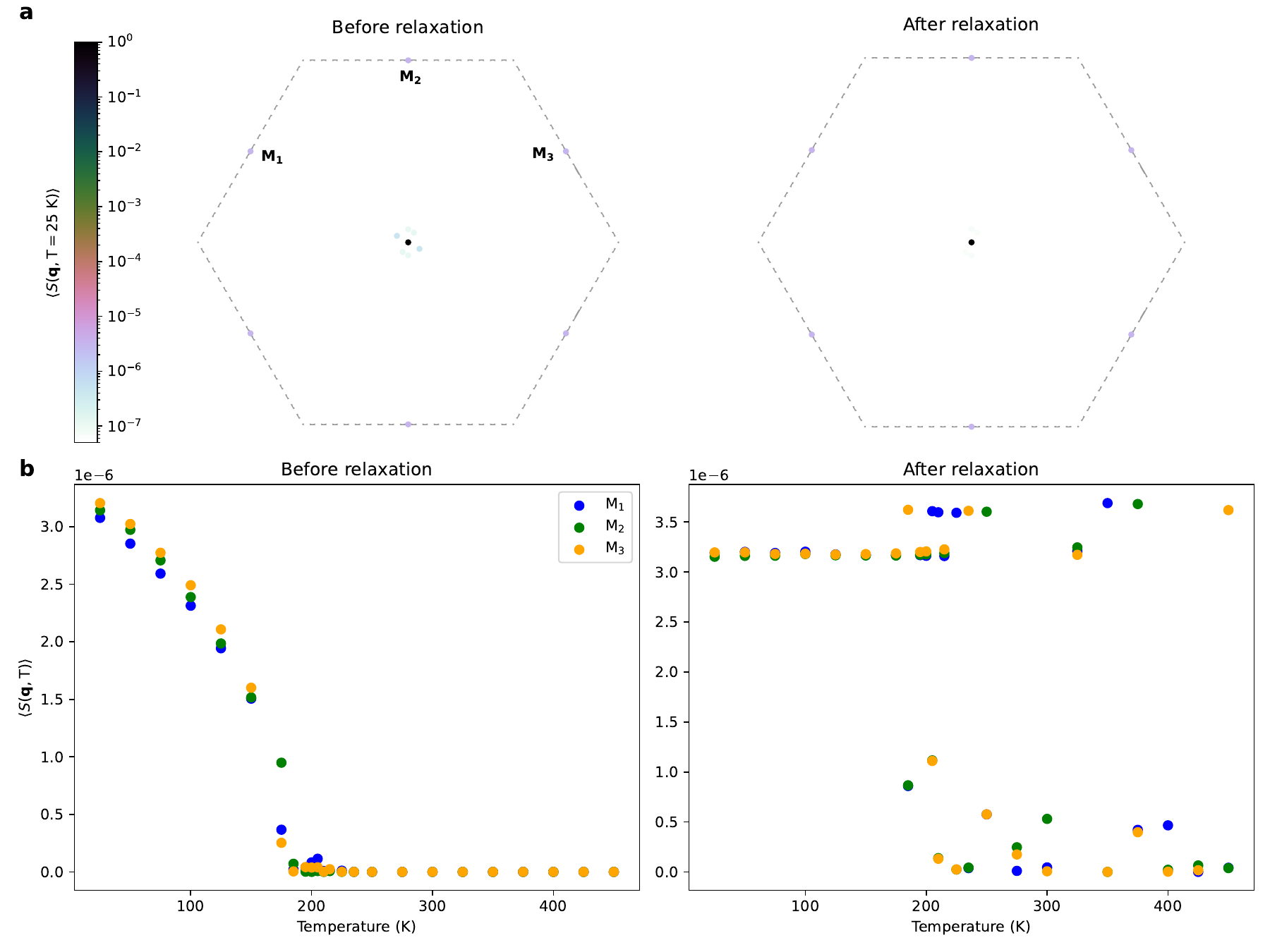}
    \caption{\textbf{a} Structure factors Eq.\,\eqref{eq:time_averaged_structure_factor_2} both before and after relaxation of the time-averaged MD structure at $25$\,K. \textbf{b} The corresponding structure factors Eq.\,\eqref{eq:time_averaged_structure_factor_2} for $3$ different $\mathbf{q}=\mathrm{M}$ points as a function of temperature, both before and after the relaxation.}
    \label{fig:struct_factors}
\end{figure}

In Fig.\,\ref{fig:positions} we show a comparison of the normal phase positions with the positions of the time-averaged structure of the MDs at $25$\,K and $195$\,K, for both Ti and Se atoms, where the atomic positions of the original $56\times56\times1$ supercell are folded onto a $4\times4\times1$ supercell and then averaged. The numbers in black indicate the amplitudes of the atomic displacements in the $xy$ plane. For better visualization, displacements of Ti atoms are enhanced by a factor of $3$ and of Se atoms by a factor of $6$ if the displacement of an atom in the $xy$ plane is greater than $0.002~\text{\AA}$, otherwise displacements of Ti atoms are enhanced by a factor of $90$ and of Se atoms by a factor of $180$. In addition, normal phase Se atoms that appear below the middle Ti layer are colored in light colors and those Se atoms that are above the middle layer in dark colors. Differently colored circles represent the pairs of equally displaced Ti atoms. At $25$\,K we have stabilization of three non-equivalent Ti pairs ($\delta\mathrm{Ti_1}> \delta\mathrm{Ti_2} > \delta\mathrm{Ti_3}$) that produces the anisotropic $3Q$ pattern, while at $195$\,K we have $2Q$ nematic order with $\delta\mathrm{Ti_1}\approx \delta\mathrm{Ti_2} > \delta\mathrm{Ti_3}$.
Furthermore, as we use the time-averaged MD structure the corresponding structure factors are calculated in the following way
\begin{align}
    S_{\mathrm{avg}}(\mathbf{q}, T) = \frac{1}{N^{2}_{\mathrm{atoms}}}\left|\sum_{j}e^{-i\mathbf{q}\cdot\langle\mathbf{R}_{j}\rangle}\right|^{2}~,\label{eq:time_averaged_structure_factor_2}
\end{align}
where $\displaystyle \langle\mathbf{R}_{j}\rangle=\frac{1}{N_{\mathrm{steps}}}\sum_{t}\mathbf{R}_{j}(t)$, which differs from Eq.\,\eqref{eq:time_averaged_structure_factor} because we first time-average atomic positions before calculating the structure factor.
%
In Fig.\,\ref{fig:average_around_M} we show the average of time-averaged structure factors of nine different $\mathbf{q}$ points around $\mathbf{q}=\mathrm{M_{3}}$ from Fig.\,1c in the main text and Fig.\,\ref{fig:struct_factors}a, including the $\mathbf{q}=\mathrm{M_{3}}$ point. From these results we can see that around $200$\,K there is a postponement in the transition between CDW and the normal phase, which is in the range between $225$\,K and $250$\,K.
In Fig.\,1d in the main text we report the structure factors for three different $\mathbf{q}=\mathrm{M}$ points as functions of temperature calculated by Eq.\,\eqref{eq:time_averaged_structure_factor} (filled circles) and by Eq.\,\eqref{eq:time_averaged_structure_factor_2} (empty circles). In the bottom panel of the same figure, we report their difference given by
\begin{equation}
    \Delta S(\mathbf{q}, T) = \langle S(\mathbf{q}, T)\rangle - S_{\mathrm{avg}}(\mathbf{q}, T)~.\label{eq:diff_struct_factors}
\end{equation}
With this we are able to quantify the intensity of the CDW fluctuations. Moreover, the peak in $\Delta S(\mathbf{q}, T)$ in the CDW fluctuating region deviates from the background increase due to thermal population of phonons at this particular wave vector.
We projected positions from $56\times56\times1$ supercell MD simulations that we ran using the PBE data based MACE interatomic potential onto $4\times4\times1$ supercell, then averaged them over time to obtain average $4\times4\times1$ structures for each MD run at a different temperature. In Fig.\,\ref{fig:energy_diff} we report energies of these average structures.
In Fig.\,\ref{fig:displ} we show distances between the time-averaged positions of each Ti atom and Ti atoms from normal phase in the $xy$ plane given by Eq.\,\eqref{eq:average_distance} for several different temperatures.
In Fig.\,\ref{fig:struct_factors}a we show structure factors Eq.\,\eqref{eq:time_averaged_structure_factor_2} both before and after relaxation at $25$\,K, where we used both Ti and Se atoms in calculations. These are the standard relaxations without thermal fluctuations at $T\approx0$\,K, where the initial structures are the corresponding ones from the MD simulations. Furthermore, in Fig.\,\ref{fig:struct_factors}b we show structure factors Eq.\,\eqref{eq:time_averaged_structure_factor_2} for $3$ different $\mathbf{q}=\mathrm{M}$ points indicated in Fig.\,\ref{fig:struct_factors}a as a function of temperature, both before and after relaxation. In Fig.\,\ref{fig:struct_factors}a and Fig.\,\ref{fig:struct_factors}b the convergence criterion of the forces on atoms during relaxation was set to $10^{-3}$ eV/\AA. It is obvious that once you remove the thermal fluctuations and relax the averaged MD structures one does not obtain the anisotropic $3Q$ order.
\newpage
We calculated the time-averaged order parameter given by Eq.\,\eqref{eq:avg_order_param} from MD data which we ran using the PBE data based MACE interatomic potential for $110000$ steps with the first $10000$ steps treated as thermalization period for each temperature. In our calculations, we used every fifth step of thermalized period from each MD. From this we extracted $\bar{A}_{\alpha}\left(\mathbf{R}_{i}^{0}\right)$ and $\bar{\phi}_{\alpha}\left(\mathbf{R}_{i}^{0}\right)$ at the positions of all atoms.

\begin{figure}[H]
    \centering
    \includegraphics[width=\textwidth]{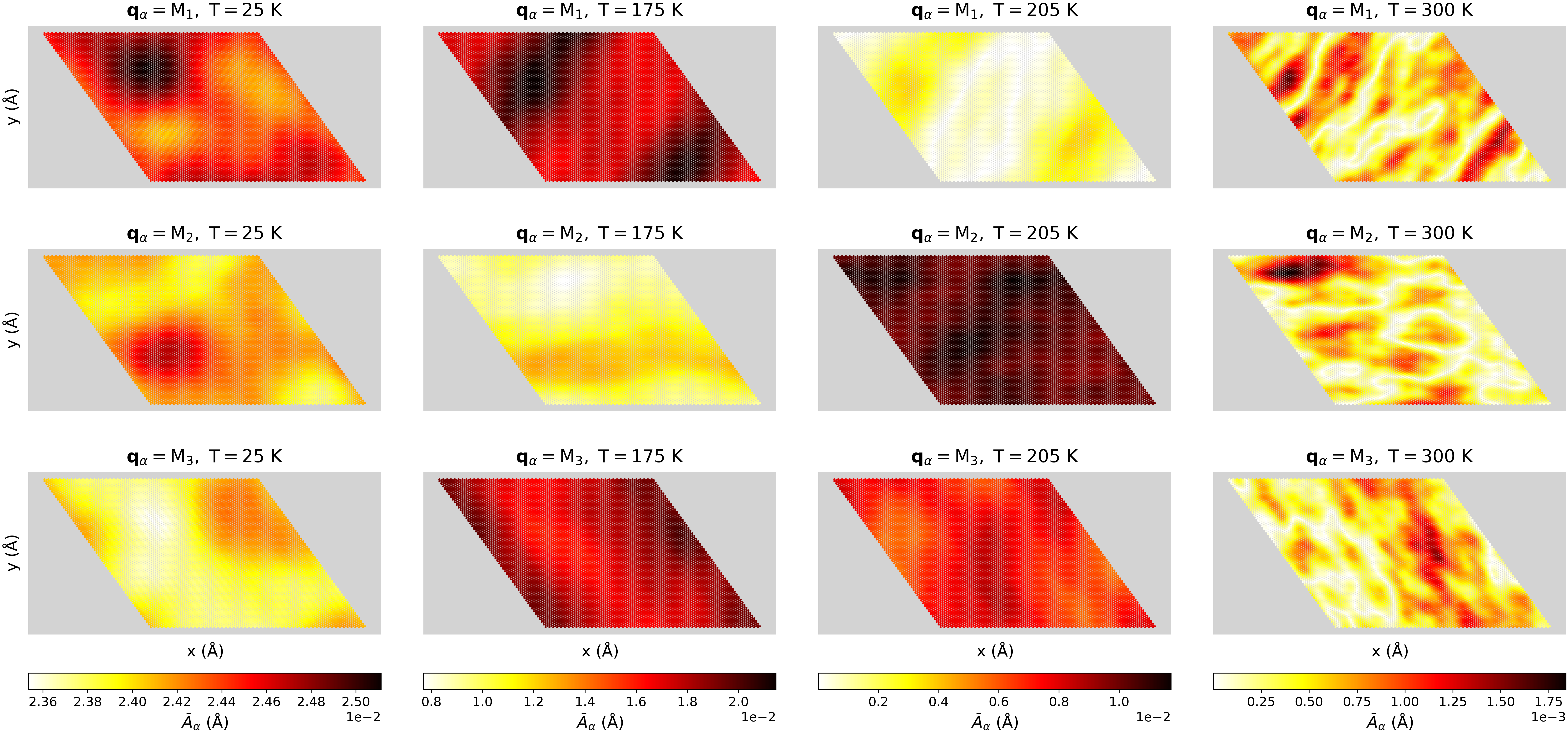}
    \caption{$\bar{A}_{\alpha}\left(\mathbf{R}_{i}^{0}\right)$ calculated from PBE based MD data for three different $\mathbf{q}=\mathrm{M}$ points for several different temperatures. Notice that a different color scale was used than in Fig.\,3a in the main text.}
    \label{fig:subplots_amplitudes}
\end{figure}

\begin{figure}[H]
    \centering
    \includegraphics[width=\textwidth]{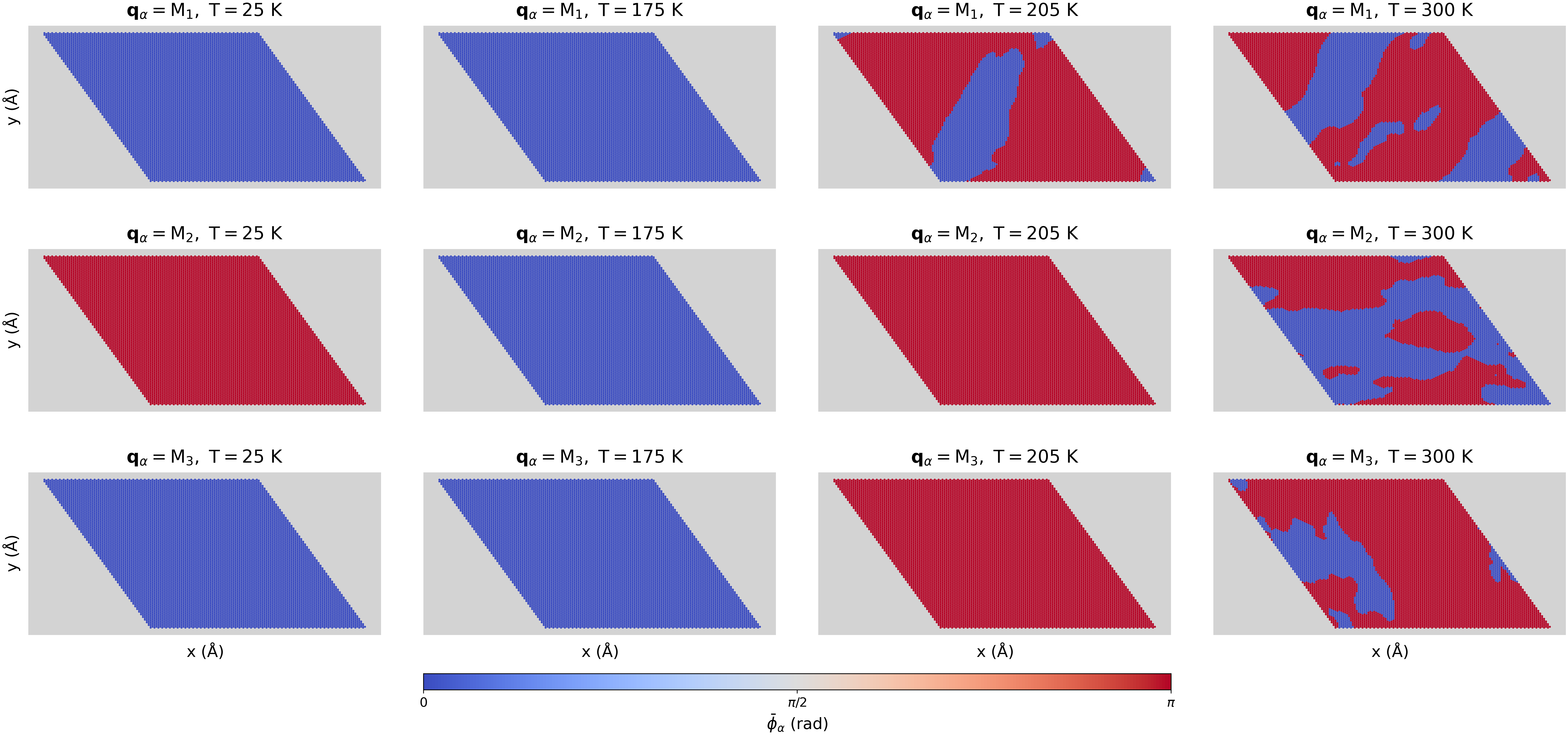}
    \caption{$\bar{\phi}_{\alpha}\left(\mathbf{R}_{i}^{0}\right)$ calculated from PBE based MD data for three different $\mathbf{q}=\mathrm{M}$ points for several different temperatures. The phase takes only the values $0$ and $\pi$.}
    \label{fig:subplots_phases}
\end{figure}

\begin{figure}[H]
    \centering
    \includegraphics[width=0.65\textwidth]{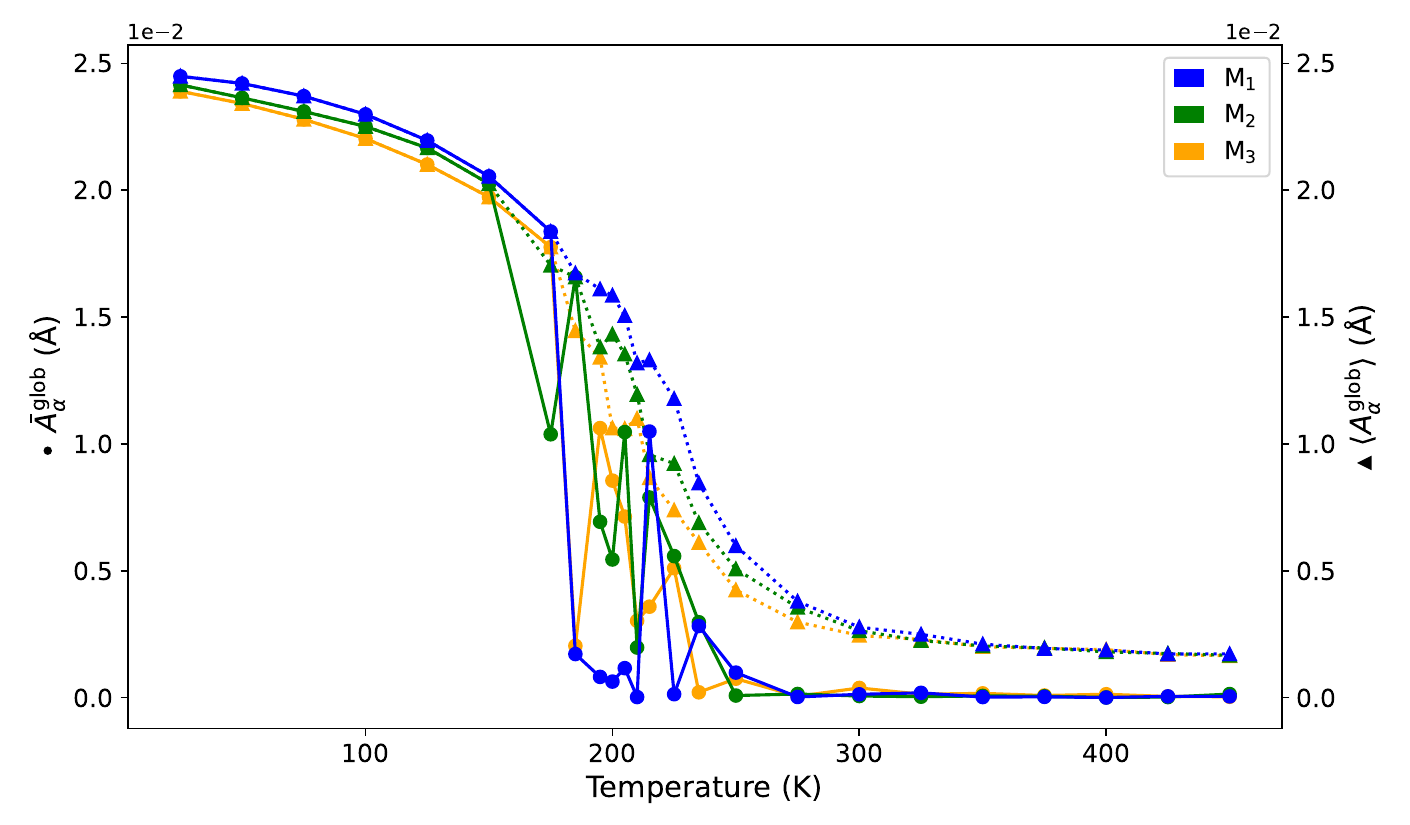}
    \caption{Amplitudes of the global order parameter Eq.\,\eqref{eq:global_order_par} calculated without the Gaussian as functions of temperature for three different $\mathbf{q}=\mathrm{M}$ points. Dots show $\bar{A}^{\mathrm{glob}}_{\alpha}$ and triangles show $\langle A^{\mathrm{glob}}_{\alpha}\rangle$ given by Eq.\,\eqref{eq:global_avg_ampl}.}
    \label{fig:amplitude_avg_no_gaussian}
\end{figure}

\begin{figure}[H]
    \centering
    \includegraphics[width=\textwidth]{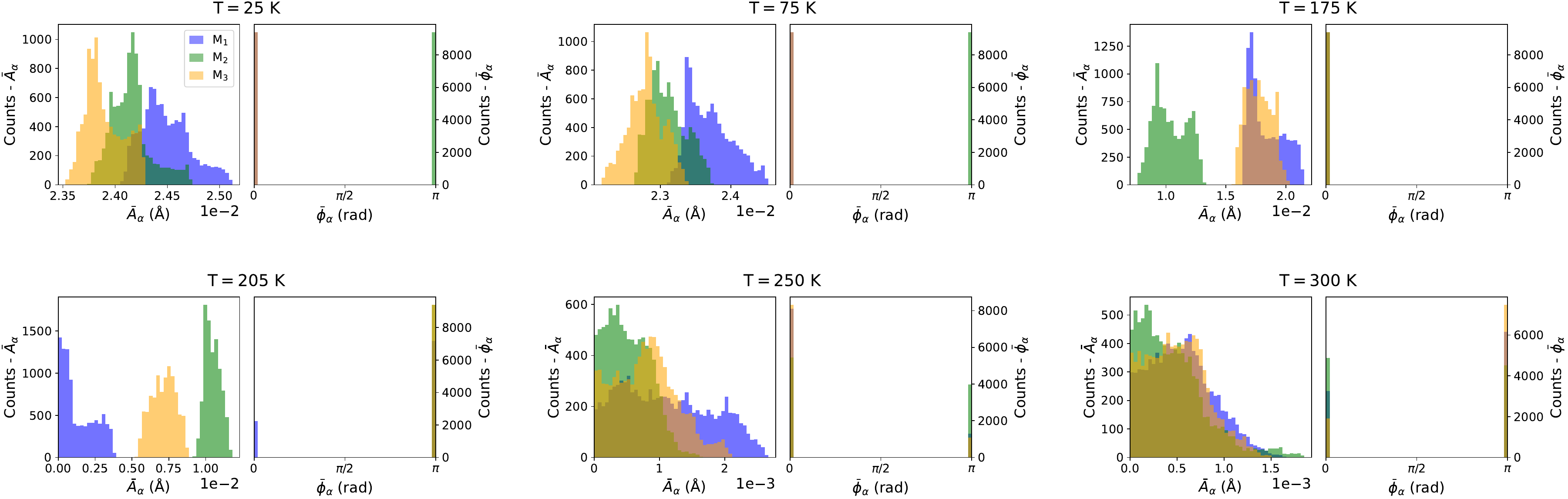}
    \caption{Distributions of $\bar{A}_{\alpha}\left(\mathbf{R}_{i}^{0}\right)$ and $\bar{\phi}_{\alpha}\left(\mathbf{R}_{i}^{0}\right)$ for all three $\mathbf{q}=\mathrm{M}$ points for several temperatures. Since the phase takes only the values $0$ and $\pi$, its distribution consists of two bins.}
    \label{fig:subplots_histograms}
\end{figure}

\begin{figure}[H]
    \centering
    \includegraphics[width=\columnwidth]{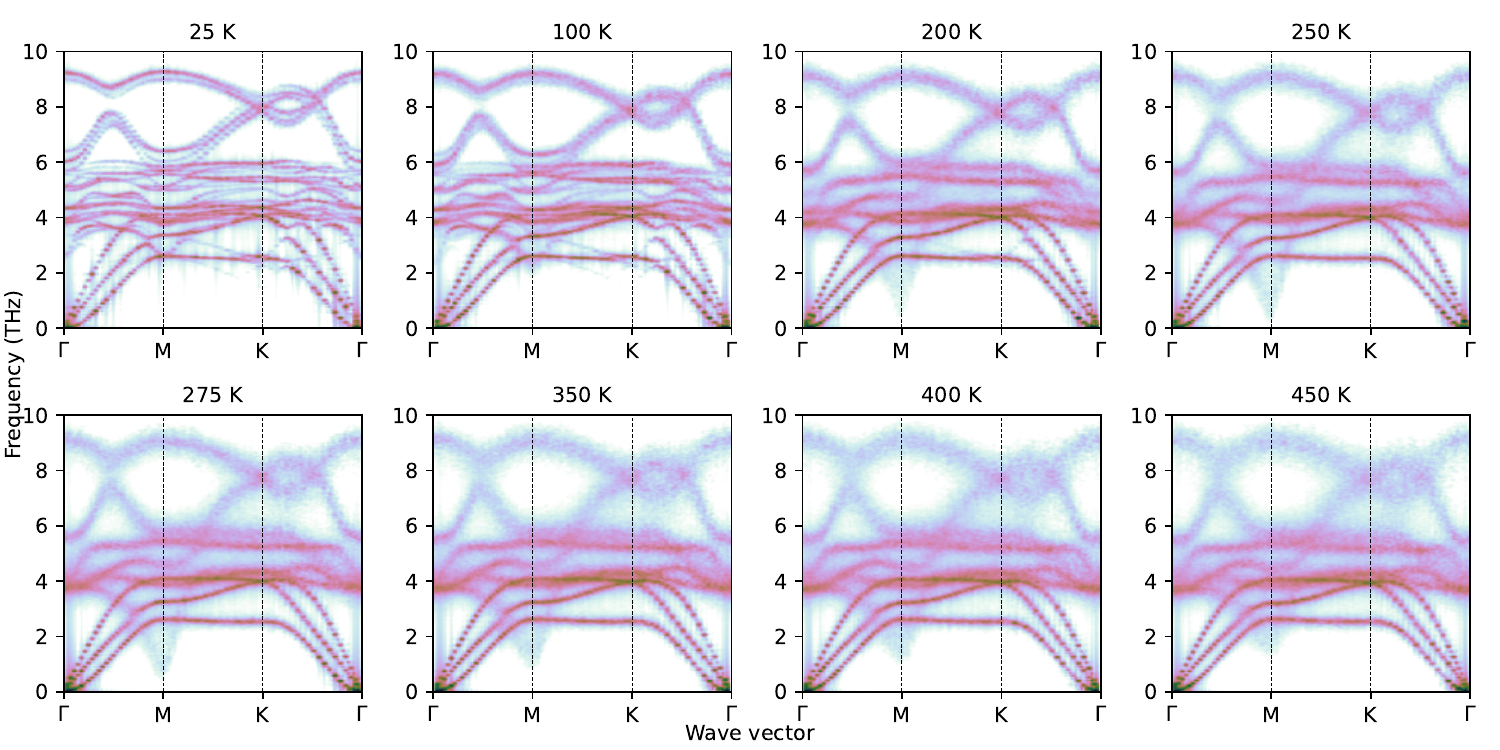}
    \caption{Phonon dispersions calculated from the MD data for several temperatures along the $\Gamma-\mathrm{M}-\mathrm{K}-\Gamma$ path using Eq.\,\eqref{eq:power_spectrum}.}
    \label{fig:phonon_dispersions}
\end{figure}

In Fig.\,\ref{fig:subplots_amplitudes} we show $\bar{A}_{\alpha}\left(\mathbf{R}_{i}^{0}\right)$ for all three $\mathbf{q}=\mathrm{M}$ points for several temperatures. In all images, we used both Ti and Se atoms. We see that at $25$\,K the amplitudes are uniform across the supercell. With an increase in temperature the amplitudes of the three components become different and spatially inhomogeneous, CDW ordering breaks down and we have domain formation. The lines along which $\bar{A}_{\alpha}\left(\mathbf{R}_{i}^{0}\right)$ vanishes are the domain walls between the regions with the opposite sign of the order parameter (compare with Fig.\,\ref{fig:subplots_phases}).
\newpage
In Fig.\,\ref{fig:subplots_phases} we show $\bar{\phi}_{\alpha}\left(\mathbf{R}_{i}^{0}\right)$ for all three $\mathbf{q}=\mathrm{M}$ points for several temperatures. In all images, we used both Ti and Se atoms. At $25$\,K and $175$\,K each component has a single value of the phase ($0$ or $\pi$) across the whole supercell. At $205$\,K the $\mathrm{M}_{1}$ component is split into domains with the phases $0$ and $\pi$, while the $\mathrm{M}_{2}$ and $\mathrm{M}_{3}$ components still have a single value of the phase, and at $300$\,K we have domain formation for all three components.
In Fig.\,\ref{fig:amplitude_avg_no_gaussian} we show the amplitudes of the global order parameter Eq.\,\eqref{eq:global_order_par}, which is calculated without the Gaussian, as functions of temperature for three different $\mathbf{q}=\mathrm{M}$ points. The results for $\bar{A}^{\mathrm{glob}}_{\alpha}$ closely follow the results for $\bar{A}_{\alpha}$ obtained from the local order parameter in Fig.\,3c in the main text. On the other hand, $\langle A^{\mathrm{glob}}_{\alpha}\rangle$ decreases to small values above $T_{\mathrm{CDW}}$, while $\langle A_{\alpha}\rangle$ obtained from the local order parameter remains finite, because the domains of opposite sign cancel in the global order parameter and contribute to the local one.
In Fig.\,\ref{fig:subplots_histograms} we show distributions of $\bar{A}_{\alpha}\left(\mathbf{R}_{i}^{0}\right)$ and $\bar{\phi}_{\alpha}\left(\mathbf{R}_{i}^{0}\right)$ for all three $\mathbf{q}=\mathrm{M}$ points for several temperatures. The results indicate that at low temperatures the amplitudes of the three components have narrow distributions centered at slightly different values, and the phase of each component is locked to a single value ($0$ or $\pi$) for all atoms. As the temperature increases the distributions of the amplitudes become broader and shift to lower values differently for different components, and this locking of the phases breaks, so that both values of the phase are present.
To calculate the phonon dispersions from the MD data we used the DynaPhoPy framework\,\cite{dynaphopy}, and details about the calculations are in the text above. In Fig.\,\ref{fig:phonon_dispersions} we show phonon dispersions obtained for several temperatures. We can see that the Kohn anomaly at $\mathbf{q}=\mathrm{M}$ softens as temperature increases and goes to $0$ THz around $250$\,K, after which it hardens and slowly disappears as the temperature increases. This agrees with all the results we report and indicates that our MD simulations predict $T_{\mathrm{CDW}}\approx250$\,K.

\begin{figure}[H]
    \centering
    \includegraphics[width=0.8\columnwidth]{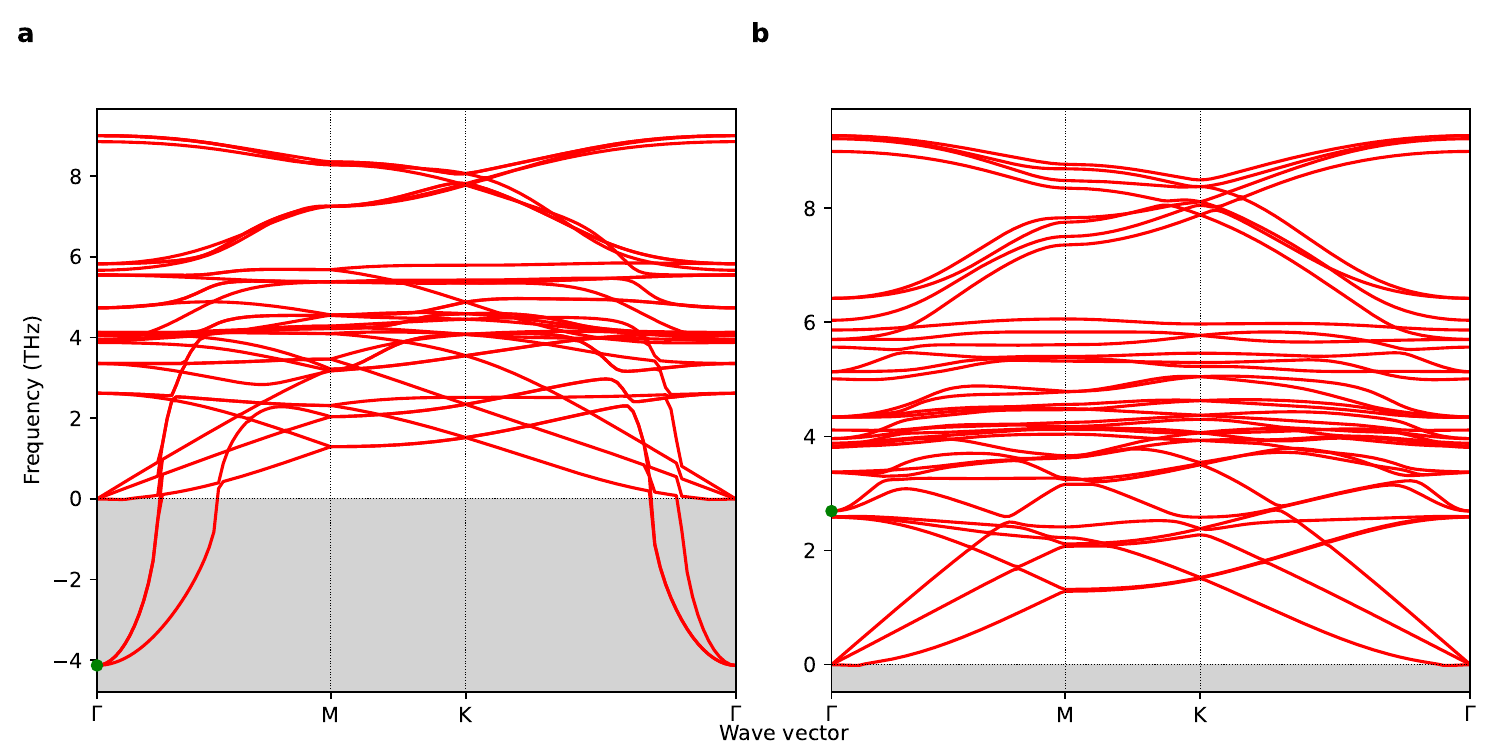}
    \caption{\textbf{a} Harmonic phonon dispersions for the $2\times2\times1$ normal phase structure without periodic lattice distortions. The green point is located at $-4.1278$ THz. \textbf{b} Harmonic phonon dispersions for the $2\times2\times1$ CDW phase structure (with periodic lattice distortions). The green point is located at $2.6909$ THz.}
    \label{fig:structure_2x2x1_harmonic_phonons}
\end{figure}

\begin{figure}[H]
    \centering
    \includegraphics[scale=0.5]{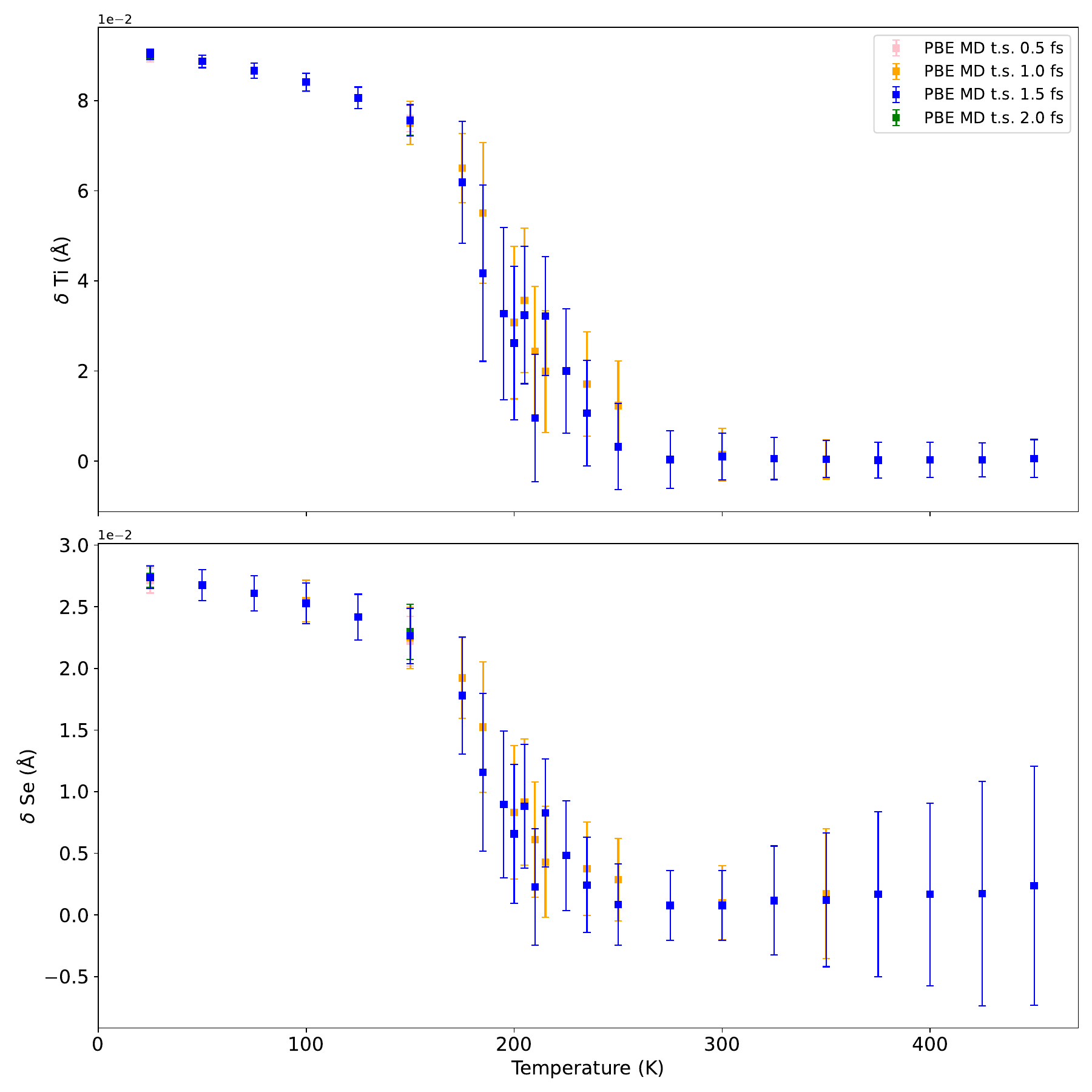}
    \caption{Average displacements and their standard deviations of Ti and Se atoms calculated from MD simulations run with the PBE data based MACE interatomic potential for different time steps.}
    \label{fig:diff_timesteps}
\end{figure}

\begin{figure}[H]
    \centering
    \includegraphics[width=0.65\columnwidth]{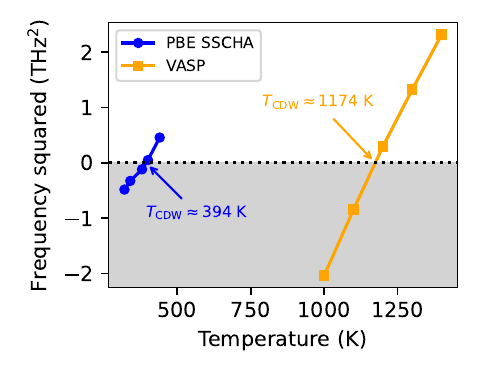}
    \caption{$\omega^{2}$ versus temperature comparison between DFT and SSCHA results, with $T_{\mathrm{CDW}}$ indicated in both cases.}
    \label{fig:omega_squared}
\end{figure}

\begin{figure}[H]
    \centering
    \includegraphics[width=0.55\columnwidth]{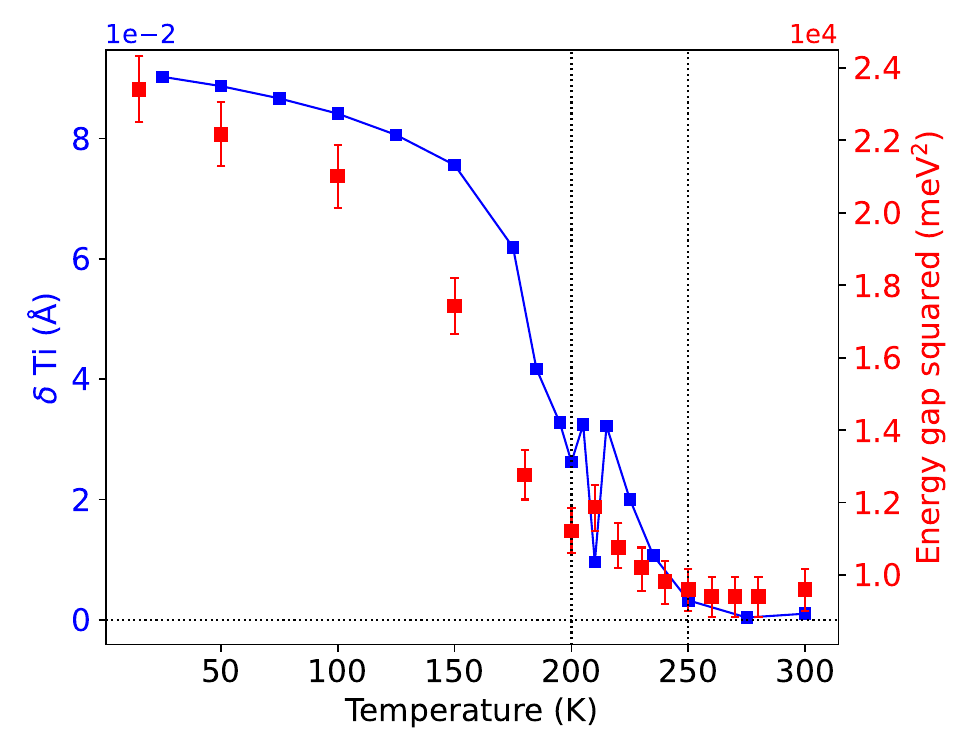}
    \caption{Comparison of the average displacements of Ti atoms in the $xy$ plane and the ARPES band gap measurements\,\cite{chen2016}.}
    \label{fig:dti_vs_gap}
\end{figure}

In Fig.\,\ref{fig:structure_2x2x1_harmonic_phonons} we show the results of harmonic phonon dispersions calculations in a $2\times2\times1$ cell for both normal and CDW phases. The green points at $\mathbf{q}=\mathrm{\Gamma}$ at $-4.1278$ THz in Fig.\,\ref{fig:structure_2x2x1_harmonic_phonons}a and at $2.6909$ THz in Fig.\,\ref{fig:structure_2x2x1_harmonic_phonons}b indicate the CDW amplitude mode. The result for the CDW phase with periodic lattice distortions agrees with the Raman measurements\,\cite{sugai1980} shown in Fig.\,4. This indicates that our MACE interatomic potential trained on the PBE DFT data correctly reproduces the frequency of the CDW amplitude mode in the low-temperature CDW phase.
In Fig.\,\ref{fig:diff_timesteps} we report average displacements and their standard deviations of Ti and Se atoms calculated by the method explained in Sec.\,\ref{subsec:average_atomic_displ} from MD simulation data obtained with the PBE data based MACE interatomic potential for different time steps. Blue points correspond to the setting with the time step set to $1.5$ fs. Additionally we ran MDs with time steps of $0.5$ fs and $2.0$ fs at $25$ K and $150$ K, and MD with a time step of $1.0$ fs was run with $13$ temperatures in total.
The transition temperature for monolayer TiSe$_{2}$ between the CDW phase and the normal phase is reported as $T_{\mathrm{CDW}}\approx 440$\,K\,\cite{bianco_tise2}. We used our PBE data based MACE interatomic potential to estimate $T_{\mathrm{CDW}}$ using SSCHA. We performed calculations in $8\times8\times1$ supercell made from normal phase unit cell. Most of the SSCHA parameters were kept at their default values. The initial harmonic dynamical matrices for the SSCHA calculations were obtained with the ASE, with a finite difference step of $0.01$ \AA, where we have relaxed the initial normal phase unit structure with the convergence criterion of the forces on atoms during relaxation set to $10^{-4}$ eV/\AA. Relaxation in SSCHA was performed with $2500$ configurations and, to generate ensembles, we used $10000$ configurations. The meaningful factor was set to $10^{-5}$. Both the dynamical matrix and the structure minimization steps were set to a value of $0.05$. From Fig.\,\ref{fig:omega_squared} we see that running SSCHA with our MACE interatomic potential gives $T_{\mathrm{CDW}}\approx 400$\,K, whereas VASP calculations give $T_{\mathrm{CDW}}\approx 1170$\,K.
To further corroborate that the phase transition in monolayer TiSe$_{2}$ is not of purely second order, in Fig.\,\ref{fig:dti_vs_gap} we compare the average displacements of Ti atoms in the $xy$ plane calculated from MD simulations which we ran using the PBE data based MACE interatomic potential and the ARPES band gap measurements\,\cite{chen2016}. We can see that experimental band gap results also show deviation from the purely second order transition below $T_{\mathrm{CDW}}$.
%

\bibliography{ref}